\documentclass[preprintnumbers,amsmath,amssymb]{revtex4}
\usepackage{amsmath}    
\usepackage{graphicx}   
\usepackage{color}
\newcommand{\compl}{{\mathbb C}}
\newcommand{\real}{{\mathbb R}}
\pdfoutput=1
\begin{document}

\title{The extended Bloch representation of quantum mechanics\\
and the hidden-measurement solution to the measurement problem
}

\author{Diederik Aerts}
\affiliation{Center Leo Apostel for Interdisciplinary Studies and Department of Mathematics, \\
Brussels Free University, Brussels, Belgium}\date{\today}
\email{diraerts@vub.ac.be}  
\author{Massimiliano Sassoli de Bianchi}
\affiliation{Laboratorio di Autoricerca di Base, 6914 Lugano, Switzerland}\date{\today}
\email{autoricerca@gmail.com}

\begin{abstract}

\noindent A generalized Bloch sphere, in which the states of a quantum entity of arbitrary dimension are geometrically represented, is investigated and further extended, to also incorporate the measurements. This extended representation constitutes a general solution to the measurement problem, inasmuch it allows to derive the Born rule as an average over hidden-variables, describing not the state of the quantum entity, but its interaction with the measuring system. According to this modelization, a quantum measurement is to be understood, in general, as a tripartite process, formed by an initial deterministic decoherence-like process, a subsequent indeterministic collapse-like process, and a final deterministic purification-like process. We also show that quantum probabilities can be generally interpreted as the probabilities of a first-order non-classical theory, describing situations of maximal lack of knowledge regarding the process of actualization of potential interactions, during a measurement. 
\\


\end{abstract}

\maketitle

\section{Introduction}
\label{Introduction}

One of the major problems of quantum mechanics, since its inception, has been that of explaining the origin of the statistical regularities predicted by its formalism. Simplistically, we could say that two diametrically  opposite approaches to this problem stand out: the \emph{instrumentalist} and the \emph{realist}. According to the former, the solution of the problem is equivalent to its elimination: quantum probabilities are not required to be further explained, as what really matters in a physical theory is its predictive power, expressed by means of a \emph{rule of correspondence} between the formalism of the theory and the results of the measurements, performed in the laboratories; and quantum mechanics is equipped with an extremely effective rule of this kind: the so-called \emph{Born rule}, first stated by Max Born in the context of scattering theory~\cite{Born1926}.

While for the instrumentalist (by virtue of necessity and because of the difficulty of finding a coherent picture) it is unnecessary, if not wrong, to explain the predictive power of the Born rule, for the realist explanation must precede prediction, and one cannot settle for simply checking that the Born rule makes excellent correspondences: one also has to explain the reason of such success, possibly deriving the rule from first principles, even if this is at the price of having to postulate the existence of new \emph{elements of reality}, which so far have remained \emph{hidden} to our direct observation, in accordance with Chatton's anti-razor principle: ``no less than is necessary.''~\cite{Smailing2005}

The main way to do this, is to create a \emph{model}, in which the different terms of the quantum formalism possibly find a correspondence, receiving in this way a better interpretation and explanation; and if the additional explanations contained in the model are  able to  produce new predictions, the model can also become a candidate for an upgraded version of the theory, providing a more refined  correspondence with the experiments, through which in turn the model can be tested and possibly refuted. 

Among the major obstacles that have prevented the development of new explicative models for quantum mechanics, and more specifically for quantum probabilities, there are the famous \emph{no-go theorems} about hidden-variables, which  restrict the permissible hidden-variable models explaining the origin of  quantum randomness~\cite{Neumann1932,Bell1966,Gleason1957,Jauch1963,Kochen1967,Gudder1970}. So much so that, over time, this has led many physicists to believe that the nature of quantum probabilities would be \emph{ontological}, and not \emph{epistemic}, that is, that they would be quantities not explainable as a condition of lack of knowledge about an objective  deeper  reality.

The no-go theorems, which all draw their inspiration from von Neumann's original proof~\cite{Neumann1932}, affirm that quantum probabilities cannot reflect a lack of knowledge about ``better defined states'' of a quantum entity, so that quantum observables would be interpretable as averages over the physical quantities deterministically associated with these hypothetical better defined states (much in the spirit of classical statistical mechanics). As a consequence, if quantum probabilities are explainable as a lack of knowledge about an underlying reality, such reality cannot be associated with an improved specification of the actual states of the quantum entities. 

Therefore, to bypass the obstacle of the no-go theorems, one must think of the hidden-variables not as elements of reality that would make a quantum mechanical state a more ``dispersion free'' state, but as something describing a different aspect of the reality of a quantum entity interacting with its environment, and in particular with a measuring system. This possibility was explored by one of us, in the eighties of the last century, by showing that if hidden-variables are associated, rather than with the state of the quantum entity, with its  interaction with the measuring system, one can easily derive the Born rule of correspondence and render useless the idea that quantum probabilities would necessarily have an ontological nature~\cite{Aerts1986}.

This preliminary 1986 study has generated over the years a number of works (see~\cite{Aerts1994,Coecke1995,Coecke1995b,Coecke1995c,Aerts1995,AertsAerts1997,AertsEtAl1997,Aerts1998,Aerts1998b,Aerts1999,AertsAerts2004,SvenAerts2005} and the references cited therein) further exploring the explicative power contained in this approach to the  measurement problem, today known as the \emph{hidden-measurement approach}, or \emph{hidden-measurement interpretation}. More precisely, the very natural idea that was brought forward at that time, and subsequently developed, is that in a typical quantum measurement the experimenter is in a situation of \emph{lack of knowledge} regarding the specific measurement interaction which is selected at each run of the measurement. And since these different potential measurement interactions would not  in general be equivalent, as to the change they induce on the state of the measured entity, they can produce different outcomes, although each individual interaction can be considered to act deterministically (or almost deterministically, and we will specify in the following in detail what we mean by `almost deterministically').  

We emphasize that this condition of lack of knowledge is not to be understood in a subjective sense, as it results from an \emph{objective} condition of \emph{lack of control} regarding the way a potential interaction is actualized during a measurement, as a consequence of the irreducible fluctuations inherent to the experimental context, and of the fact that the operational definition of the measured physical quantity  doesn't allow the experimental protocol to be altered, in order to reduce them~\cite{Sassoli2014}. 

The purpose of the present article is to put forward, for the first time, a complete self-consistent hidden-measurement modelization of a quantum measurement process, valid for arbitrary $N$-dimensional quantum entities, which will fully highlight the explicative power contained in the hidden-measurement interpretation. But to fully appreciate the novel aspects contained in this work, it will be useful to first recall what has been proven in the past, and what are the points that still needed to be clarified and elaborated. 

What was initially proved in~\cite{Aerts1986,Aerts1987}, is that hidden-measurement models could in principle be constructed for arbitrary quantum mechanical entities of finite dimension, and the possibility of constructing hidden-measurement models for infinite-dimensional entities was afterwards demonstrated by Coecke~\cite{Coecke1995b}. However, these proofs, although general, were only about that aspect of a measurement that we may call the ``naked measurement,'' corresponding to the description of the pure ``potentiality region'' of contact between the states of the entity under investigation and those describing the measuring apparatus. A measurement, however, is known to contain much more structure than just that associated with such ``potentiality region.'' 

What we are here referring to is the \emph{structure of the set of states} of the measured entity (which is Hilbertian for quantum entities, but could be non-Hilbertian for entities of a more general nature~\cite{AertsSassoli2014a,AertsSassoli2014b}), and how these states relate, \emph{geometrically}, to those describing the measuring system. This is what in the standard Hilbertian formalism is described by means of the so-called (Dirac) \emph{transformation theory}, which allows to calculate, for a given state, not only the probabilities associated with a single observable, but also those associated with all possible observables one may choose to measure. And of course, to obtain a complete description of a measurement process, also this additional geometric information, associated with the ``generalized rotations in Hilbert space,'' needed to be taken into account, and incorporated in the mathematical modelization. 

This, however, was only possible to do (until the present work) in the special situation of two-dimensional entities, like spin-${1\over 2}$ entities, and for higher-dimensional entities it was not at all obvious to understand how to transform the state relative to a given measurement context (defined by a given observable), when a different measurement context (defined by a different observable) was considered. 

This ``transformationally complete'' two-dimensional model has been extensively studied over the years, and is today known by different names. One of these names is \emph{spin quantum-machine}, with the term ``machine'' referring to the fact that the model is not just an abstract construct, but also the description of a macroscopic object that can be in principle constructed in reality, thus allowing to fully visualize how quantum and quantum-like probabilities arise. Another name for the model is \emph{$\epsilon$-model}~\cite{Aerts1998,Aerts1999,Sassoli2013b}, where the $\epsilon$ refers to a parameter in the model that can be continuously varied, describing the transition between quantum and classical measurements, passing through  measurement situations which are neither quantum nor classical, but truly intermediary. A third name is \emph{sphere-model}~\cite{AertsEtAl1997}, where the term ``sphere'' refers to the \emph{Bloch sphere}, the well known geometrical representation of the state space of a two-dimensional quantum entity (qubit).

In fact, the possibility of representing the full measurement process (not just its ``naked part'') of two-dimensional entities, in terms of hidden-measurement interactions, is related to the existence of a complete representation of the complex quantum states (the vectors in the two-dimensional Hilbert space ${\cal H} =\compl^2$) in a real two-dimensional unit sphere, or in a three-dimensional unit ball, if also density operators are considered. Such representation wasn't available for higher dimensional entities, and this was the reason why a complete representation for the full measurement process was still lacking.

In retrospect, we can say that this technical difficulty did not favor the spread of the hidden-measurement ideas, and possibly promoted a certain suspicion about the true reach of this interpretation, as a candidate to solve the measurement problem. In this regard, we can mention the fact that when presenting the spin machine-model to an audience, the objection was sometimes raised  that this kind of models could only be conceived for  two-dimensional quantum entities, because of Gleason's theorem~\cite{Gleason1957} and an article by Kochen and Specker~\cite{Kochen1967}. Indeed, Gleason's theorem is only valid for a Hilbert space with more than two dimensions, hence not for the two-dimensional complex Hilbert space that is used in quantum mechanics to describe the spin of a spin-${1\over 2}$ entity. And in addition to that, Kochen and Specker constructed in the above mentioned work a spin model for the spin of a spin-${1\over 2}$ entity, proposing also a real macroscopic realization for it, but also pointing out, on different occasions, that such a real model could only be constructed for a quantum entity with a Hilbert space of dimension not larger than two. 

Afterwards, some effort was given to clarify this dimensionality issue, and counter act the prejudice about the impossibility of a hidden-measurement model beyond the two-dimensional situation. In~\cite{Aertsetal1997}, for example, a mechanistic model was proposed for a macroscopic physical entity whose measurements give rise to a description in a three-dimensional (real) Hilbert space, a situation where Gleason's theorem is already fully applicable. However, although certainly sufficient to make the point of the \emph{non sequitur} of the no-go theorems in a simple and explicit example, the model was admittedly not particularly elegant, and a bit \emph{ad hoc}, and this may have prevented a full recognition of its consequences, as to the status of the hidden-measurement interpretation. 

In the same period, Coecke also proposed a more general approach, showing that a complete representation of the measurement process, and not just of its ``naked'' part, was possible also for a general $N$-dimensional quantum entity~\cite{Coecke1995}. This was undoubtedly an important progress, as for the first time it was possible to affirmatively answer the question about the existence of a generalization of the two-dimensional sphere-model to an arbitrary number of dimensions. However, although Coecke could successfully show that an Euclidean real representation of the complex states of a quantum entity was possible, and that in such representation the hidden-measurements could also be incorporated, the number of dimensions he used to do this was not optimal. Indeed, he represented a $N$-dimensional complex Hilbert space in a $N^2$-dimensional real Euclidean space, and for the $N=2$ case this gave an Euclidean representation in $\real^4$, whereas the Bloch sphere lives in $\real^3$. So, strictly speaking, Coecke's model wasn't the natural generalization of the sphere-model, but a different model whose mathematics was less immediate and the physics less transparent. 

To complete this short overview, a more recent work of Sven Aerts~\cite{SvenAerts2005} should also be mentioned, in which the author successfully formalized the hidden-measurement approach within the  general ambit of an \emph{interactive probability model}, showing how to characterize,  in a complex Hilbert space, the hidden-measurement scheme, deriving the Born rule from a \emph{principle of consistent interaction}, used to partition the apparatus' states,.

Now, for those physicists who from the beginning evaluated in a positive way the explicative power contained in the hidden-measurement interpretation, all the mentioned results incontrovertibly showed that there was a way to go to find more advanced models. But we can also observe that the approach remained difficult to evaluate by those who were less involved in these developments, mainly for the lack of a natural higher-dimensional generalization of the $N=2$ sphere-model representation, and the fact that it was known that the two-dimensional situation was, in a sense, a ``degenerate'' one, as it excluded the possibility of sub-measurements, and Gleason's theorem did not apply.

This situation started to change recently. Indeed, in the ambit of so-called \emph{quantum models of cognition and decision} (an emerging transdisciplinary field of research where quantum mechanics is intensively used and investigated~\cite{BusemeyerBruza2012,Aerts2009}) we could provide a very general mechanistic-like modelization of the ``naked part'' of a measurement process, including the possibility of describing degenerate observables, which is something that was not done in the past~\cite{AertsSassoli2014a,AertsSassoli2014b}. In that context, we also succeeded to show that the uniform average over the measurement interactions, from which the Born rule was derived,  could  be replaced by a much ampler averaging process, describing a much more general condition of lack of knowledge in a measurement, in what was called a \emph{universal measurements}. In other terms, what we could prove is that quantum measurements are interpretable as universal measurements having a Hilbertian structure, which in part could explain the great success of the quantum statistics in the description of a large class of phenomena (like for instance those associated with human cognition~\cite{Aerts2009,BusemeyerBruza2012,AertsSassoli2014a,AertsSassoli2014b}).

Once we completed this more detailed analysis of the ``potentiality region'' of a measurement process (which, as a side benefit, allowed us to propose a solution to the longstanding Bertrand's paradox~\cite{AertsSassoli2014c}), we became aware of the existence of some very interesting mathematical results, exploiting the generators of $SU(N)$ (the special unitary group of degree $N$) to generalize the Bloch representation of the states of a quantum entity to an arbitrary number $N$ of dimensions~\cite{Arvind1997, Kimura2003, Byrd2003, Kimura2005, Bengtsson2006, Bengtsson2013}. This was precisely the missing piece of the puzzle that we needed in order to complete the modelization of a quantum measurement process, by also including the entire structure of the state space. Contrary to the model proposed by Coecke, this generalized Bloch representation was carried out in a $(N^2-1)$-dimensional real Euclidean space, that is, a space with an optimal number of dimensions, which reduces exactly to the standard Bloch sphere (or ball) when $N=2$. In other terms, it is the natural generalization of the two-dimensional Bloch sphere representation.

Bringing together our recent results regarding the modelization of the ``naked part'' of a measurement process~\cite{AertsSassoli2014a,AertsSassoli2014b}, with the new mathematical results on the generalized Bloch  representation~\cite{Arvind1997, Kimura2003, Byrd2003, Kimura2005, Bengtsson2006, Bengtsson2013},  we are in a position to present, in this article, what we think is the natural $N$-dimensional generalization of the sphere-model, providing a self-consistent and complete modelization of a general finite-dimensional quantum measurement, also incorporating the full Hilbertian structure of the set of states, and the description of how the quantum entity enters into contact with the  ``potentiality region'' of  the measuring system, and subsequently  remerges from it, thus producing an outcome. To our opinion, the modelization has now reached a very clear physical and mathematical expression,  describing what possibly happens -- ``behind the macroscopic scene'' --  during a quantum measurement process, thus offering a challenging solution to the central (measurement) problem of quantum theory.

Before describing how the article is organized, a last remark is in order. The hidden-measurement interpretation can certainly be understood as a hidden-variable theory. However, it should not be understood as a tentative to resurrect classical physics. Quantum mechanics is here to stay, and cannot be replaced by classical mechanics. However, we also think that there are aspects of the theory which can, and need to, be demystified, and that only when this is done the truly deeper aspects of what the theory reveals to us, about our physical reality, can be fully appreciated. When hidden-measurements are used to explain how probabilities enter quantum mechanics, the measurement problem can be solved in a convincing way, and an explanation is given for that part of quantum physics. This, however, requires us to accept that quantum observations cannot be understood only as processes of pure discovery, and that the \emph{non-locality} of elementary quantum entities is in fact a manifestation of a more general condition of \emph{non-spatiality}, as it will be explained in the following of the article.

In Sec.~\ref{Operator-states}, we start by recalling some basic facts of the standard quantum formalism, emphasizing the difference between vector and operator states, and between  measurements of degenerate and non-degenerate observables. In Sec.~\ref{generators}, we introduce the generators of $SU(N)$ and recall their properties, whereas in Sec.~\ref{The generalized}, we explain how to use these generators to generalize the Bloch real space representation of the complex quantum states. 

In Sec.~\ref{Transition}, we analyze how the transition probabilities are expressed in terms of the real vectors of the generalized Bloch representation, shedding some light into the structure of the set of states, which contrary to the $N=2$ case, only corresponds to a small convex portion of the  $(N^2-1)$-dimensional unit ball. In Sec.~\ref{unitary}, we then show how  the vectors representative of states are affected by a deterministic -- norm preserving -- unitary  evolution.

The above sections can be considered as a preparation for the subsequent ones, where the \emph{generalized} Bloch representation will be \emph{extended} so as to  include also the measurements. In Sec.~\ref{HiddenmeasurementsN=2}, we start by describing the $N=2$ situation. This will allow us to introduce some of the important concepts, thus facilitating the understanding of the general $N$-dimensional situation. This more general situation will be presented in two steps: first, in Sec.~\ref{non-degenerate situation}, we analyze the simpler case of a non-degenerate observable, and then, in Sec.~\ref{degenerate-situation}, we generalize the analysis to also include degenerate outcomes. 

In Sec.~\ref{tripartite},  we summarize the obtained results, and emphasize that a quantum measurement process, when viewed from the hidden-measurement perspective, reveals a tripartite structure, which could not be evidenced in the standard quantum formalism, and which in principle is experimentally testable. We also emphasize in this section that if we take seriously the hidden-measurement interpretation and modelization, then density operators should also admit in quantum mechanics an interpretation as pure states, not  always referable to a statistical mixture of vector states. 

In Sec.~\ref{Nonspatiality}, we comment about the reach and richness of the hidden-measurement explanation, also with respect to the notion of \emph{non-spatiality}, that is implied by it, and which represents the truly novel aspect of the quantum revolution. 

In Sec.~\ref{Non-uniform-fluctuations}, we show that our extended Bloch representation can be further generalized to also admit experimental contexts characterized by \emph{non-uniform fluctuations}, giving rise to probability models which are neither Kolmogorovian nor Hilbertian. However, as we show in Sec.~\ref{universal}, when all these non-uniform forms of fluctuations are in turn averaged out, in what is called a \emph{universal measurement}, they produce an effective uniform distribution of potential hidden-measurement interactions, which yields back the Hilbertian Born rule, thus showing that the latter can be  generally interpreted as giving the probabilities of a first-order non-classical theory, describing situations of maximal lack of knowledge regarding the process of actualization of potential interactions.

Finally, in Sec.~\ref{The-infinite-dimensional-limit}, we briefly discuss the status of the hidden-measurement interpretation for infinite-dimensional entities, and in Sec.~\ref{Concluding} some final remarks are given. The content of the above sections gives shape to a rather long article. However, its length is justified, we think, by the importance of providing the readers with all those elements, technical and conceptual, that will enable them to fully appreciate the great clarification offered by the hidden-measurement interpretation, which we are convinced  contains some key ingredients for  understanding our physical reality, as it is revealed to us through our observations.

\section{Operator-states and L\"uders-von Neumann formula}
\label{Operator-states}

In standard quantum mechanics, the states of an entity  are the vectors $|\psi\rangle$ of a complex vector space ${\cal H}$, the so-called Hilbert space, which are normalized to unit and defined up to an arbitrary  global phase factor. Unless otherwise indicated, in this article we shall only consider finite-dimensional Hilbert spaces ${\cal H}\equiv{\mathbb C}^N$, with $N\geq 2$. All vectors of the form $e^{i\alpha}|\psi\rangle$, $\alpha\in {\mathbb R}$, belonging to a same \emph{ray} of ${\cal H}$,  describe the same  state, and are in correspondence with a one-dimensional (rank-one) orthogonal projection operator $P_\psi =|\psi\rangle\langle\psi|$, which is \emph{self-adjoint}, $P_\psi^\dagger = P_\psi$, \emph{idempotent}, $P_\psi^2 = P_\psi$, and of \emph{unit trace}: 
\begin{equation}
{\rm Tr}\, P_\psi = \sum_{i=1}^N \langle b_i|P_\psi|b_i\rangle = \sum_{i=1}^N \langle b_i|\psi\rangle\langle\psi|b_i\rangle =\sum_{i=1}^N |\langle b_i|\psi\rangle|^2 = \|\psi\|^2 = 1,
\end{equation}
where $\{|b_1\rangle,\dots,|b_N\rangle\}$ denotes an arbitrary orthonormal basis of ${\mathbb C}^N$. Clearly, $P_\psi$ is also \emph{positive semidefinite}, $\langle\phi| P_\psi|\phi\rangle =|\langle\phi|\psi\rangle|^2 \geq 0$, $\forall |\phi\rangle$, and its square is also of unit trace: ${\rm Tr}\, P_\psi^2 = {\rm Tr}\, P_\psi =1$.

\emph{Density operators} $D$, also called \emph{density matrices} and \emph{operator-states}, are a generalization of one-dimensional orthogonal projection operators $P_\psi$, in the sense that they can be written as  \emph{convex linear combinations} of an arbitrary number $n$ of one-dimensional orthogonal projection operators:
\begin{equation}
D=\sum_{i=1}^n p_i P_{\psi_i} = \sum_{i=1}^n p_i |\psi_i\rangle\langle\psi_i|, \quad \sum_{i=1}^n p_i =1, \quad p_i\geq 0, \,\,\forall i\in\{1,\dots,n\},
\label{stateoperator}
\end{equation}
where the normalized vectors $|\psi_i\rangle$ are not necessarily mutually orthogonal. A density operator $D$ defined by (\ref{stateoperator}) is manifestly a self-adjoint operator and, like one-dimensional orthogonal projection operators, it is of unit trace: 
\begin{equation}
{\rm Tr}\, D = {\rm Tr}\, \sum_{i=1}^n p_i P_{\psi_i} =  \sum_{i=1}^n p_i {\rm Tr}\, P_{\psi_i}= \sum_{i=1}^n p_i =1.
\end{equation}
It is also positive semidefinite, $\langle\phi| D|\phi\rangle \geq 0$, $\forall |\phi\rangle$, however, different from $P_\psi$, it is not  in general idempotent, so that ${\rm Tr}\, D^2 \leq 1$ (the minimum possible value of ${\rm Tr}\, D^2$ being ${1\over N}$). 

When not idempotent (i.e., when not a one-dimensional orthogonal projection operator), a density operator (\ref{stateoperator}) is usually interpreted as a \emph{classical statistical mixture} of states, describing a situation where there is  lack of knowledge about the specific state in which the entity is, and only the probabilities $p_i$ of finding it in a given state $P_{\psi_i}$ would be known. Definitely, an operator written as a convex linear combination of one-dimensional orthogonal projection operators admits such an interpretation. However, it cannot be taken in a too literal sense, considering that a same density operator can have infinitely many different representations  as  a mixture of one-dimensional projection operators. 

Just to give some simple examples, consider the two-dimensional ($N=2$) case, and the density operator $D={1\over 2}{\mathbb I}$, where ${\mathbb I}$ is the identity operator. Clearly, $D$ can be written as a convex linear combination of any pair of one-dimensional orthogonal projection operators producing a resolution of the identity. But it can also be written as a convex linear combination of three one-dimensional orthogonal projections; for instance: $D={1\over 3}(P_{\psi_1}+P_{\psi_2}+P_{\psi_3})$, with $|\psi_1\rangle =  |b_1\rangle$, $|\psi_2\rangle =  {1\over 2}|b_1\rangle + {\sqrt{3}\over 2}|b_2\rangle$, $|\psi_3\rangle =  {1\over 2}|b_1\rangle - {\sqrt{3}\over 2}|b_2\rangle$, with $\{|b_1\rangle,|b_2\rangle\}$ an arbitrary basis of ${\mathbb C}^2$. Another possibility is: $D={281\over 900}P_{\psi_1}+{97\over 450}P_{\psi_2}+{17\over 36}P_{\psi_3}$, with $|\psi_1\rangle =  {9\over \sqrt{281}}|b_1\rangle + {10\sqrt{2} i\over \sqrt{281}}|b_2\rangle$, $|\psi_2\rangle =  {12\over \sqrt{194}}|b_1\rangle +  {5\sqrt{2}i \over \sqrt{194}}|b_2\rangle$ and $|\psi_3\rangle =  {3i\over \sqrt{17}}|b_1\rangle + {2\sqrt{2}\over \sqrt{17}}|b_2\rangle$, and one can easily construct convex linear combinations involving arbitrarily many one-dimensional projections, with unequal probabilities~\cite{Hughston1993}.

The existence of an infinity of different possible representations for a same density operator $D$ immediately suggests that an interpretation of $D$ only as a classical mixture of states (usually referred to as a ``real mixture,'' or simply as a ``mixture of states'') is in general inappropriate. This becomes even more clear if one considers a compound entity $S$ formed by two sub-entities $S_1$ and $S_2$. Then, the Hilbert space ${\cal H}$ associated with $S$ is  the tensor product of the Hilbert spaces associated with the two sub-entities: ${\cal H}= {\cal H}_1\otimes {\cal H}_2$. If $|\psi\rangle \in {\cal H}$ is the state of $S$, and we are only interested in the description of, say, sub-entity $S_1$, irrespective of its possible correlations with sub-entity $S_2$, we can take the \emph{partial trace} of $P_\psi=|\psi\rangle\langle\psi|$ with respect to  ${\cal H}_2$, so defining the density operator $D_1 = {\rm Tr}_2\, P_\psi$, describing the state of sub-entity $S_1$, irrespective of its possible relations with sub-entity $S_2$. $D_1$ is clearly a density operator, considering that ${\rm Tr}_1D_1={\rm Tr}_1{\rm Tr}_2\, P_\psi= {\rm Tr}\, P_\psi=1$,  that it is manifestly self-adjoint and positive semidefinite, so that, by virtue of the spectral theorem, it can always be written as a convex linear combination of one-dimensional orthogonal projection operators (we recall that we limit our considerations to finite-dimensional Hilbert spaces).

Now, if we exclude the hypothetical state describing the entire physical reality, we observe that every entity is by definition a sub-entity of a larger composite entity, as is clear that every entity is in principle in contact with its environment. Thus, for all practical purposes, density operators are necessarily to be understood as \emph{approximate} states, as after all isolated entities only exist in an idealized sense. However, one of the interesting features of the measurement model that will be presented and analyzed in this article, is that it indicates that an interpretation of density operators as bona fide \emph{non-approximate pure states}, exactly like rank-one projection operators, is required if one wants to have a full representation of what goes on, in structural terms, when an entity is subjected to a measurement process. Therefore, in the present work we shall not a priori distinguish, in an ontological sense, states $P_\psi$, described by one-dimensional orthogonal projection operators, which will be called \emph{vector-states}, from density operators $D$, i.e., convex linear combinations of vector-states,  which  will be called \emph{operator-states} (a vector-state being of course a special case of an operator-state).  

Consider now an observable $A$ (a self-adjoint operator), which we assume for the moment being \emph{non-degenerate}. We can  write:
\begin{equation}
A=\sum_{i=1}^N a_i P_{a_i}, \quad P_{a_i} =|a_i\rangle \langle a_i|, \quad A|a_i\rangle  = a_i|a_i\rangle, \quad \langle a_i|a_j\rangle=\delta_{ij}, \quad i,j\in \{1,\dots,N\}.
\label{nondegenerate}
\end{equation}
The probability ${\cal P}(A=a_k|D)$, that the outcome of a measurement of $A$ is $a_k$, given that the entity was prepared in state $D$, is given by:
\begin{eqnarray}
\lefteqn{{\cal P}(A=a_k|D) = {\rm Tr}\, DP_{a_k} =  {\rm Tr}\,  \sum_{i=1}^n p_i P_{\psi_i} P_{a_k}= \sum_{i=1}^n p_i{\rm Tr}\,P_{\psi_i} P_{a_k} = \sum_{i=1}^n p_i {\rm Tr}\, |\psi_i\rangle\langle\psi_i| a_k\rangle\langle a_k|}\\
&&= \sum_{i=1}^n p_i  \langle\psi_i|a_k\rangle {\rm Tr}\, |\psi_i\rangle\langle a_k| = \sum_{i=1}^n p_i \langle \psi_i|a_k\rangle\sum_{j=1}^N \langle a_j|\psi_i\rangle\langle a_k|a_j\rangle = \sum_{i=1}^n p_i |\langle a_k|\psi_i\rangle|^2,
\label{probA=anondegenerate}
\end{eqnarray}
and of course, if $D=P_\psi$, the above reduces to: ${\cal P}(A=a_k|P_{\psi})=|\langle a_k|\psi\rangle|^2$.

According to \emph{L\"uders-von Neumann projection formula}, if the eigenvalue $a_k$ is obtained, we know that the measurement  has provoked a transition from the operator-state $D$, to the vector-state:
\begin{equation}
\label{Luder}
D_{a_k}\equiv  {P_{a_k} D P_{a_k}\over {\rm Tr}\, DP_{a_k}} = P_{a_k},
\end{equation}
where the equality follows from:
\begin{equation}
P_{a_k}DP_{a_k} =  \sum_{i=1}^n p_i | a_k\rangle\langle a_k|\psi_i\rangle\langle\psi_i| a_k\rangle\langle a_k|= \left(\sum_{i=1}^n p_i |\langle a_k|\psi_i\rangle|^2\right) | a_k\rangle\langle a_k| =  \left({\rm Tr}\, DP_{a_k}\right)P_{a_k}.
\end{equation}
This means that the probability ${\cal P}(A=a_k|D)$, that the outcome of a measurement of $A$ is $a_k$, can also be understood as the probability ${\cal P}(D\to D_{a_k})$, that the measurement context associated with $A$ provokes the state transition $D\to D_{a_k}$, and we  notice that, in accordance with von Neumann's ``first kind condition,'' we have: ${\cal P}(A=a_k|D_{a_k}) =1$.

If the observable $A$ is \emph{degenerate}, then a same eigenvalue can be associated with more than one eigenstate. To describe this situation, we consider $m$ disjoint subsets $I_{k}$ of $\{1,\dots,N\}$,  $k=1\dots,M$,  having $M_k$ elements each, with $0\leq M_k\leq N$, and $\sum_{k=1}^{M} M_k=N$, so that $\cup_{k=1}^{M} I_{k} =\{1,\dots,N\}$. We then assume that the eigenvectors $|a_i\rangle$, whose indexes belong to a same set $I_{k}$, are all associated with a same eigenvalue $a_{I_{k}}$,  $M_k$ times degenerate. Therefore, defining the $M_k$-dimensional orthogonal projector $P_{I_{k}}\equiv \sum_{i\in I_{k}} P_{a_i}$, projecting onto the $M_k$-dimensional eigenspace associated with the eigenvalue  $a_{I_{k}}$, we can write:
\begin{equation}
\label{degenerate}
A =\sum_{k=1}^{M} \sum_{i\in I_{k}} a_{I_{k}} P_{a_i} = \sum_{k=1}^{M} a_{I_{k}} P_{I_{k}},
\end{equation}
and of course (\ref{degenerate}) gives back (\ref{nondegenerate}) when each of the sets $I_{k}$ is a singleton $\{k\}$, i.e., a set containing the single element $k$, and consequently $M=N$. For a degenerate $A$, the probability that the outcome of a measurement is $a_{I_{k}}$,  given that the entity was prepared in state $D$, is then given by:
\begin{equation}
{\cal P}(A=a_{I_{k}}|D)  = {\rm Tr}\, DP_{I_k} =\sum_{i=1}^n p_i \sum_{j\in I_{k}}  |\langle a_j|\psi_i\rangle|^2.
\label{probA=adegenerate}
\end{equation}
The L\"uders-von Neumann projection formula also holds in the degenerate situation: if the  eigenvalue $a_{I_{k}}$ is obtained, we know that the measurement has provoked the transition from the operator-state $D$ to the state:
\begin{eqnarray}
\label{Luder-deg}
D_{I_k}\equiv  {P_{I_k} D P_{I_k}\over {\rm Tr}\, DP_{I_k}},
\end{eqnarray}
so that also in this case the probability ${\cal P}(A=a_{I_{k}}|D)$ corresponds to the probability ${\cal P}(D\to D_{I_k})$ that the measurement context associated with the degenerate observable $A$ provokes the state transition $D\to D_{I_k}$. This time, however, the post-measurement state $D_{I_k}$ will not in general be a vector-state, but an operator-state, unless the pre-measurement state is itself a vector-state, i.e., $D=P_\psi$. Indeed,  in this case  ${\rm Tr}\, DP_{I_k}=\|P_{I_k}|\psi\rangle\|^2$, and:
\begin{eqnarray}
D_{I_k} ={P_{I_k} |\psi\rangle\langle\psi| P_{I_k}\over \|P_{I_k}|\psi\rangle\|^2}=\left({P_{I_k} |\psi\rangle\over \|P_{I_k}|\psi\rangle\|}\right)\left({\langle\psi| P_{I_k}\over \|P_{I_k}|\psi\rangle\|}\right)=|\psi_{I_k}\rangle\langle\psi_{I_k}| =P_{\psi_{I_k}}.
\end{eqnarray}

It is worth emphasizing that the above description of a measurement, although formulated in Hilbert space and not explicitly mentioning the hidden-interactions, is fully compatible with the logic of the hidden-measurement interpretation. Indeed, a measurement context, associated with a given observable, can be understood as a collection of potential interactions, which once selected  (actualized) can bring a given initial state into a predetermined final state, corresponding to the outcome of the measurement. In other terms, the hidden-interactions are those elements of reality producing the quantum transition, so that, in a sense, we can say that the standard Hilbert space formulation of quantum mechanics already contains, in embryo, the hidden-measurement modelization.

We conclude this preliminary section by observing that a convex linear combination of operator-states is again an operator-state. Indeed, if $D = \lambda_1 D_1 + \lambda_2 D_2$, with $\lambda_1+\lambda_2 = 1$,  $\lambda_1,\lambda_2\geq 0$, and $D_1$ and $D_2$ are two operator-states, then $D$ is clearly self-adjoint, positive semidefinite and of unit trace: ${\rm Tr}\, D = \lambda_1 {\rm Tr}\,D_1 + \lambda_2 {\rm Tr}\,D_2 = \lambda_1+\lambda_2 = 1.$  In other terms, the set of operator-states of a quantum entity is \emph{convex}, and it is not difficult to show that its extremal points are necessarily the vector-states, i.e., the one-dimensional orthogonal projection operators.

\section{The generators of $SU(N)$}
\label{generators}

Since operator-states are bounded operators, they are \emph{Hilbert-Schmidt operators} (this is of course trivially so in a finite Hilbert space) and the square of their Hilbert-Schmidt norm is: $\|D\|_{\rm HS}^2\equiv \sum_{i=1}^N\|D|b_i\rangle\|^2=\sum_{i=1}^N\langle b_i|D^\dagger D|b_i\rangle  =  {\rm Tr}\, D^2 \leq 1$, where the equality is reached when $D=P_\psi$. This means that the convex set of states of a quantum entity is localized in a complex ball $B_1(\compl^{2N})$ of unit radius in the Hilbert space ${\cal H}_{\rm HS}=\compl^{2N}$ of Hilbert-Schmidt operators acting in ${\cal H}=\compl^{N}$ (isomorphic to ${\cal H}\otimes{\cal H}$), whose surface corresponds to the vector-states, and interior to the non-idempotent operator-states. Of course, not all the operators inside such unit ball are states, as for this they also need to be positive semidefinite. In other terms, only a portion of $B_1(\compl^{2N})$ is associated with states. 

In this section we will exploit the properties of the generators of the group $SU(N)$ to map that (convex) portion of the $2N$-dimensional \emph{complex} ball $B_1(\compl^{2N})$, which contains the states, onto a convex portion of a $(N^2-1)$-dimensional \emph{real} ball $B_1(\real^{N^2 -1})$. The main advantage of this is that within the real ball it will be possible to provide a full description of the quantum measurement process, by representing not only the states, but also the measurement  interactions, and how these interactions can produce an indeterministic change of the state of the entity. 

The reason for the number $N^2-1$, for the dimension of the real ball, is easily explained. An operator-state is a $N\times N$ complex matrix. Being self-adjoint, the $N$ elements forming its diagonal (in a given basis) are real numbers. These numbers must  sum up to $1$, so  we only need $N-1$ real parameters to specify the diagonal elements. Regarding the off-diagonal elements, the upper ones being the complex conjugate of the lower ones, we only have to determine ${1\over 2}(N^2-N)$ complex numbers, i.e., $N^2-N$ real numbers. Thus, the number of real parameters needed to specify a general operator-state $D$ is: $N^2-N + (N-1)= N^2-1$.

Consider now a $N\times N$ complex matrix $U$, describing a symmetry transformation. Conservation of the scalar product under symmetry transformations requires that: $U^\dagger U = UU^\dagger={\mathbb I}$. Since $\det U^\dagger = (\det U)^*$, this implies that $|\det U|=1$. An overall phase factor being not observable, one can fix the phases of the transformation matrices $U$ by requiring that $\det U=1$. Then, one can easily prove that the ensemble of matrices with the above properties forms a group, called the \emph{special unitary group of degree $N$}, denoted $SU(N)$. 

A $N\times N$ complex matrix requires $2N^2$ real numbers to be specified. However, for $SU(N)$ matrices, the unitary condition plus the determinant condition produce $N^2+1$ relations, so that one only needs $2N^2 -(N^2+1)=N^2-1$ real parameters to specify a $SU(N)$ matrix, precisely the same number of real parameters needed to specify an operator-state. 

Since unitary operators can be written in exponential form, a general $SU(N)$ matrix can be written as  $U=\exp(i \,{\bf u}\cdot{\bf \Lambda})=\exp(i \sum_{i=1}^{N^2-1}u_i \Lambda_i)$, where  $ {\bf u}$ is a $(N^2-1)$-dimensional real vector (vectors are denoted in bold) and ${\bf \Lambda}$  is a $(N^2-1)$-dimensional vector whose components are the self-adjoint $N\times N$ matrices $\Lambda_i$. Considering that $\det U=\det (\exp i \,{\bf u}\cdot{\bf \Lambda}) = \exp (i\,{\rm Tr}\, {\bf u}\cdot{\bf \Lambda})=\exp (i\sum_{i=1}^{N^2-1}u_i {\rm Tr}\, \Lambda_i)  =1$, we must have ${\rm Tr}\, \Lambda_i =0$, for all $i=1,\dots,N^2-1$. 

The $N^2-1$ traceless self-adjoint $N\times N$ matrices $\Lambda_i$ are called the \emph{generators} of $SU(N)$. It is convenient to choose them to be mutually \emph{orthogonal}, in the sense of the Hilbert-Schmidt scalar product: $(\Lambda_i |\Lambda_j)_{\rm HS}={\rm Tr}\, \Lambda_i\Lambda_j=0$, if $i\neq j$. Also, considering that the $\Lambda_i$ are traceless, we have: $(\Lambda_i |{\mathbb I})_{\rm HS}={\rm Tr}\, \Lambda_i=0$, for all $i$. This means that the $N^2$ matrices $\{{\mathbb I},\Lambda_1,\dots,\Lambda_{N^2-1}\}$ form an \emph{orthogonal basis} for the set of all linear operators on ${\mathbb C}^N$. Conventionally, their normalization is usually chosen to be: $\|\Lambda_i\|_{\rm HS}^2  =  {\rm Tr}\, \Lambda_i^2 = 2$, for all $i=1,\dots,N^2-1$. 

We can observe that the commutators $[\Lambda_i,\Lambda_j]_-$, and anticommutators $[\Lambda_i,\Lambda_j]_+$, are  self-adjoint operators, and therefore can be expanded on the basis of the generators, so that we can write: 
\begin{equation}
[\Lambda_i,\Lambda_j]_-=2 i \sum_{k=1}^{N^2-1} f_{ijk}\Lambda_k, \quad [\Lambda_i,\Lambda_j]_+={4\over N}\delta_{ij} \mathbb{I} +2 \sum_{k=1}^{N^2-1} d_{ijk}\Lambda_k.
\label{commutationNxN}
\end{equation}
Multiplying these equations by $\Lambda_k$, then taking the trace and using ${\rm Tr}\, \Lambda_i\Lambda_j=2\delta_{ij}$, we obtain that the structure constants are given by:
\begin{equation}
f_{ijk}={1\over 4i} {\rm Tr}\, [\Lambda_i,\Lambda_j]_- \Lambda_k, \quad d_{ijk}={1\over 4} {\rm Tr}\, [\Lambda_i,\Lambda_j]_+ \Lambda_k,
\label{structureconstants}
\end{equation}
implying that the $f_{ijk}$ (resp. the $d_{ijk}$) are elements of a totally antisymmetric (resp. symmetric) tensor.

Considering the above properties of the $\Lambda_i$ matrices, it is not difficult to systematically construct them, for an arbitrary $N$. Given an orthonormal basis $\{|b_1\rangle,\dots,|b_N\rangle\}$, a convenient construction is~\cite{Hioe1981, Alicki1987, Mahler1995}: $\{\Lambda_i\}_{i=1}^{N^2-1}=\{U_{jk},V_{jk},W_{l}\}$, where:
\begin{eqnarray}
\label{rgeneratorsN}
&&U_{jk}=|b_j\rangle\langle b_k| + |b_k\rangle\langle b_j|, \quad V_{jk}=-i(|b_j\rangle\langle b_k| - |b_k\rangle\langle b_j|),\\
&&W_l =-\sqrt{2\over l(l+1)}\left(l|b_{l+1}\rangle\langle b_{l+1}| - \sum_{j=1}^l |b_j\rangle\langle b_j|\right),\label{generatorsW}\\
&&1\leq j < k\leq N, \quad 1\leq l\leq N-1.
\end{eqnarray}

It is important to observe that,  even if we have fixed their normalization and  chosen them to be orthogonal, the generators $\Lambda_i$ are not for this uniquely determined. Indeed, if $Q\in O(N^2-1)$ is a $(N^2-1)\times (N^2-1)$ orthogonal matrix, then the $\Lambda'_i$ obtained through an orthogonal transformation ${\bf \Lambda}' = Q {\bf \Lambda}$, are also orthogonal generators of $SU(N)$, as they are clearly self-adjoint (orthogonal matrices have real entries), ${\rm Tr}\,\Lambda'_i=\sum_{j=1}^{N^2-1}Q_{ij} {\rm Tr}\,\Lambda_j =0$, and ${\rm Tr}\,\Lambda'_i\Lambda'_j=\sum_{k,l=1}^{N^2-1}Q_{ik} Q_{jl}{\rm Tr}\,\Lambda_k\Lambda_l =2\sum_{k=1}^{N^2-1}Q_{ik} Q_{jk}=2(QQ^T)_{ij}=2\delta_{ij}$.

We also observe that, given the canonical basis $\{{\bf n}_1^c,\dots,{\bf n}_{N^2-1}^c\}$ of $\mathbb{R}^{N^2-1}$, with components $({\bf n}_i^c)_j=\delta_{ij}$,  we can write: $\Lambda_i ={\bf \Lambda}\cdot {\bf n}_i^c\equiv \Lambda_{{\bf n}_i^c}$. Then, if we consider a basis  $\{{\bf n}'_1,\dots,{\bf n}'_{N^2-1}\}$, with ${\bf n}'_i =Q^T {\bf n}_i^c$, we have: $\Lambda'_i =Q {\bf \Lambda}\cdot {\bf n}_i^c = {\bf \Lambda}\cdot Q^T {\bf n}_i^c\equiv \Lambda_{{\bf n}'_i}$. So, to every basis of $\mathbb{R}^{N^2-1}$, we can associate a specific determination of the generators, such that each generator is obtained by performing the scalar product of ${\bf \Lambda}$ with a different vector of the basis. And this in particular means that given an arbitrary unit vector  ${\bf n}$, the component $\Lambda_{{\bf n}}={\bf \Lambda}\cdot {\bf n}$ can always be considered as an element of a set of generators of $SU(N)$.

For the two-dimensional case ($N=2$), the generators are $2^2-1=3$, and they correspond to the well-known Pauli's spin matrices, usually denoted $\sigma_i$, $i=1,2,3$:
 \begin{equation}
\sigma_1=
\left[ \begin{array}{cc}
0 & 1 \\
1 & 0 \end{array} \right],\quad
\sigma_2=
\left[\begin{array}{cc}
0 & -i \\
i & 0 \end{array} \right],\quad 
\sigma_3=
\left[ \begin{array}{cc}
1 & 0 \\
0 & -1 \end{array} \right],
\end{equation}
which obey the commutation relations: $[\sigma_i,\sigma_j]_-=2i \sum_{k=1}^3\epsilon_{ijk} \sigma_k$, and $[\sigma_i,\sigma_j]_+=2\mathbb{I}\delta_{ij}$, where $\epsilon_{ijk}$ is the  \emph{Levi-Civita} symbol. 
For the three-dimensional case ($N=3$), the generators are $3^2-1=8$, and correspond to the so-called \emph{Gell-Mann} matrices, usually denoted $\lambda_i$, $i=1,\dots,8$:
\begin{eqnarray}
&&\lambda_1 = \begin{bmatrix} 0 & 1 & 0 \\ 1 & 0 & 0 \\ 0 & 0 & 0 \end{bmatrix},\quad \lambda_2 = \begin{bmatrix} 0 & -i & 0 \\ i & 0 & 0 \\ 0 & 0 & 0 \end{bmatrix},\quad
\lambda_3 =\begin{bmatrix} 1 & 0 & 0 \\ 0 & -1 & 0 \\ 0 & 0 & 0 \end{bmatrix},\quad \lambda_4 = \begin{bmatrix} 0 & 0 & 1 \\ 0 & 0 & 0 \\ 1 & 0 & 0 \end{bmatrix},\\
&&\lambda_5 = \begin{bmatrix} 0 & 0 & -i \\ 0 & 0 & 0 \\ i & 0 & 0 \end{bmatrix},\quad
\lambda_6 = \begin{bmatrix} 0 & 0 & 0 \\ 0 & 0 & 1 \\ 0 & 1 & 0 \end{bmatrix},\quad \lambda_7 = \begin{bmatrix} 0 & 0 & 0 \\ 0 & 0 & -i \\ 0 & i & 0 \end{bmatrix},\quad
\lambda_8 = \frac{1}{\sqrt{3}} \begin{bmatrix} 1 & 0 & 0 \\ 0 & 1 & 0 \\ 0 & 0 & -2 \end{bmatrix},
\label{generatorsN=3}
\end{eqnarray}
with non-vanishing independent structure constants:
\begin{eqnarray}
&f_{123}=1,\, f_{458}= f_{678} = {\sqrt{3}\over 2},\,\, f_{147}=f_{246}=f_{257}=f_{345}=f_{516}=f_{637}={1\over 2},\\
&d_{118}=d_{228}=d_{338} = -d_{888} = {1\over\sqrt{3}}, \,\, d_{448}=d_{558}= d_{668}=d_{778}-{1\over 2\sqrt{3}},\\
&d_{146}=d_{157}=-d_{247}=d_{256}=d_{344}=d_{355}=-d_{366}=-d_{377}={1\over 2}.
\end{eqnarray}

\section{The generalized Bloch sphere}
\label{The generalized}

As we have seen, the $N^2$ matrices $\{{\mathbb I},\Lambda_1,\dots,\Lambda_{N^2-1}\}$ form an \emph{orthogonal basis} for the set of all linear operators on ${\mathbb C}^N$. This means that any state $D$ can be written as a linear combination:
\begin{equation}
D({\bf r}) =  {1\over N}\left(\mathbb{I} + c_N\sum_{i=1}^{N^2-1} r_i \Lambda_i\right) = {1\over N}\left(\mathbb{I} +c_N\, {\bf r}\cdot\mbox{\boldmath$\Lambda$}\right),
\label{formulaNxN}
\end{equation}
where we have defined $c_N\equiv \sqrt{N(N-1)\over 2}$. We observe that if all the components $r_i$ of the $(N^2-1)$-dimensional vector ${\bf r}$ are chosen to be real numbers, then $D({\bf r})$ is manifestly a self-adjoint operator. Also, considering that the $\Lambda_i$ are traceless, we have ${\rm Tr}\, D({\bf r}) = {\rm Tr}\,{1\over N}\mathbb{I}=1$. This however is not sufficient to guarantee that an operator $D({\bf r})$, written as the real linear combination (\ref{formulaNxN}), is an operator-state. Indeed, for this it must also be positive semidefinite, but this will not be the case for any choice of a vector ${\bf r}\in\mathbb{R}^{N^2-1}$. Just to give an example, consider a unit vector such that $r_i=0$, for all $i=1,\dots,N^2-2$, and $r_{N^2-1}=1$. Then, $D(0,\dots,0,1)= {1\over N}\left(\mathbb{I} + c_N\Lambda_{N^2-1}\right)$, and according to (\ref{generatorsW}) $\Lambda_{N^2-1}=W_{N-1}$, so that:
\begin{equation}
D(0,\dots,0,1) =  -{N-2\over N} |b_N\rangle\langle b_N|+ {2\over N} \sum_{j=1}^{N-1} |b_j\rangle\langle b_j|,
\label{negativeeigenvalue}
\end{equation}
which for $N\geq 3$ is clearly a matrix with a strictly negative eigenvalue $-{N-2\over N}$; thus, it is not positive semidefinite and cannot be representative of an operator-state.

To characterize the set of vectors ${\bf r}\in\mathbb{R}^{N^2-1}$ which are representative of bona fide states, it is useful to define the following two vector products in $\mathbb{R}^{N^2-1}$:
\begin{equation}
({\bf u}\star {\bf v})_i = {c_N\over N-2}\sum_{j,k=1}^{N^2-1}d_{ijk} u_jv_k, \quad ({\bf u}\wedge {\bf v})_i = \sum_{j,k=1}^{N^2-1}f_{ijk} u_jv_k,
\label{productsnxn}
\end{equation}
which are clearly symmetric (${\bf u}\star {\bf v} = {\bf v}\star {\bf u}$) and antisymmetric (${\bf u}\wedge {\bf v}=-{\bf v}\wedge {\bf u}$), considering that the structure constants $d_{ijk}$ and $f_{ijk}$ are totally symmetric and  antisymmetric tensors, respectively. Let us assume that $D({\bf r})$ is a vector-state, i.e., a one-dimensional orthogonal projection operator. Considering that it is of unit trace, this is equivalent to asking that it is idempotent: $D^2({\bf r})=D({\bf r})$. According to (\ref{formulaNxN}), we have: 
\begin{eqnarray}
\lefteqn{D^2({\bf r}) = {1\over N^2}\left(\mathbb{I} + c_N\, {\bf r}\cdot\mbox{\boldmath$\Lambda$}\right)\left(\mathbb{I} + c_N\, {\bf r}\cdot\mbox{\boldmath$\Lambda$}\right)= {1\over N^2}\mathbb{I} +  {2\over N^2} c_N\,{\bf r}\cdot\mbox{\boldmath$\Lambda$} +{c_N^2\over N^2}\,({\bf r}\cdot\mbox{\boldmath$\Lambda$})^2}\nonumber \\
&=&{1\over N}\left(\mathbb{I} + c_N\, {\bf r}\cdot\mbox{\boldmath$\Lambda$}\right)  -  {N-1\over N^2}\left(\mathbb{I} +{N-2\over N-1}c_N\,{\bf r}\cdot\mbox{\boldmath$\Lambda$} -{N\over 2}  ({\bf r}\cdot\mbox{\boldmath$\Lambda$})^2 \right).
\label{Dsquare}
\end{eqnarray}

In order for $D({\bf r})$ to be a one-dimensional orthogonal projector, the second term of (\ref{Dsquare}) has to vanish, i.e., we must have: $({\bf r}\cdot\mbox{\boldmath$\Lambda$})^2 = {2\over N}\mathbb{I} +{N-2\over c_N}{\bf r}\cdot\mbox{\boldmath$\Lambda$}$. According to (\ref{commutationNxN}) and (\ref{productsnxn}), we can write:
\begin{eqnarray}
({\bf r}\cdot\mbox{\boldmath$\Lambda$})^2&=&\sum_{i,j=1}^{N^2-1}r_ir_j\Lambda_i\Lambda_j=\sum_{i,j=1}^{N^2-1}r_ir_j\left(i \sum_{k=1}^{N^2-1} f_{ijk}\Lambda_k+ \sum_{k=1}^{N^2-1} d_{ijk}\Lambda_k +{2\over N}\delta_{ij} \mathbb{I}  \right)\\
&=&{2\over N}\sum_{i,j=1}^{N^2-1}r_ir_j\delta_{ij}\,\mathbb{I} + i\sum_{k=1}^{N^2-1} \Lambda_k \sum_{i,j}f_{ijk}r_ir_j + \sum_{k=1}^{N^2-1} \Lambda_k \sum_{i,j=1}^{N^2-1}d_{ijk}r_ir_j\\
&=&{2\over N} \|{\bf r}\|^2\, \mathbb{I} + i\sum_{k=1}^{N^2-1} \Lambda_k ({\bf r}\wedge{\bf r})_k + {N-2\over c_N}\sum_{k=1}^{N^2-1} \Lambda_k ({\bf r}\star {\bf r})_k\\
&=& {2\over N} \|{\bf r}\|^2\, \mathbb{I} + {N-2\over c_N} ({\bf r}\star {\bf r}) \cdot \mbox{\boldmath$\Lambda$}.
\end{eqnarray}
So, $D({\bf r})$ is a vector-state iff its representative vector in $\mathbb{R}^{N^2-1}$ is a unit vector, $\|{\bf r}\|=1$, obeying  ${\bf r}\star {\bf r} =  {\bf r}$. In other terms, state-vectors all live in a portion of the surface of a $(N^2-1)$-dimensional unit ball $B_1(\real^{N^2-1})$, i.e., on a portion of a $(N^2-2)$-dimensional unit sphere which is the generalization of the $2$-dimensional Bloch sphere (see below). Also, taking the trace of (\ref{Dsquare}), we find that ${\rm Tr}\, D^2({\bf r})=1-{N-1\over N}(1-\|{\bf r}\|^2)$, which means that  operator-states which are not vector-states, i.e., such that ${\rm Tr}\, D^2({\bf r})<1$, are associated with vectors of length $\|{\bf r}\|<1$.

According to the above, we see that it is possible to map the  quantum states, which are contained in (a convex portion of) the $2N$-dimensional \emph{complex} ball $B_1(\mathbb{C}^{2N})$, of unit radius, with the vector-states located at the surface and the operator-states  inside, into a $(N^2-1)$-dimensional \emph{real} ball $B_1(\mathbb{R}^{N^2-1})$, also of unit radius, with the vector-states also at the surface and the operator-states inside. For $N=2$, the real ball is completely filled with states (see below), but for $N\geq 3$ only a portion of it can be representative of states. Also, considering that the structure constants of the generators of $SU(N)$ have no rotational invariance, such portion will not be rotationally invariant. 

However, the set of vectors representative of states within $B_1(\mathbb{R}^{N^2-1})$ has the property of being a closed \emph{convex} set of vectors. To see this, let ${\bf r} = a_1\, {\bf r}_1 + a_2\, {\bf r}_2$, with $a_1+a_2=1$, $a_1,a_2\geq0$, and assume that  ${\bf r}_1$ and ${\bf r}_2$ are two vectors representing two states. Then, from (\ref{formulaNxN}) we have: $D({\bf r})=a_1D({\bf r}_1)+a_2D({\bf r}_2)$, with $D({\bf r}_1)$ and $D({\bf r}_2)$ two operator-states. And since a convex linear combination of operator states is an operator state (see the last paragraph of Sec.~\ref{Operator-states}), we deduce that an operator $D({\bf r})$ associated with a vector ${\bf r}$ which is a convex combination of vectors representative of states, is also representative of a bona fide state.

\begin{figure}[!ht]
\centering
\includegraphics[scale =.75]{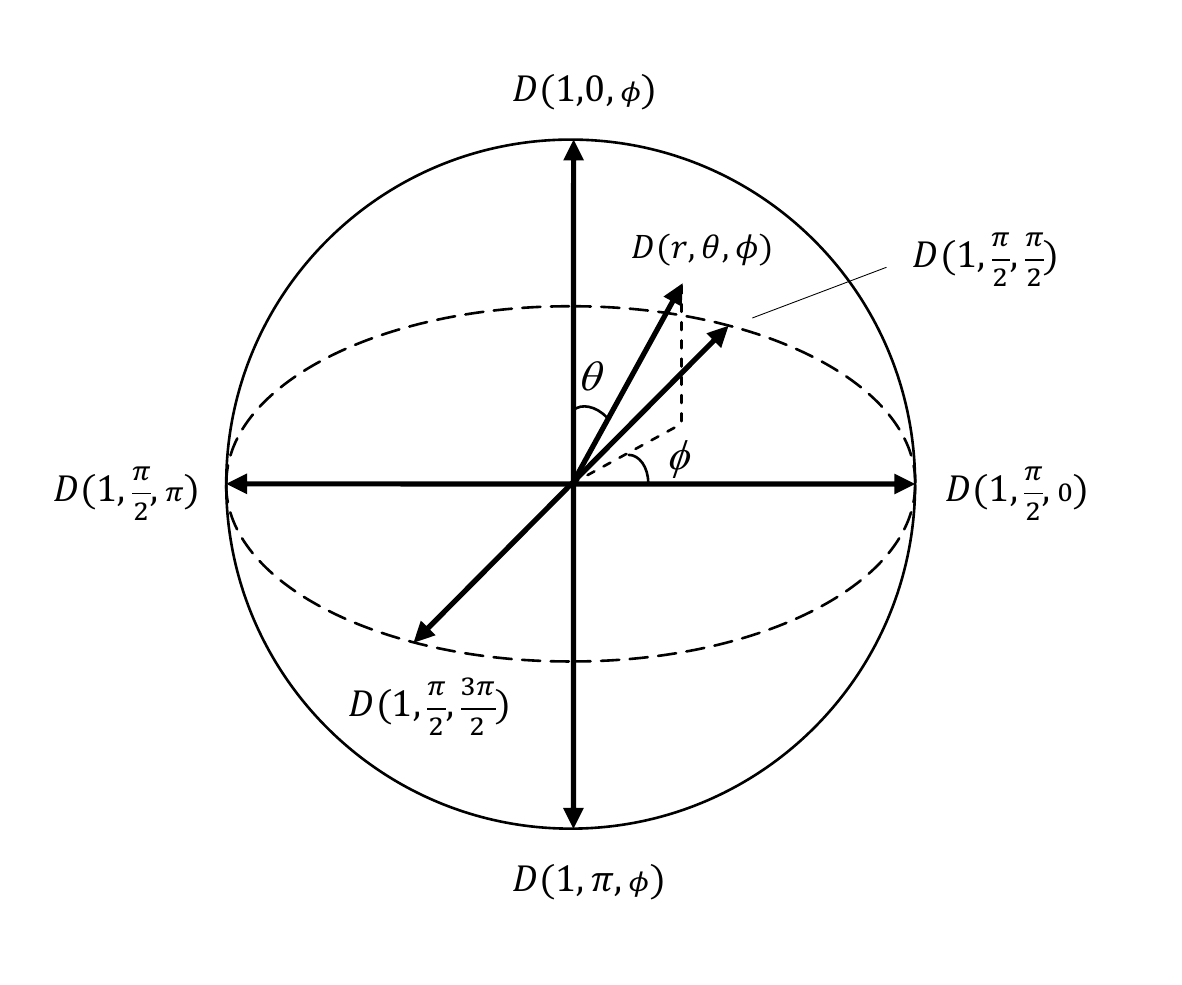}
\caption{The Bloch representation of the states $D(r,\theta,\phi)= {1\over 2}\left(\mathbb{I} + {\bf r}(r,\theta,\phi)\cdot\mbox{\boldmath$\sigma$}\right)$ of a two-dimensional quantum system, with ${\bf r}(r,\theta,\phi)=r(\sin\theta\cos\phi, \sin\theta\sin\phi,\cos\theta)$, $0\leq r\leq 1$, $0\leq \theta\leq\pi$, $0\leq \phi\leq 2\pi$. 
\label{Blochsphere}}
\end{figure}

More will be said in the next section regarding the properties of the convex set of vectors in $B_1(\mathbb{R}^{N^2-1})$. We conclude the present section observing that, for the special $N=2$ case, (\ref{negativeeigenvalue}) is positive. Indeed, the generators of $SU(2)$ are the three Pauli's spin matrices, and (\ref{formulaNxN}) becomes: 
\begin{equation}
D({\bf r}) =  {1\over 2}\left(\mathbb{I} + {\bf r}\cdot\mbox{\boldmath$\sigma$}\right).
\label{formula2X2}
\end{equation}
One of the remarkable properties of the Pauli's matrices is to obey the multiplication rule $({\bf u}\cdot \mbox{\boldmath$\sigma$})({\bf v}\cdot \mbox{\boldmath$\sigma$})={\bf u}\cdot{\bf v}\,\mathbb{I} +({\bf u}\times{\bf v})\cdot \mbox{\boldmath$\sigma$}$, implying that $({\bf r}\cdot \mbox{\boldmath$\sigma$})^2 =\|{\bf r}\|^2 \,\mathbb{I}$. Therefore, when $\|{\bf r}\| =1$, $({\bf r}\cdot \mbox{\boldmath$\sigma$})^2 =\mathbb{I}$, and for the spin-${1\over 2}$ operator $S_{\bf r}={1\over 2} {\bf r}\cdot\mbox{\boldmath$\sigma$}$  (we have set $\hbar =1$), oriented along direction ${\bf r}$, we have $S_{\bf r}D({\pm\bf r})=\pm{1\over 2}D({\pm\bf r})$. In other terms, all vector-states are spin eigenstates, and all points on the surface of  $B_1(\mathbb{R}^{3})$ are representative of vector-states. Also, when $\|{\bf r}\| <1$, considering that the Pauli matrices $\sigma_i$ have eigenvalues $\pm 1$, $D({\bf r})$ has eigenvalues ${1\over 2}(1\pm\|{\bf r}\|)$, and therefore is positive definite. Thus, all points inside $B_1(\mathbb{R}^{3})$ are  representative of operator-states. This bijection, between states $D({\bf r})$ and points of  $B_1(\mathbb{R}^{3})$, is known as the Bloch geometrical representation of the state space of a two-level quantum  system (qubit), and is an expression of the well-known $SU(2)$-$SO(3)$ homomorphism (see Fig.~\ref{Blochsphere}).

\section{Transition probabilities}
\label{Transition}

We now describe the transition probabilities, and their expression in terms of the vectors of $B_1(\mathbb{R}^{N^2-1})$, in the generalized Bloch representation of the set of states. For this, let $\{P_{a_1},\dots,P_{a_N}\}$ be the basis of eigenvector-states associated with an arbitrary observable $A$. To each of them, we can associate a unit vector ${\bf n}_i$, such that  $P_{a_i} = {1\over N}\left(\mathbb{I} +c_N\, {\bf n}_i\cdot\mbox{\boldmath$\Lambda$}\right)$, $i=1,\dots,N$. Then, the transition probability from an arbitrary state $D({\bf r})$ to an eigenstate $P_{a_i}$, is given by:
\begin{eqnarray}
\lefteqn{{\cal P}(D({\bf r})\to P_{a_i}) = {\rm Tr}\, D({\bf r})P_{a_i} =  {\rm Tr}\,  {1\over N^2}\left(\mathbb{I} +c_N\, {\bf r}\cdot\mbox{\boldmath$\Lambda$}\right)\left(\mathbb{I} + c_N\,{\bf n}_i\cdot\mbox{\boldmath$\Lambda$}\right)}\nonumber\\
&=& {\rm Tr}\, {1\over N^2}\left[\mathbb{I} + c_N\,({\bf r}\cdot\mbox{\boldmath$\Lambda$}+{\bf n}_i\cdot\mbox{\boldmath$\Lambda$}) +c_N^2\,({\bf r}\cdot\mbox{\boldmath$\Lambda$})({\bf n}_i\cdot\mbox{\boldmath$\Lambda$})\right] = {1\over N} +{c_N^2\over N^2}\,{\rm Tr}\,({\bf r}\cdot\mbox{\boldmath$\Lambda$})\, ({\bf n}_i\cdot\mbox{\boldmath$\Lambda$})\nonumber\\
&=&{1\over N} + {c_N^2\over N^2}\sum_{j,k=1}^{N^2-1} r_j[n_i]_k {\rm Tr}\, \Lambda_j\Lambda_k = {1\over N} + {2c_N^2\over N^2}\sum_{j,k=1}^{N^2-1}r_j[n_i]_k \delta_{jk}\nonumber\\
&=& {1\over N} \left[1+ (N-1)\,{\bf r}\cdot {\bf n}_i\right]={1\over N} \left[1+ (N-1)\,\|{\bf r}\|\cos\theta({\bf r},{\bf n}_i)\right],\label{transitiongeneralNxN}
\end{eqnarray}
where $\theta({\bf r},{\bf n}_i)$ is the angle between ${\bf r}$ and ${\bf n}_i$. 

The above formula allows us to gain some insight into the structure of the set of states in $B_1(\mathbb{R}^{N^2-1})$. If $D({\bf r}) =D({\bf n}_j)=P_{a_j}$, then ${\cal P}(P_{a_j}\to P_{a_i}) = {\rm Tr}\, P_{a_j}P_{a_i}=\delta_{ji}$. This means that $\cos\theta({\bf n}_j,{\bf n}_i)=-{1\over N-1}$, for $i\neq j$, that is: $\theta({\bf n}_j,{\bf n}_i) = \theta_N$, with $\theta_N\equiv \cos^{-1} (-{1\over N-1})$, for $i\neq j$. Since all the vector-states $\{P_{a_1},\dots,P_{a_N}\}$ are mutually orthogonal, the angle subtended by any two of their associated vectors $\{{\bf n}_1,\dots,{\bf n}_N\}$ must be $\theta_N$. And since the angle subtended by any two vertices of a $(N-1)$-dimensional \emph{simplex}, through its center, is precisely $\theta_N$, the ${\bf n}_i$ must necessarily be the $N$ vertex-vectors of a $(N-1)$-simplex $\triangle_{N-1}$, with edges of length $\|{\bf n}_i -{\bf n}_j\| = \sqrt{2(1-\cos\theta_N)} = \sqrt{2N\over N-1}$, whose center coincides with the center of the ball $B_1(\mathbb{R}^{N^2-1})$ in which it is inscribed. Also, considering that $\triangle_{N-1}$ is a convex set of vectors, it immediately follows that all points contained in it are representative of  operator-states, in accordance with the fact that the states in $B_1(\mathbb{R}^{N^2-1})$ form a closed convex subset. 

We open a parenthesis to draw the reader's attention on how these simplexes emerge from the generalized Bloch representation of states, as the natural geometric structure representing the reality of a measurement context, associated with a given observable. Simplexes were also used in the first modelization of the hidden-measurement interpretation~\cite{Aerts1986}, to conveniently encode the statistical information of the different states, relative to a single observable, and as we will see in the following they  play a crucial role in the dynamical description of the measurement process. However, what was just adopted as a convenient probabilistic representation in~\cite{Aerts1986}, now naturally follows from the general geometry of Hilbert spaces, and applies not only to single measurement situations, but also to situations where different measurements (different simplexes) can be jointly  represented, within a same unit ball, compatibly with Dirac's transformation theory. 

As we emphasized in the previous section, not all ${\bf r}\in  B_1(\mathbb{R}^{N^2-1})$ are representative of states. This is in particular the case for all vectors ${\bf r}$ such that ${\rm Tr}\, D({\bf r})P_{a_i} <0$, as they would give rise to unphysical negative  transition probabilities. This is generally the case if $\cos\theta({\bf r},{\bf n}_i)<-{1\over \|{\bf r}\|(N-1)}$. However, if $\|{\bf r}\|\leq r_N\equiv {1\over N-1}$, the inequality becomes $\cos\theta({\bf r},{\bf n}_i)<-1$, which can never be satisfied. This reflects the fact that within the ball $B_1(\mathbb{R}^{N^2-1})$ there is a smaller ball $B_{r_N}(\mathbb{R}^{N^2-1})$, of radius $r_N = {1\over N-1}$, which is completely filled with states~\cite{Kimura2005}. This is the ball inscribed in every simplex $\triangle_{N-1}$ representative of a basis (see Appendix~\ref{simplexes}). Of course, for the special case $N=2$, we have $r_N =1$, so that the inner and outer spheres coincide in this case.

Let us take  advantage of (\ref{transitiongeneralNxN}) to further clarify the structure of the set of vectors representative of the states. For this, consider the $2$-dimensional plane containing the two unit vectors ${\bf n}_i$ and ${\bf n}_j$, $i\neq j$, associated with two orthonormal vector-states $P_{a_i}$ and $P_{a_j}$, respectively. In this plane, it is possible to consider a unit vector ${\bf n}^\alpha$, obtained by rotating ${\bf n}_i$ by an angle $\alpha$, around the origin, so that ${\bf n}^0={\bf n}_i$ and ${\bf n}^{\theta_N}={\bf n}_j$. We want to show that none of the unit vectors ${\bf n}^\alpha$, for $\alpha\in (0,\theta_N)$, can be representative of a state. For this, it is sufficient to show that,  for all $\alpha\in (0,\theta_N)$, ${\rm Tr}\, D({\bf n}^\alpha)P_{a_j}<0$. To do so, we observe that ${\bf n}^\alpha$ can be written in the form: 
\begin{equation}
{\bf n}^\alpha=\left(\cos\alpha +{\sin\alpha\over \sqrt{N(N-2)}}\right){\bf n}_i +\sin\alpha\, {N-1\over \sqrt{N(N-2)}}\, {\bf n}_j.
\label{rotated}
\end{equation}
Considering that, for $i\neq j$, ${\bf n}_i\cdot {\bf n}_j=-{1\over N-1}$, (\ref{transitiongeneralNxN}) gives: 
\begin{equation}
{\rm Tr}\, D({\bf n}^\alpha)P_{a_j} ={1\over N}\left[1- (\cos\alpha -\sqrt{N(N-2)}\sin\alpha)\right]={1\over N}\left[1- (N-1)\sin\left(\alpha +\tan^{-1}{1\over\sqrt{N(N-2)}}\right)\right].
\end{equation}
This expression is negative if (mod $2\pi$): $0<\alpha <2\tan^{-1}\sqrt{N(N-2)}$, and since $\theta_N < 2\tan^{-1}\sqrt{N(N-2)}$, this proves that vectors (\ref{rotated}), for $\alpha\in (0,\theta_N)$, cannot be representative of vector-states. 

We will return to the above limitation on the allowed states in the last section of the article, which will find in our measurement model a very simple physical interpretation. For the moment, we just observe that, in particular, it also implies that if a unit vector ${\bf n}$ represents a vector-state, then, for $N\geq 3$,  its opposite vector $-{\bf n}$ cannot represent a vector-state. However, density states will always exist in the opposite direction, and one can prove that: if a unit vector ${\bf n}$ represents a vector-state, then in the opposite direction only operator-states can exist, represented by vectors ${\bf r}=-r\, {\bf n}$, with $r\leq r'_{N-1}$. Conversely, if in some direction ${\bf n}$ only operator-states ${\bf r}=r\, {\bf n}$, with $r\leq r'_{N-1}$, exist, then $-{\bf n}$ will be representative of a vector-state~\cite{Kimura2005}.

Another interesting feature of the structure of the  set of states, which is revealed by (\ref{transitiongeneralNxN}), and is worth mentioning, is the following. As we have seen, a relation of orthogonality of two vector-states in the complex ball $B_1(\mathbb{C}^{2N})$ does not translate into a relation of orthogonality of their representative vectors in the real ball $B_1(\mathbb{R}^{N^2-1})$. But then, to what kind of property a relation of orthogonality between unit vectors of $B_1(\mathbb{R}^{N^2-1})$ does correspond to? To answer this question, we observe that if two unit vectors ${\bf n}_i$ and ${\bf m}_j$, representative of two vector-states $P_{a_i}$ and $P_{b_j}$, respectively,  are orthogonal, that is ${\bf n}_i \cdot {\bf m}_j =0$, it  follows from (\ref{transitiongeneralNxN}) that ${\cal P}(P_{a_i}\to P_{b_j}) = {\rm Tr}\, P_{a_i}P_{b_j} ={1\over N}$. Vectors with such property are called \emph{mutually unbiased}, and two bases $\{P_{a_1},\dots,P_{a_N}\}$ and $\{P_{b_1},\dots,P_{b_N}\}$ such that all transition probabilities are equal to ${1\over N}$ are called \emph{mutually unbiased bases}. Measurements associated with mutually unbiased bases are as uncorrelated as possible, which is the reason why they are the most suitable bases to choose to reconstruct a state from the probabilities of different measurements~\cite{Durt2010}. Thus, we find that this condition of minimal correlation of two mutually unbiased bases simply translates, in the generalized Bloch representation, in a condition of orthogonality of the subspaces associated with the bases' simplexes. 

We conclude this section by deriving a more explicit and simple expression for the transition probabilities (\ref{transitiongeneralNxN}). For this, we write ${\bf r}$ as the sum ${\bf r} = {\bf r}^\perp + {\bf r}^\parallel$, where ${\bf r}^\parallel$ is the vector obtained by orthogonally projecting ${\bf r}$ onto the $(N-1)$-dimensional subspace generated by the simplex $\triangle_{N-1}$, with vertex-vectors  $\{{\bf n}_1,\dots,{\bf n}_N\}$. Since by definition ${\bf r}^\perp\cdot {\bf n}_i =0$, (\ref{transitiongeneralNxN}) becomes:
\begin{equation}
{\cal P}(D({\bf r})\to P_{a_i}) = {1\over N} \left[1+ (N-1)\,{\bf r}^\parallel\cdot {\bf n}_i\right].
\label{transitiongeneralNxN-bis}
\end{equation}
We write: 
\begin{equation}
{\bf r}^\parallel =\sum_{i=1}^{N} r^\parallel_i \,{\bf n}_i,
\label{rparallelexpansion}
\end{equation}
and observe that since the angle between the vectors ${\bf n}_i$ is $\theta_N$, we have:
\begin{equation}
{\bf n}_i \cdot {\bf n}_j = -{1\over N-1}+\delta_{ij}\left(1+{1\over N-1}\right) = -{1\over N-1}+\delta_{ij} {N\over N-1},
\label{scalar-n}
\end{equation}
and consequently: 
\begin{equation}
 {\bf r}^\parallel \cdot {\bf n}_i =\sum_{j=1}^{N} r^\parallel_j \,{\bf n}_j \cdot {\bf n}_i =\sum_{j=1}^{N} r^\parallel_j\left( -{1\over N-1}+\delta_{ij} {N\over N-1}\right).
\label{rparallelnj}
\end{equation}
Inserting (\ref{rparallelnj}) into (\ref{transitiongeneralNxN-bis}), the transition probabilities become:
\begin{equation}
{\cal P}(D({\bf r})\to P_{a_i}) = r^\parallel_i + {1\over N}\left(1-\sum_{j=1}^{N} r^\parallel_j\right).
\label{trans-general-g}
\end{equation}
At this point, we introduce the vector ${\bf r}_R = \sum_{i=1}^{N}{\cal P}(D({\bf r})\to P_{a_i}) \,{\bf n}_i$, associated with the fully reduced density matrix: $D({\bf r}_R) = \sum_{i=1}^{N}{\cal P}(D({\bf r})\to P_{a_i}) P_{a_i}$. Using (\ref{trans-general-g}), we have: 
\begin{eqnarray}
{\bf r}_R &=& \sum_{i=1}^{N}\left[r^\parallel_i + {1\over N}\left(1-\sum_{j=1}^{N} r^\parallel_j\right)\right] {\bf n}_i= \sum_{i=1}^{N} r^\parallel_i {\bf n}_i + \sum_{i=1}^{N} {1\over N}\left(1-\sum_{j=1}^{N} r^\parallel_j\right) {\bf n}_i\nonumber\\
&=& {\bf r}^\parallel + {1\over N}\left(1-\sum_{j=1}^{N} r^\parallel_j\right) \sum_{i=1}^{N} {\bf n}_i={\bf r}^\parallel,
\label{R2}
\end{eqnarray}
where for the last equality we have used the fact that the sum of the vertex vectors of a simplex is zero: $\sum_{i=1}^{N} {\bf n}_i=0$. Since by construction ${\bf r}_R\in\triangle_{N-1}$, we obtain that ${\bf r}^\parallel\in\triangle_{N-1}$. This means that $\sum_{j=1}^{N} r^\parallel_j=1$, and (\ref{trans-general-g}) becomes: 
\begin{equation}
{\cal P}(D({\bf r})\to P_{a_i})  =  r^\parallel_i.
\label{trans-general}
\end{equation}

The convex combination (\ref{rparallelexpansion}) is then unique, as is clear from equality (\ref{trans-general}) which connects the coefficients $r^\parallel_i$ to the physical probabilities. To show this directly, assume that there would exist another convex combination ${\bf r}^\parallel =\sum_{i=1}^{N} \tilde{r}^\parallel_i \,{\bf n}_i$; then we would have $\sum_{i=1}^{N} (\tilde{r}^\parallel_i -r^\parallel_i )\,{\bf n}_i=0$, and by multiplying this expression by ${\bf n}_j$, then using (\ref{scalar-n}),  it  immediately follows that $\tilde{r}^\parallel_j =r^\parallel_j$, for all $j=1,\dots,N$, so proving that the coefficients of the convex linear combination are uniquely determined.

\section{Unitary evolution}
\label{unitary}

To complete our description of the generalized Bloch representation, and before we address the main issue of the present article, which is that of providing a full description of the measurement process, in terms of hidden-interactions, we briefly analyze in this section how the deterministic change of quantum entities, as described by the \emph{Schr\"odinger equation}, manifests at the level of the vectors in $B_1(\mathbb{R}^{N^2-1})$. To begin with, we explain how to calculate the components of a real vector ${\bf r}$, representative of a given operator-state $D({\bf r})$ in $B_1(\mathbb{R}^{N^2-1})$. In view of (\ref{formulaNxN}) and (\ref{commutationNxN}), we can write:  
\begin{equation}
{1\over 2}\left[D({\bf r})\Lambda_j + \Lambda_j D({\bf r})\right] =  {1\over N}\left(\Lambda_j + {c_N\over 2}\sum_{i=1}^{N^2-1} r_i [\Lambda_i,\Lambda_j]_+\right) = {1\over N}\left(\Lambda_j +   {2c_N\over N} r_j \mathbb{I} + c_N\sum_{i,k=1}^{N^2-1}r_i d_{ijk}\Lambda_k\right).
\label{reconstruction1}
\end{equation}
Taking the trace of the above expression, we obtain: 
\begin{equation}
r_j =\sqrt{N\over 2(N-1)} \,{\rm Tr}\, D({\bf r})\Lambda_j, \quad j=1,\dots,N^2-1.
\label{reconstruction2tris}
\end{equation}
This equality allows one to calculate the $N^2-1$ components of the real vector associated with an arbitrary quantum state. Also, it shows how to reconstruct the state on the basis of the data obtained from the measurement of the $\Lambda_j$ observables, which thus form a set of \emph{informationally complete} observables.  

\begin{figure}[!ht]
\centering
\includegraphics[scale =.35]{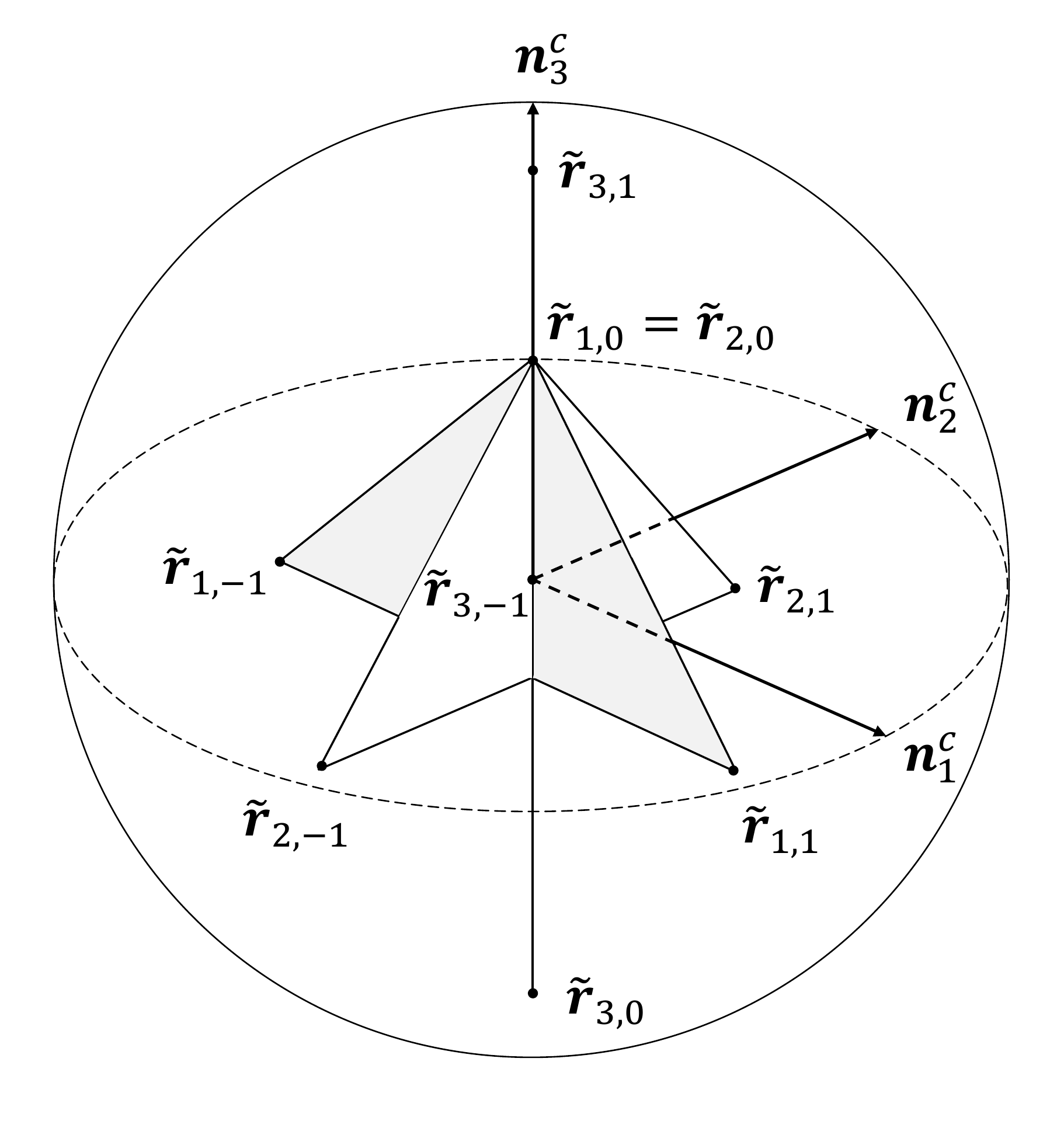}
\caption{The projections $\tilde{\bf r}_{i,\mu}$ of the vectors ${\bf r}_{i,\mu}$ representative of the eigenvectors of the spin operators $S_i$, $i=1,2,3$, onto the 3-dimensional sub-ball of $B_1(\mathbb{R}^{8})$, generated by the first three canonical vectors ${\bf n}_1^c=(1,0000000)^T$, ${\bf n}_2^c=(0,1000000)^T$ and ${\bf n}_3^c=(0,0100000)^T$.  Note that the two eigenvectors $D(S_1=0)$ and $D(S_2=0)$ are represented by a same point in the sub-ball, and that the $2$-simplex associated with $S_3$ reduces to a line segment along the  ${\bf n}_3^c$-axis. Note also that the angle between the projected 2-simplexes associated with $S_1$ and $S_2$, represented in gray and white, respectively, does not correspond to their actual angle in $B_1(\mathbb{R}^{8})$.
\label{Blochspheresection}}
\end{figure}

To give an example, and help us develop our intuition on how vectors representative of states are arranged in $B_1(\mathbb{R}^{N^2-1})$, let us calculate, for the $N=3$ case, the 9 vectors associated with the eigenvector-states of the 3 spin observables, oriented along the three Cartesian axes:
\begin{equation}
S_1 = {\lambda_1+\lambda_6\over\sqrt{2}}=\begin{bmatrix} 0 & {1\over \sqrt{2}} & 0 \\ {1\over \sqrt{2}} & 0 & {1\over \sqrt{2}} \\ 0 & {1\over \sqrt{2}} & 0 \end{bmatrix}, \quad S_2 =  {\lambda_2+\lambda_7\over\sqrt{2}}= \begin{bmatrix} 0 & {-i\over\sqrt{2}} & 0 \\ {i\over\sqrt{2}} & 0 & {-i\over\sqrt{2}} \\ 0 & {i\over\sqrt{2}} & 0 \end{bmatrix}, \quad S_3 = {\lambda_3+\sqrt{3}\lambda_8\over 2}= \begin{bmatrix} 1 & 0 & 0 \\ 0 & 0 & 0 \\ 0 & 0 & -1\end{bmatrix}.
\end{equation} 
Denoting $|S_i=\mu\rangle$ the eigen-ket of the spin operator $S_i$, for the eigenvalue $\mu$,  with $i=1,2,3$, and  $\mu=1,0,-1$, we have:  
\begin{eqnarray}
&|S_1=1\rangle ={1 \over 2}(1,\sqrt{2},1)^T, \,\,\, |S_1=0\rangle ={1 \over \sqrt{2}}(-1, 0,1)^T, \,\,\, |S_1=-1\rangle ={1 \over 2}(1, -\sqrt{2},1)^T,\\
&|S_2=1\rangle ={i \over 2}(-1, -i\sqrt{2},1)^T, \,\,\, |S_2=0\rangle ={-i \over \sqrt{2}}(1, 0,1)^T,  \,\,\, |S_2=-1\rangle ={i \over 2}(-1, i\sqrt{2},1)^T,\\
&|S_3=1\rangle =(1, 0,0)^T, \,\,\, |S_3=0\rangle =(0, 1,0)^T, \,\,\, |S_3=-1\rangle =(0, 0,1)^T.
\end{eqnarray}
Using (\ref{reconstruction2tris}), after some calculations one obtains the following components for the $8$-dimensional representative ${\bf r}_{i,\mu}$ of the eigenvector-states $D(S_i=\mu)=|S_i=\mu\rangle\langle S_i=\mu|$:
\begin{eqnarray}
&{\bf r}_{1,1} = ({1\over 2}\sqrt{3\over 2}, 0, -{\sqrt{3}\over 8}, {\sqrt{3}\over 4}, 0, {1\over 2}\sqrt{3\over 2}, 0, {1\over 8})^T, \quad 
{\bf r}_{1,0} =(0, 0, {\sqrt{3}\over 4}, -{\sqrt{3}\over 2}, 0, 0, 0,  -{1\over 4})^T,\\
&{\bf r}_{1,-1} = (-{1\over 2}\sqrt{3\over 2}, 0, -{\sqrt{3}\over 8}, {\sqrt{3}\over 4},  0, -{1\over 2}\sqrt{3\over 2}, 0, {1\over 8})^T,\quad {\bf r}_{2,1} = (0, {1\over 2}\sqrt{3\over 2}, -{\sqrt{3}\over 8}, 
-{\sqrt{3}\over 4},  0,  0, {1\over 2}\sqrt{3\over 2}, {1\over 8})^T,\\
&{\bf r}_{2,0} =(0,0,{\sqrt{3}\over 4},{\sqrt{3}\over 2},0,0,0,-{1\over 4})^T,\quad 
{\bf r}_{2,-1} = (0,-{1\over 2}\sqrt{3\over 2},-{\sqrt{3}\over 8},-{\sqrt{3}\over 4},0,0,-{1\over 2}\sqrt{3\over 2},{1\over 8})^T,\\
&{\bf r}_{3,1} =(0,0,{\sqrt{3}\over 2},0,0,0,0,{1\over 2})^T,\quad 
{\bf r}_{3,0} =(0,0,-{\sqrt{3}\over 2},0,0,0,0,{1\over 2})^T,\quad {\bf r}_{3,-1} =(0,0,0,0,0,0,0,-1)^T.
\end{eqnarray}
For $N=3$, each basis is associated with a $2$-simplex, i.e., with a $2$-dimensional \emph{equilateral triangle}. To visualize these bases, a possibility is to project them onto different $3$-dimensional sub-balls of $B_1(\mathbb{R}^{8})$. An example  of such projection is given in Figure~\ref{Blochspheresection}. 

Having clarified how vectors can be constructed by calculating the average values of the generators $\Lambda_i$, let us consider a vector ${\bf r}$ representing a given state at time $t=0$. We want to determine the continuous dynamical change of such vector when subjected to a \emph{deterministic unitary evolution}. For this, let $U_{t}=\exp(-iHt)$ be the unitary evolution operator, obeying the Schr\"odinger equation $i\partial_t U_{t} = HU_{t}$, where $H$ is the Hamiltonian of the system (assumed to be time-independent) and we have set $\hbar =1$. If $D({\bf r})$ is the state of the entity at time $t=0$, at time $t$ its state will be $D_t({\bf r}) = U_{t}D({\bf r}) U_{t}^\dagger$, and by deriving with respect to $t$, we obtain: $i\partial_tD_t({\bf r}) = [H,D_t({\bf r})]$, which is the well known \emph{Liouville-von Neumann equation} (which has a sign difference with respect to the \emph{Heisenberg equation} for the unitary evolution of observables). 

In view of (\ref{formulaNxN}), we  have: $D_t({\bf r}) =  {1\over N}[\mathbb{I} +c_N\, {\bf r}\cdot U_{t}\mbox{\boldmath$\Lambda$}U_{t}^\dagger]$. Therefore, if we set $D({\bf r}(t)) \equiv D_t({\bf r})$, we deduce that:  ${\bf r}(t)\cdot \mbox{\boldmath$\Lambda$} = {\bf r}\cdot U_{t}\mbox{\boldmath$\Lambda$}U_{t}^\dagger$. Multiplying this equation by $\Lambda_k$, taking the trace and using ${\rm Tr}\, \Lambda_j\Lambda_k =2\delta_{jk}$, we then obtain: 
\begin{equation}
r_k(t) = {1\over 2}\,{\bf r}\cdot {\rm Tr}\,\Lambda_k(t)\mbox{\boldmath$\Lambda$} = \sum_{j=i}^{N^2-1} V_{kj}(t) \, r_j = \left[V(t){\bf r}\right]_k
\label{evolutionmatrix}
\end{equation}
where $\Lambda_k(t)=U_{t}^\dagger\Lambda_k U_{t}$, and we have defined the $(N^2-1)\times (N^2-1)$ evolution matrix $V_{kj}(t)\equiv {1\over 2} {\rm Tr}\,\Lambda_k(t)\Lambda_j$, which is  an element of $SO(N^2-1)$, the group of orthogonal matrices with unit determinant, which is the symmetry group of $B_1(\mathbb{R}^{N^2-1})$, whose dimension is $(N^2-1)[(N^2-1)-1]/2$. Note however that the matrices $V(t)$ are determined by the $N\times N$ unitary matrices $U_t$, which are elements of the group $SU(N)$,  whose dimension is only $N^2-1$. Thus, the $V(t)$ only constitute a very small portion of $SO(N^2-1)$, in a accordance with the fact that only a small convex portion of $B_1(\mathbb{R}^{N^2-1})$ contains states. 

As an illustration,  consider the precession of a spin in a uniform magnetic field. The evolution operator is $U_{t}= \exp{(i{\gamma\over 2} B\sigma_3 t)}$, with $B>0$ the field strength and $\gamma$ the gyromagnetic ratio. Using the commutation relations of the spin components $[S_3,S_\pm]=\pm S_\pm$, $[S_+,S_-]=2S_3$, where $S_\pm =S_1\pm i S_2$, we have: $S_\pm(t)=S_\pm e^{\pm i\omega t}$, with $\omega\equiv -\gamma B$ the so-called Larmor frequency. In the $N=2$ case, (\ref{evolutionmatrix}) becomes: $r_k(t)= {1\over 2}\,{\bf r}\cdot {\rm Tr}\,\mbox{\boldmath$\sigma$}\sigma_k(t)=2{\bf r}\cdot {\rm Tr}\,{\bf S}S_k(t)$, so that $r_\pm (t) = e^{\pm i\omega t} r_\pm$.  In other terms:
\begin{equation}
\begin{bmatrix}
  r_1(t)\\
  r_2(t)\\
  r_3(t)
 \end{bmatrix} =
\begin{bmatrix}
  \cos\omega t& -\sin\omega t  & 0 \\
  \sin\omega t & \phantom{-}\cos\omega t & 0 \\
  0 & 0 & 1
 \end{bmatrix}
\begin{bmatrix}
   r_1\\
  r_2\\
  r_3
 \end{bmatrix}.
\end{equation}

Let us  consider also the case of a spin-$1$ entity ($N=3$). Then, $r_k(t)= {1\over 2}\,{\bf r}\cdot {\rm Tr}\,\lambda_k(t)\mbox{\boldmath$\lambda$}$, and after some calculation one obtains the evolution matrix:
\begin{equation}
\begin{bmatrix}
r_1(t)\\
r_2(t)\\
r_3(t)\\
r_4(t)\\
r_5(t)\\
r_6(t)\\
r_7(t)\\
r_8(t)
 \end{bmatrix} =
\begin{bmatrix}
  \cos\omega t& -\sin\omega t  & 0  & 0 & 0 & 0 & 0 & 0\\
 \sin\omega t & \phantom{-}\cos\omega t & 0  & 0 & 0 & 0 & 0 & 0 \\
  0 & 0 & 1& 0 & 0 & 0 & 0 & 0 \\
0 & 0 & 0&\cos\omega t& -\sin\omega t  & 0& 0 & 0 \\
0 & 0 & 0&\sin\omega t& \phantom{-}\cos\omega t  & 0& 0 & 0 \\
0 & 0 & 0&0&0&\cos\omega t& -\sin\omega t  & 0\\
0 & 0 & 0&0&0&\sin\omega t& \phantom{-}\cos\omega t  & 0\\
0 & 0 & 0&0&0&0& 0  & 1\\
 \end{bmatrix}
\begin{bmatrix}
r_1\\
r_2\\
r_3\\
r_4\\
r_5\\
r_6\\
r_7\\
r_8
 \end{bmatrix},
\end{equation}
which is block diagonal, with the $2\times 2$ blocks which are $2\times 2$ rotation matrices, and clearly belongs to $SO(\real^8)$.

\section{Hidden-measurements: the $N=2$ case}
\label{HiddenmeasurementsN=2}

In the previous sections we have shown how in the standard quantum formalism states can be expressed, in very general terms, as positive semidefinite, unit trace self-adjoint operators, and how the portion of the complex unit ball formed by states can be mapped into a convex portion of a higher dimensional real unit ball. In the latter, unitary evolution acts by means of real isometries $V(t)$, describing  orientation and length preserving transformations of the vector representative of the state. As we will now show, the generalized Bloch representation is the natural stage to also represent the indeterministic measurement processes, and more specifically the so-called measurements of the \emph{first kind}, which are such that if a second identical measurement is repeated, immediately after the first one, the same outcome will be obtained, with certainty. This means that a measurement context ``of the first kind'' produces a stable change of the state of the entity, that no longer changes under its influence, which is clearly an ideal situation to identify an outcome of an experiment by means of an eigenstate. 

So, what we will now describe in the following of this article, in a detailed way, is a model where the measurement processes can be fully represented inside $B_1(\real^{N^2-1})$,  thus providing an extension of the standard quantum formalism. In this section, we start by describing  the special $N=2$ situation, as the Bloch representation is particularly simple in this case and can be fully visualized in a three-dimensional space. This will allow us to introduce all the important concepts, facilitating in this way the subsequent understanding of the more general and  articulated $N\geq 3$ situations. 

We thus start by considering a two-dimensional entity and a measurement associated with an observable $A =a_1 P_{a_1}+ a_2P_{a_2}$. Non trivial measurements are necessarily non-degenerate, so we can assume that $a_1\neq a_2$. We denote ${\bf n}_1$ and ${\bf n}_2$ the two unit-vectors of $B_1(\real^{3})$ representative of the vector-states $P_{a_1}$ and $P_{a_2}$, respectively. As we know, they have to make an angle $\theta$ such that $\cos\theta = -1$, which means that they are opposite vectors, i.e.,  ${\bf n}_1=-{\bf n}_2$, and that they generate a $1$-simplex $\triangle_1\equiv \triangle_1({\bf n}_1,{\bf n}_2)$, which corresponds to a one-dimensional line segment of length $2$, going from ${\bf n}_1$ to ${\bf n}_2$, passing through the center of $B_1(\real^3)$ (in other terms, it corresponds to one of the diameters of the unit ball). 

Since $\triangle_1$ is a convex subset of $B_1(\real^3)$, and  ${\bf n}_1$ and ${\bf n}_2$ are both representative of states, we also know that all points of $\triangle_1$ are representative of bona fide states of the entity under consideration (which for instance can be imagined being the spin of a spin-${1\over 2}$ entity, like an electron). Considering all couples of possible opposite vectors, and the associated $1$-simplexes, we then deduce that the entire ball $B_1(\real^3)$ is actually filled with vectors representative of states, as already emphasized in the previous sections. 

In our model, $\triangle_1$ represents the measurement context, and more precisely what we should more properly call the ``naked'' measurement context, or the measurement context \emph{per se}, i.e., that specific aspect of the measurement which is responsible for the indeterministic ``collapse'' of the state. If we make this clear, it is because, as we shall see, a measurement also involves deterministic processes, of the non-unitary (non-isometric) kind, which can therefore be distinguished, at least in principle, from the genuinely indeterministic collapse-like process. 

Let ${\bf r}$ be the  real vector in $B_1(\real^3)$ which is representative, at time $t=0$, of the state $D({\bf r})$ of the entity under consideration (which can be a vector-state, or more generally an operator-state, depending on the length of ${\bf r}$). Inside $B_1(\real^3)$, such entity can thus be visualized \emph{as if} it were a \emph{point-like particle}. We stress that we are not here saying that the entity \emph{is} a point-like particle, but just that a point-like particle is able to conveniently represent the \emph{element of reality}  associated with that specific entity. 

The measurement of the observable $A$, on the other hand, is represented inside $B_1(\real^3)$ by a substance which fills the $1$-simplex, possessing the following three remarkable properties: it is (1) \emph{attractive}, (2) \emph{unstable}, and (3) \emph{elastic}, in a way that we are going shortly to explain. More precisely, when the entity is subjected to the measurement of the observable $A$, the process can be represented in our model by the action of this substance that fills the $\triangle_1$-region, on the point-like particle. Again, we emphasize that we are not here affirming that the measurement of $A$ \emph{is} the action of such substance on that point-like object, but that the process can be conveniently represented, and fully visualized, in this way.

\begin{figure}[!ht]
\centering
\includegraphics[scale =.45]{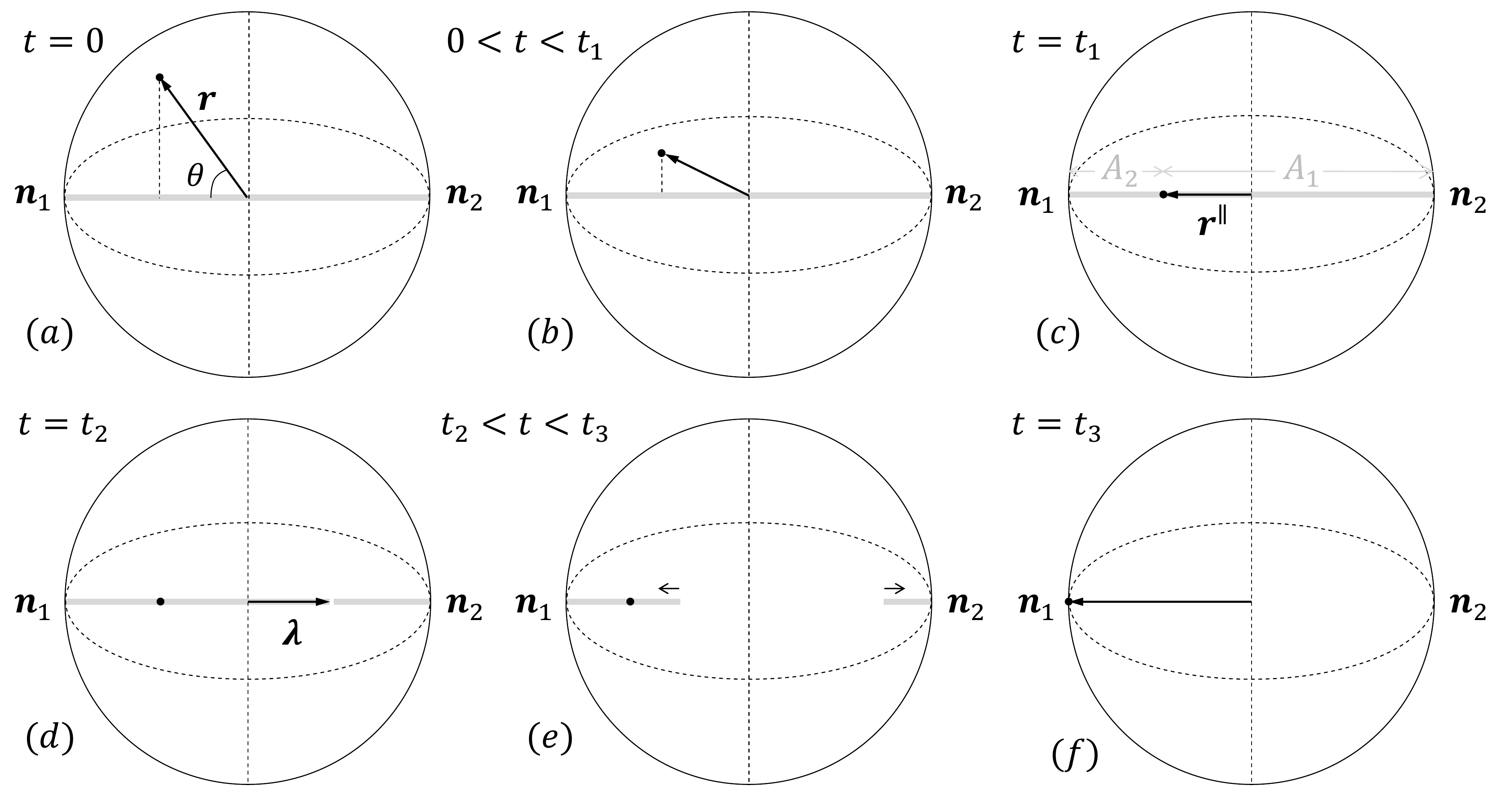}
\caption{The unfolding of the measurement process within the ball $B_1(\mathbb{R}^3)$; (a): the one-dimensional elastic structure $\triangle_1$ (in gray color) stretched along the two end points  ${\bf n}_1$ and ${\bf n}_2$, and the initial location (state) ${\bf r}$ of the entity, at time $t=0$; (b): the particle, attracted by the elastic band, orthogonally ``falling'' onto it; (c): the position ${\bf r}^\parallel$ of the particle on the elastic band, at time $t=t_1$, defining the two regions $A_1$ and $A_2$; (d): the disintegration of elastic band, at time $t=t_2$, at some unpredictable point \mbox{\boldmath$\lambda$}, here assumed to belong to $A_1$; (e): the contraction (collapse) of the elastic, drawing the point-particle toward position ${\bf n}_1$; (f): the reaching of the final destination, at ${\bf n}_1$, at time $t=t_3$, corresponding to the vector-state representative of the outcome of the quantum measurement. 
\label{2Dcollapse}}
\end{figure}

So, the measurement begins when the point-particle is subjected to the substance forming $\triangle_1$. As we mentioned, one of its properties is that it is an elastic substance. Thus, we can think of it as an elastic band, of length $2$, stretched and fixed at the two opposite end points  ${\bf n}_1$ and ${\bf n}_2$, passing through the center of the ball (see Fig.~\ref{2Dcollapse} $(a)$). When this is done, being the band attractive, it will cause the point-particle, initially at position ${\bf r}$, to move toward it. Here we assume that the physics that governs the movement of the particle inside the ball requires it to move toward $\triangle_1$ by  taking always the \emph{shortest path}. This means that when the particle deterministically approaches $\triangle_1$, it does so by following a straight line, orthogonal to $\triangle_1$. 

In other terms, during this first deterministic part of the measurement, the entity immerses itself as deeply as possible in the measurement context, establishing the deepest possible contact with it. This corresponds to the particle orthogonally ``falling''  onto the elastic band, and then firmly sticking on it (being  the attraction then exerted by the band maximal). More precisely, assuming that this movement takes place between time $t=0$ and $t=t_1$, we can write (see Fig.~\ref{2Dcollapse} $(a)$-$(c)$):
\begin{equation}
{\bf r}(t)={\bf r}+{t\over t_1}({\bf r}^\parallel - {\bf r}), \quad t\in [0,t_1],
\label{deterministicprojection}
\end{equation}
where ${\bf r}^\parallel = ({\bf r}\cdot {\bf n}_i)\, {\bf n}_i = \|{\bf r}\|\cos\theta({\bf r},{\bf n}_i)\, {\bf n}_i$, is the position that the particles reach on $\triangle_1$, at time $t=t_1$ (which can be represented by either projecting ${\bf r}$ onto  ${\bf n}_1$, or ${\bf n}_2$).

Once the point-like entity has reached $\triangle_1$, thus becoming strongly anchored to its substance, after some additional time the latter, being unstable, disintegrates. This process of disintegration occurs initially in some a priori unpredictable point $\mbox{\boldmath$\lambda$} \in\triangle_1$, so producing the splitting of the band into two halves. Being the band elastic, these two halves immediately contract toward their respective anchor points ${\bf n}_1$ and ${\bf n}_2$, drawing in this way also the particle to one of these two points, depending on whether the initial disintegration happens in the region $A_2$,  between ${\bf r}^\parallel$ and ${\bf n}_1$, or in the region $A_1$, between ${\bf r}^\parallel$ and  ${\bf n}_2$ (see Fig.~\ref{2Dcollapse} $(c)$).  So, assuming that the disintegration happens at time $t_2\geq t_1$, and that the collapse of the band brings the particle to one the two end points at exactly time $t_3\geq t_2$, we can write: 
\begin{equation}
{\bf r}(t)={\bf r}^\parallel + \frac{t-t_2}{t_3-t_2}\, \Theta(t-t_2)({\bf n}_i -{\bf r}^\parallel),\quad t\in [t_1,t_3],
\label{deterministicprojection}
\end{equation}
where $i=1$, if \mbox{\boldmath$\lambda$} belongs to $A_1$, and $i=2$, if \mbox{\boldmath$\lambda$} belongs to $A_2$, with $\Theta$  the Heaviside step function.

There is of course also the possibility that the initial point of disintegration of the band is such that $\mbox{\boldmath$\lambda$}={\bf r}^\parallel$, i.e., that it precisely coincides with the position of the point-particle. In this case, we are in a situation of unstable equilibrium, and the outcome of the collapse cannot be predicted in advance. As we will see, these exceptional values of \mbox{\boldmath$\lambda$}, being of zero measure, will not contribute to the determination of the transition probabilities.  

Before continuing our exploration of the model, it is important to say that the details of the above movements of the point-like particle representative of the entity's state inside the ball are not, as such, particularly important. Here, for simplicity, we have adopted a parameterization describing these movements only in terms of uniform speeds, but of course more general and complex movements can be imagined. For instance, we can consider that the speed of contraction of the band is not uniform, and that when the particle moves toward the band it does so in accelerated motion. What is important is that these movements, however complex,  are able to produce the conditions that we have described: the landing of the particle on $\triangle_1$, on point ${\bf r}^\parallel$, and its consequent collapse toward either ${\bf n}_1$ or ${\bf n}_2$, depending on a \emph{random} process of selection of a point  $\mbox{\boldmath$\lambda$} \in\triangle_1$.

It should also be emphasized that, considering that the unit ball in which the vector ${\bf r}(t)$ moves is not the ordinary three-dimensional Euclidean space, it is not even necessary to assume that the parameter $t$ in the above equations corresponds to the actual time coordinate, as it would be measured by a clock in the laboratory. More generally, it could simply be understood as an abstract ``order parameter,'' describing a process which, considered from our ordinary spatial perspective, could just appear to be instantaneous.

Having said this, let us now calculate the probability ${\cal P}({\bf r}\to {\bf n}_i)$ that the particle, initially located in ${\bf r}$ inside the ball, will finally reach the end point ${\bf n}_i$, $i=1,2$. To do so, we need to know how exactly the breaking point \mbox{\boldmath$\lambda$} is randomly selected. There are of course countless different ways to do this, and all these different possibilities will be considered later on in the article, when discussing the key notion of \emph{universal measurement}. For the time being, we assume that $\triangle_1$ is representative of a uniform structure, meaning that all its points have the same probability to disintegrate. This means that the probability that the initial disintegration point \mbox{\boldmath$\lambda$} belongs to the line segment $A_i$, $i=1,2$, is given by the ratio between the length of $A_i$ (the Lebesgue measure $\mu(A_i)$ of $A_i$) and the total length of the elastic band, which is $\mu(\triangle_1)=2$. For $A_1$, we  have:
\begin{equation}
\mu(A_1)= \|{\bf r}^\parallel - {\bf n}_2\|= \|{\bf r}^\parallel + {\bf n}_1\| = \|(1+ {\bf r}\cdot {\bf n}_1){\bf n}_1\| =1+ {\bf r}\cdot {\bf n}_1,
\end{equation}
and equivalently, for $A_2$, we have:
\begin{equation}
\mu(A_2)= \|{\bf n}_1 - {\bf r}^\parallel \|=\| -{\bf n}_2 - {\bf r}^\parallel \| =  \|(1+ {\bf r}\cdot {\bf n}_2){\bf n}_2\| =1+ {\bf r}\cdot {\bf n}_2.
\end{equation}
Therefore, the probability ${\cal P}(\mbox{\boldmath$\lambda$}\in A_i)$, that the elastic structure breaks first in $A_i$, $i=1,2$, is given by the ratio: 
\begin{equation}
{\cal P}(\mbox{\boldmath$\lambda$}\in A_i) = {\mu (A_i)\over \mu(\triangle_1)} = {1\over 2}(1+ {\bf r}\cdot {\bf n}_i) =  {1\over 2}[1+\|{\bf r}\|\cos\theta({\bf r},{\bf n}_i)],
\label{plambdainAi}
\end{equation}
which is precisely the quantum mechanical probability (\ref{transitiongeneralNxN}), for $N=2$. And since, by definition, the transition probability ${\cal P}({\bf r}\to {\bf n}_i)$ is precisely the probability ${\cal P}(\mbox{\boldmath$\lambda$}\in A_i)$, we find that the described two-step process of (1) a deterministic inwards movement, producing the connection of the point-particle with the unstable elastic structure $\triangle_1$, and (2) its subsequent being driven toward one of the final positions ${\bf n}_1$ or ${\bf n}_2$, as a consequence of the disintegration and contraction of the former, provides a full representation of the quantum mechanical measurement process. 

Thus, the standard Bloch representation of a two-level quantum mechanical entity (qubit) can be completed by also representing inside of it, in a consistent way, the different possible measurement processes. Before proceeding to the next section, where the model will be generalized to the situation of an arbitrary number $N$ of outcomes, the following remarks are in order. 

What we have described is clearly a measurement of the first kind, as is clear that a point-particle in position ${\bf n}_i$, if subjected again to the same measurement, being already located in one of the two end points of the elastic structure, we will have ${\bf r} = {\bf r}^\parallel = {\bf n}_i$, so that its position cannot further be changed by the  collapse of the elastic substance.  

The vectors \mbox{\boldmath$\lambda$}, associated with the possible disintegration points of the elastic, can be interpreted as the variables specifying the \emph{measurement interactions}. Thus, the model provides a consistent hidden-measurement interpretation of the quantum probabilities, which therefore admit a clear epistemic characterization, in terms of lack of knowledge (or of control) regarding the interaction \mbox{\boldmath$\lambda$} between the entity and the measuring system which is actualized during the measurement process.

Apart from the exceptional (zero measure) circumstance $\mbox{\boldmath$\lambda$}={\bf r}^\parallel$, each  \mbox{\boldmath$\lambda$} gives rise to a deterministic process, changing the state of the entity from ${\bf r}^\parallel$ to either ${\bf n}_1$ or ${\bf n}_2$, depending whether $\mbox{\boldmath$\lambda$}\in A_1$, or $\mbox{\boldmath$\lambda$}\in A_2$. However, being the selection of the variable \mbox{\boldmath$\lambda$} the result of the random disintegration of the elastic substance forming $\triangle_1$, the transition ${\bf r}^\parallel \to {\bf n}_i$, as a whole, is genuinely indeterministic, and corresponds to what we have called the ``naked'' part of the measurement, which is preceded by a deterministic inwards movement of the point-particle, bringing it from point ${\bf r}$ to point ${\bf r}^\parallel$. The full measurement process, as we have seen, is the combination of these two processes, and produces the overall transition ${\bf r}\to {\bf n}_i$. As we shall see in the following, a third deterministic process will be needed in the more articulate situation of a degenerate measurement. 

It is worth mentioning that it is precisely this non-classical element of change of the state of the entity, from its initial state  ${\bf r}$ to a final state ${\bf n}_i$, combined with the lack of knowledge regarding the process of actualization of the measurement interaction, which confers to the measurement model its \emph{non-Kolmogorovity}. The fact that such non-Kolmogorovian probability model is precisely Hilbertian, depends however on our hypothesis of a uniform probability distribution for the way the \mbox{\boldmath$\lambda$} are randomly selected. Once this uniform assumption is relaxed, one can construct more general probability models, which are neither classical nor quantum (Hilbertian), but truly intermediate. More will be said about this in the sequel of the article. 

Although we have described the measurement process in abstract terms, emphasizing that the point-like particle is just a mathematical representation of the state of the real physical entity under consideration, and that, similarly, the segment-like entity $\triangle_1$ is also just a mathematical representation of the experimental context to which the physical entity is subjected, it is also perfectly clear that all these mathematical objects admit a physical realization in terms of concrete three-dimensional objects. Indeed, it is certainly possible to construct a machine using uniform breakable and sticky elastic bands, and small material corpuscles, which will operate exactly in the way we have described, so as to imitate a quantum measurement process. Such a macroscopic -- room temperature -- machine would thus be a \emph{quantum machine}, whose quantum behavior would not be a consequence of its internal coherence, but of the specific way we would have decided to actively experiment with it, by means of predetermined protocols.

\section{Hidden-measurements: the general non-degenerate situation}
\label{non-degenerate situation}

We consider now the general situation of a $N$-dimensional quantum entity, subjected to the measurement of a non-degenerate observable $A=\sum_{i=1}^N a_i P_{a_i}$, with $a_i\neq a_j$, if $i\neq j$ (the degenerate situation will be described in the next section). Let $\{{\bf n}_1,\dots, {\bf n}_N\}$ be the $N$ unit-vectors of $B_1(\real^{N^2-1})$, representative of the base vector-states $\{P_{a_1},\dots, P_{a_N}\}$. As they all make, with respect to one another, an angle $\theta_N$, such that $\cos\theta_N = -{1\over N-1}$, they form a $(N-1)$-simplex $\triangle_{N-1}\equiv \triangle_{N-1}({\bf n}_1,\dots, {\bf n}_N)$, whose center coincides with the center of $B_1(\real^{N^2-1})$,  which is only made of vectors representative of states, being a simplex a convex set of vectors (see  Sec.~\ref{The generalized}).

\begin{figure}[!ht]
\centering
\includegraphics[scale =.6]{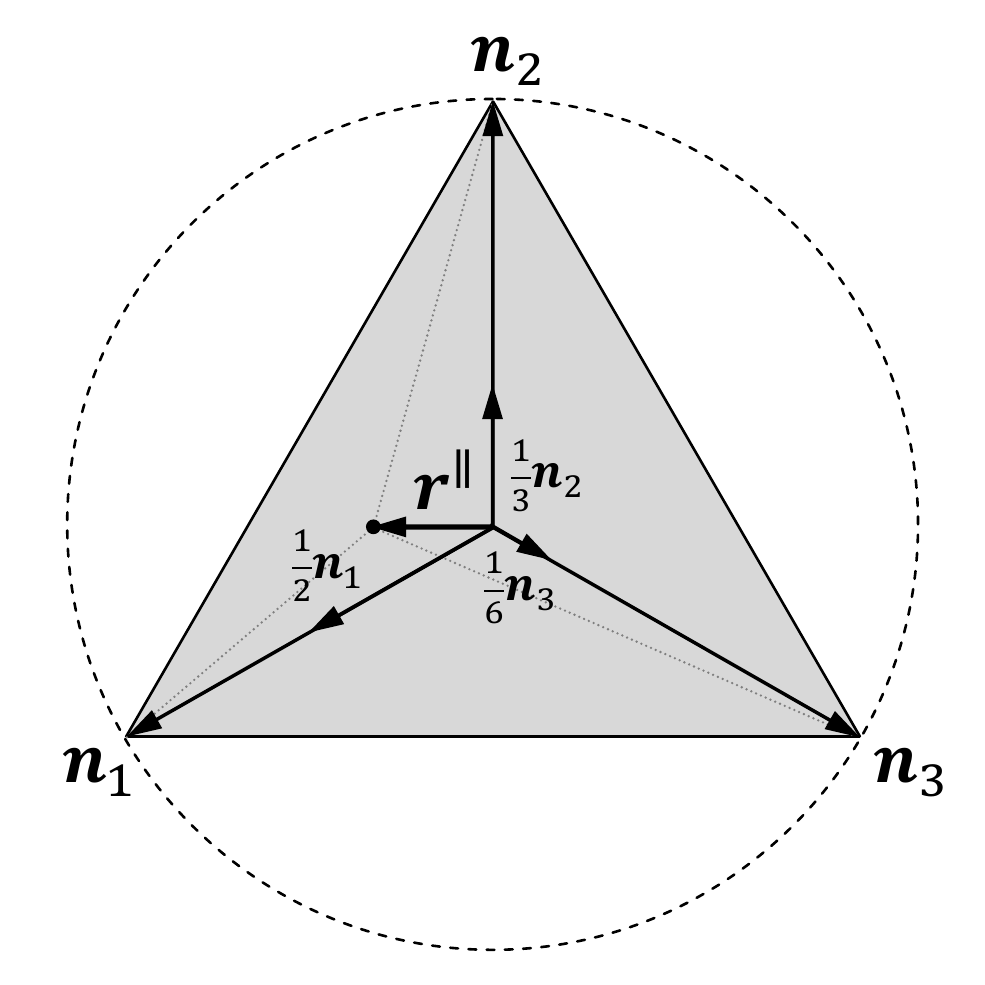}
\caption{A triangular $2$-membrane (in gray color), representing a measurement of a $3$-dimensional quantum entity ($N=3$), inscribed in a unit disk which is a section of the $8$-dimensional unit ball. The orthogonally projected vector ${\bf r}^\parallel$ can always be uniquely written as a convex linear combination of the vertex unit vectors: ${\bf r}^\parallel =r_1^\parallel {\bf n}_1 + r_2^\parallel {\bf n}_2+ r_3^\parallel {\bf n}_3$. In the specific example of the figure: $r_1^\parallel={1\over 2}$, $r_2^\parallel={1\over 3}$, $r_3^\parallel={1\over 6}$. The presence of the particle defines three convex regions, delimited in the picture by the gray dashed lines.
\label{Singletriangle}}
\end{figure}

\begin{figure}[!ht]
\centering
\includegraphics[scale =.55]{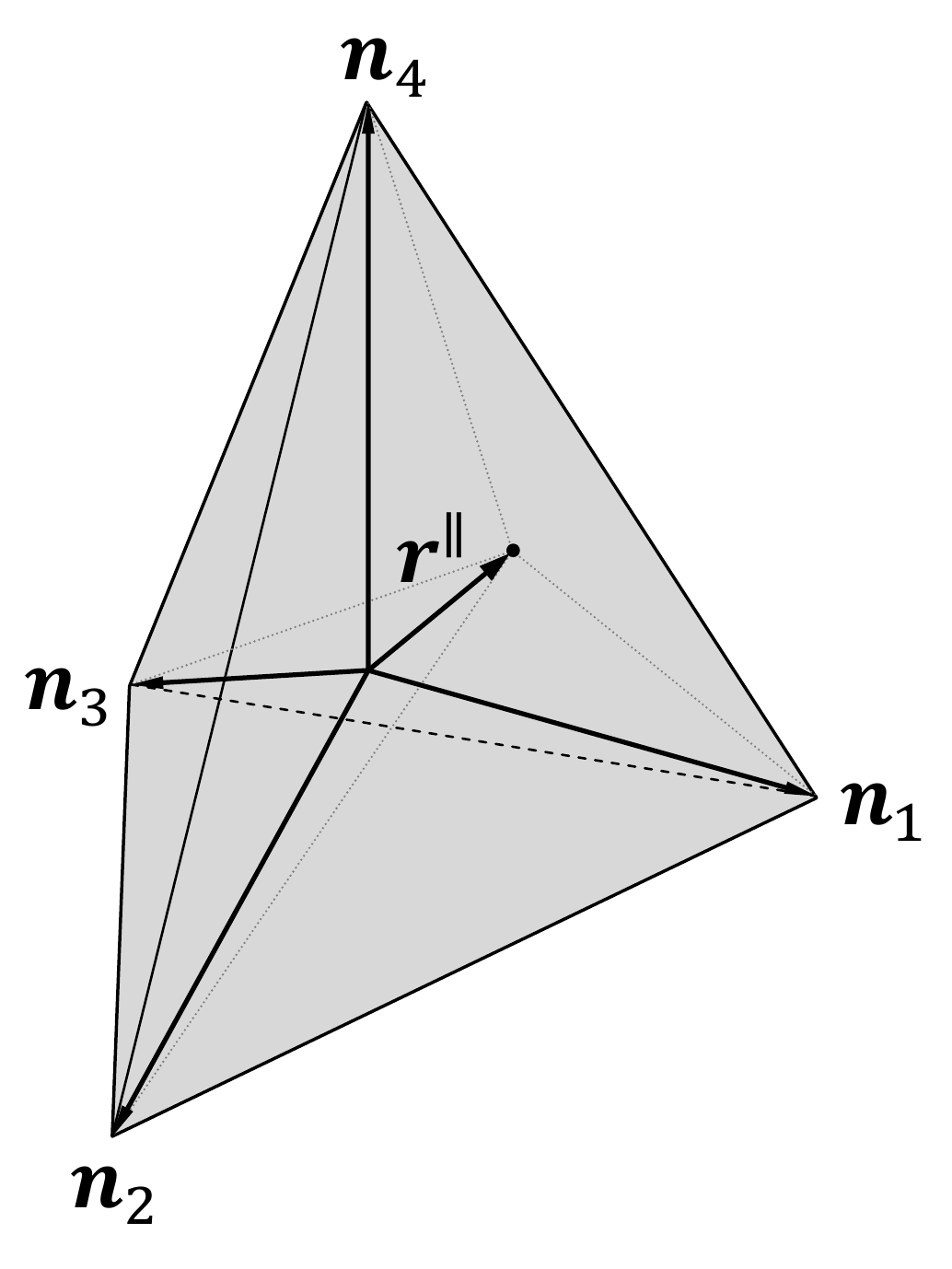}
\caption{The elastic $3$-membrane (in gray color) representing a measurement of a 4-dimensional quantum entity ($N=4$), which can be represented as a substance forming a tetrahedron. The point-particle, orthogonally ``falling'' onto the $3$-membrane, will generally occupy a point inside the tetrahedron, and defines in this way four different convex regions, delimited in the picture by the gray dashed lines. 
\label{Thetraedron}}
\end{figure}

As we did for the $N=2$ case, the measurement of the observable $A$ is represented in $B_1(\real^{N^2-1})$ by a substance  filling the $(N-1)$-simplex, possessing the three properties of being attractive, unstable and elastic, and whose action on the point-like particle describes the effects of the measurement process. For the $N=3$ case, $\triangle_2$ is an \emph{equilateral triangle} inscribed in a $8$-dimensional unit ball, so that the measurement simplex can be described by a $2$-dimensional elastic \emph{membrane} stretched over the three vertex-points ${\bf n}_i$, $i=1,2,3$ (see Fig.~\ref{Singletriangle}).
For the $N=4$ case, $\triangle_3$ is a \emph{tetrahedron} inscribed in a $15$-dimensional unit ball, so that the measurement simplexes can still be visualized as a $3$-dimensional elastic structure  stretched over the four vertexes ${\bf n}_i$, $i=1,2,3,4$ (see Fig.~\ref{Thetraedron}), which we will call a \emph{$3$-membrane}. For $N\geq 5$, however, the \emph{$(N-1)$-membrane} $\triangle_{N-1}$ cannot anymore be fully visualized inside our ordinary three-dimensional space.

\begin{figure}[!ht]
\centering
\includegraphics[scale =.75]{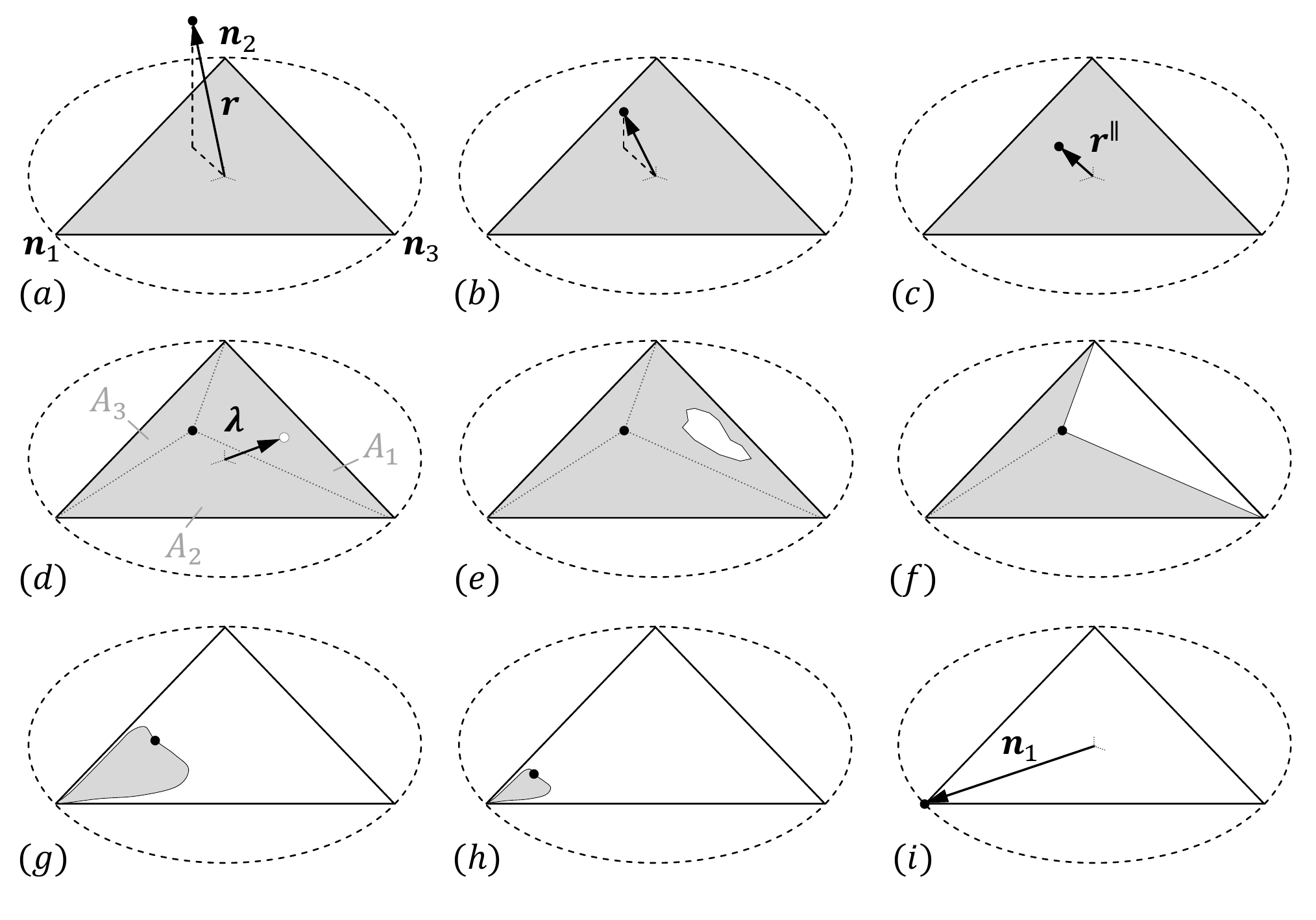}
\caption{The unfolding of a non-degenerate measurement having three distinguishable outcomes, within the ball $B_1(\mathbb{R}^8)$, which cannot be represented in the picture; (a): the $2$-membrane $\triangle_2$ (in gray color) is stretched over the three vertex points  ${\bf n}_1$, ${\bf n}_2$ and ${\bf n}_3$, with the point-particle representative of the state of the entity initially positioned in ${\bf r}$; note that the equilateral triangle, and the disk in which it is inscribed, are here represented in slight perspective; (b): the particle, attracted by the membrane, orthogonally ``falls'' onto it; (c) the particle reaches the on-membrane position ${\bf r}^\parallel$; (d): this gives rise to three distinct convex regions $A_1$, $A_2$ and $A_3$; here it is region $A_1$ which disintegrates first, with the initial point of disintegration indicated by the vector \mbox{\boldmath$\lambda$}; (e) and (f): the  region $A_1$ fully disintegrates; (g) and (h): following the simultaneous detachment of the two anchor points of $A_1$, the membrane shrinks toward point ${\bf n}_1$, bringing with it the point particle; (i): the particle reaches its final location, in ${\bf n}_1$, representative of the outcome of the quantum measurement. 
\label{TheNMeasurement}}
\end{figure}

The measurement consists in the point-particle being subjected to the action of the $(N-1)$-membrane $\triangle_{N-1}$ (see Fig.~\ref{TheNMeasurement} $(a)$) which, being attractive, causes it to move deterministically toward it, following the shortest path, from its initial position ${\bf r}= {\bf r}^\perp +{\bf r}^\parallel$, to position ${\bf r}^\parallel$ (see Fig.~\ref{TheNMeasurement} (a)-(c)), entering in this way into direct contact with the $(N-1)$-membrane and firmly sticking onto it. Assuming that the movement takes place between the instants $t=0$ and $t=t_1$, this ``downward'' movement can  be described by Eq.~(\ref{deterministicprojection}), which remains valid also in the general case. 

Different from the simpler $N=2$ situation, we know that in the general $N\geq 3$ situation not all points of the ball are representative of states. However, considering that, as can be seen from (\ref{deterministicprojection}), all points ${\bf r}(t)$ are convex combinations of the initial point ${\bf r}$ and the final point ${\bf r}^\parallel$ on the membrane, for all $t\in [0,t_1]$, the  path followed by the particle, when it ``orthogonally drops'' onto  $\triangle_{N-1}$, is made of good vectors which are all representative of states, as it should be the case if what we are describing is a physically realizable process.   

Once sticking on $\triangle_{N-1}$, the point-particle gives rise to $N$ line segments connecting its position ${\bf r}^\parallel$ to the different vertex points ${\bf n}_1, \dots, {\bf n}_{N}$, defining in this way $N$ disjoint regions $A_i$, such that $\triangle_{N-1}=\cup_{i\in I_{N}}A_i$, where the $A_i$ are the convex closures of $\{{\bf n}_1, \dots, {\bf n}_{i-1}, {\bf r}^\parallel, {\bf n}_{i+1}, \dots, {\bf n}_{N}\}$, $i=1,\dots,N$, respectively (see Fig.~\ref{TheNMeasurement} (d)). Then, at some time $t_2\geq t_1$, the $(N-1)$-membrane starts disintegrating, initially in a point {\mbox{\boldmath$\lambda$}}. If $\mbox{\boldmath$\lambda$}\in A_i$, this will produce the consequent disintegration of the entire region $A_i$ (see Fig.~\ref{TheNMeasurement} (d)-(f)), so causing the detachment of all its $N-1$ anchor points ${\bf n}_j$, $j\neq i$.  

Here we can think that the physics of the $(N-1)$-membrane is such that the presence of the particle makes it more resistant in those points lying at the border between the different regions $A_i$, so confining, at first, the disintegrative process in the specific region where it begins. We recall that the $(N-1)$-membrane is not a conventional object, but an abstract multidimensional entity describing the effective overall interaction between the measured entity and the measuring system, taking into account  the fluctuations which are present in the experimental context.   

When the $N-1$ anchor points associated with region $A_i$ detach, the $(N-1)$-membrane, being elastic, immediately contracts toward point ${\bf n}_{i}$, that is, toward the only point where it remained attached, pulling in this way  the particle into that position (see Fig.~\ref{TheNMeasurement} (g)-(i)), which corresponds to the outcome of the measurement. 

In other terms, the measurement process consists first in the deterministic immersive movement, from position ${\bf r}$ to position ${\bf r}^\parallel$, from time $t=0$ to $t=t_1$, then in the indeterministic selection, at some instant $t_2$, of a disintegration point \mbox{\boldmath$\lambda$}, which gives rise to an almost deterministic interaction, producing the movement of the particle to its final point ${\bf n}_i$, which is reached at time $t_3$, as described by Eq.~(\ref{deterministicprojection}), which remains  also valid in the general case (provided the measurement is non-degenerate; see the next section).

If we have said that the interaction \mbox{\boldmath$\lambda$} is \emph{almost} deterministic, and not fully deterministic, it is because for those exceptional  \mbox{\boldmath$\lambda$} lying at the boundaries of two regions, it is not a priori defined which one of the regions will disintegrate first, so that the final outcome remains indeterminate. These exceptional points, however, cannot contribute to the evaluation of the transition probability ${\cal P}({\bf r}\to {\bf n}_i)$, which corresponds to the probability ${\cal P}(\mbox{\boldmath$\lambda$}\in A_i)$ that it is region $A_i$ which disintegrates first.

To prove that the process we have described exactly yields the quantum Born rule, we will assume, as we did before for the two-dimensional case, that the disintegration of the $(N-1)$-membrane is described by a \emph{uniform} probability density, that is, that all points of the $(N-1)$-membrane have an equal probability to initiate the disintegration. Then, what we need to show is that the expression:
\begin{equation}
{\cal P}({\bf r}\to {\bf n}_i)\equiv {\mu (A_i)\over \mu(\triangle_{N-1})} = {(N-1)! \over \sqrt{N-1}}\left({N-1\over N}\right)^{\!\!{N\over 2}}  \!\!\! \mu (A_i),
\label{hiddenformula}
\end{equation}
is identical to the Born rule (\ref{trans-general}), where for the second equality we have used: $\mu(\triangle_{N-1})={\sqrt{N-1}\over (N-1)!}\left({N\over N-1}\right)^{N\over 2}$ (see Appendix~\ref{simplexes}). The rest of this section will be dedicated to the proof of the equivalence between (\ref{hiddenformula}) and (\ref{trans-general}). Being this just a technical calculation, the reader can skip it in a first reading of the article and, if so desired, proceed directly to the next section, as this will not compromise the understanding of the rest of the article. 

So, what we need to do is to calculate $\mu (A_i)$, i.e., the Lebesgue measure of a generic region $A_i$. For this, we can use the generalization, for a convex hull, of the formula for the computation of the area of a triangle (as the product of the length of its base times its height times ${1\over 2}$), which in the case of a $(N-1)$-dimensional convex hull $A_i$ becomes: 
\begin{equation}
\label{Ai}
\mu(A_i) = {1\over N-1}\,\mu(\tilde{\triangle}^i_{N-2})\, h^i({\bf r}^\parallel)={1\over N-1} {\sqrt{N-1}\over (N-2)!}\left(N\over N-1\right)^{\!\!{N-2\over 2}} \!\!\!h^i({\bf r}^\parallel).
\end{equation}
In the above formula, $\tilde{\triangle}^i_{N-2}$ is the $(N-2)$-dimensional simplex generated by the $N-1$ unit vectors $\{{\bf n}_1, \dots, {\bf n}_{i-1},{\bf n}_{i+1}, \dots{\bf n}_{N}\}$, $h^i({\bf r}^\parallel)$ is the smallest Euclidean distance between ${\bf r}^\parallel$ and $\tilde{\triangle}^i_{N-2}$, and for the second equality we have used: $\mu(\tilde{\triangle}^i_{N-2})={\sqrt{N-1}\over (N-2)!}\left(N\over N-1\right)^{N-2\over 2}$ (see Appendix~\ref{simplexes}). Replacing (\ref{Ai}) into (\ref{hiddenformula}), we obtain, after a simplification: 
\begin{equation}
{\cal P}({\bf r}\to {\bf n}_i) = {N-1\over N}h^i({\bf r}^\parallel).
\label{toshow-bis}
\end{equation}
In view of (\ref{trans-general}), we thus need to show that: 
\begin{equation}
h^i({\bf r}^\parallel) = {N\over N-1}  \, r_i^\parallel.
\label{toshow-tris}
\end{equation}
To calculate $h^i({\bf r}^\parallel)$, we observe that any point of $\tilde{\triangle}^i_{N-2}$ can be written as: ${\bf y}^{i} =\sum_{j=1\atop j\neq i}^{N}y_j^i \,{\bf n}_j$. 
Considering (\ref{rparallelexpansion}), the vector ${\bf r}^\parallel-{\bf y}^{i}$, on the line connecting ${\bf r}^\parallel$ and ${\bf y}^{i}$, can be written: 
\begin{equation}
{\bf r}^\parallel-{\bf y}^{i}=\sum_{j=1}^{N}b_j \,{\bf n}_j,\quad b_j\equiv r_j^\parallel - y_j^i(1-\delta_{ji}), \,\,\, j=1,\dots,N.
\end{equation}

To find the specific ${\bf y}^{i}$ for which the distance $\|{\bf r}^\parallel-{\bf y}^{i}\|$ is minimal, i.e., for which $\|{\bf r}^\parallel-{\bf y}^{i}\|=h^i({\bf r}^\parallel)$, we observe that, for such vector, ${\bf r}^\parallel-{\bf y}^{i}$ must be orthogonal to all vectors of the form ${\bf n}_j-{\bf n}_k$, with $j,k\neq i$, that is, $({\bf r}^\parallel-{\bf y}^{i})\cdot ({\bf n}_j-{\bf n}_k)=0$, for all $j,k\neq i$. More explicitly, in view of (\ref{scalar-n}), we have, for $j,k\neq i$: 
\begin{equation}
({\bf r}^\parallel-{\bf y}^{i})\cdot ({\bf n}_j-{\bf n}_k)=\sum_{\ell =1}^Nb_\ell\, ({\bf n}_\ell\cdot {\bf n}_j-{\bf n}_\ell\cdot {\bf n}_k)={N\over N-1}\sum_{\ell =1}^Nb_\ell\, (\delta_{\ell j}-\delta_{\ell k}) = {N\over N-1} \left(b_j - b_k\right)=0.
\end{equation}
This  implies that $b_j= b_k$, for all $j,k\neq i$, so that all the coefficients $b_j$, $j\neq i$, must be equal to a same constant $c$. Considering that $\sum_{j=1\atop j\neq i}^{N}{\bf n}_i \cdot {\bf n}_j=-{1\over N-1}\sum_{j=1\atop j\neq i}^{N}=-1$, so that $\sum_{j=1\atop j\neq i}^{N}  {\bf n}_j=-{\bf n}_i$, we obtain: 
\begin{equation}
{\bf r}^\parallel-{\bf y}^{i}=\sum_{j=1}^{N}b_j {\bf n}_j=r_i^\parallel \,{\bf n}_i + c\sum_{j=1\atop j\neq i}^{N} {\bf n}_j=\left(r_i^\parallel -c\right)\,{\bf n}_i,
\end{equation}
and consequently $h^i({\bf r}^\parallel) =\| {\bf r}^\parallel-{\bf y}^{i}\|=|r_i^\parallel -c|\,\|{\bf n}_i\|=|r_i^\parallel -c|$. Therefore, all we need to do is to determine the constant $c$. For this, we observe that  for $k\neq i$, we have $({\bf y}^{i} - {\bf n}_k)\cdot {\bf n}_i =0$, that is: $\sum_{j=1\atop j\neq i}^{N}y_j^i \, {\bf n}_j \cdot {\bf n}_i ={\bf n}_k\cdot {\bf n}_i$, so that, since ${\bf n}_j \cdot {\bf n}_i = {\bf n}_k \cdot {\bf n}_i = -{1\over N-1}$, for $j,k\neq i$, we have: $\sum_{j=1\atop j\neq i}^{N}y_j^i =1$. The $b_j$, for $j\neq i$, being constant, we have $\sum_{j=1\atop j\neq i}^{N}b_j = c\sum_{j=1\atop j\neq i}^{N}=c\,(N-1)$. In other terms:
\begin{equation}
c={1\over N-1}\sum_{j=1\atop j\neq i}^{N}b_j={1\over N-1}\sum_{j=1\atop j\neq i}^{N}\left[r_j^\parallel  - y_j^i (1-\delta_{ji}) \right]= {1\over N-1}\sum_{j=1\atop j\neq i}^{N-1} (r_j^\parallel -y_j^i) ={1\over N-1}\left(1- r_i^\parallel -1\right) = -{r_i^\parallel\over N-1}.
\end{equation}
Thus, we finally obtain: $r_i^\parallel  -c =r_i^\parallel + {r_i^\parallel\over N-1} = {N\over N-1}  r_i^\parallel$,  which is precisely (\ref{toshow-tris}), thus proving that hidden-measurements, selected according to a uniform probability density, produce (here in the situation of a non-degenerate measurement) transition probabilities in perfect accordance with the Born rule.

\section{Hidden-measurements: the general degenerate situation}
\label{degenerate-situation}

In this section, we generalize the previous derivation to also include the possibility of the measurements of degenerate observables. As we recalled in Section~\ref{Operator-states}, this corresponds to  situations where we have $M$ different  disjoint subsets $I_{k}$ of $\{1,\dots,N\}$,  $k=1\dots,M$,  having $M_k$ elements each, with $0\leq M_k\leq N$, and $\sum_{k=1}^{M} M_k=N$, so that the  observable can be written as the sum: $A = \sum_{k=1}^{M} a_{I_{k}} P_{I_{k}}$, with $P_{I_{k}}= \sum_{i\in I_{k}} P_{a_i}$. This means that the experimenter will not be in a position to discriminate between the different eigenvector-states $P_{a_i}$, belonging to a same subset $I_{k}$. 

Such situation of lack of discrimination regarding  certain outcome-states, does not correspond, however, to a non-degenerate situation in which certain outcomes would have simply been identified. If this would be the case, then a degenerate measurement would just be a sub-experiment of a non-degenerate measurement, changing the state of the entity in exactly the same way as the latter does. This, however, is not how things work in quantum mechanics: quantum degenerate measurements do not arise through a mere procedure of identification of certain outcomes, as is clear that they also change the state of the measured entity in a different way than non-degenerate measurements. In other terms, they are not sub-experiments of non-degenerate measurements, but measurements of a genuine different kind.

Note that this change of state aspect of degenerate measurements, which make them processes of a fundamentally different nature than mere sub-measurements, is often overlooked in textbooks of quantum mechanics, and surprisingly has also been overlooked in certain axiomatic approaches to quantum theory, inspired from probability theory, where outcomes were taken as the basic concepts (see the discussion in~\cite{Aerts2002}, and the references cited therein). This is also to point out that the mechanism we propose here, for the description of a degenerate measurement (which in part was explained in~\cite{AertsSassoli2014a,AertsSassoli2014b}),  has not been identified in the previous hidden-measurement approaches, and therefore constitutes one of the innovative contributions of the present model.

So, from the perspective of what happens at the level of the $(N-1)$-membrane, a degenerate measurement has to be \emph{operationally} defined in a different way than a non-degenerate one, and a mere identification of the outcomes will not be sufficient to reproduce the L\"uders-von Neumann projection processes (\ref{Luder-deg}), and the associated transition probabilities (\ref{probA=adegenerate}). Let us see, then, how a degenerate measurement must unfold. 

\begin{figure}[!ht]
\centering
\includegraphics[scale =.75]{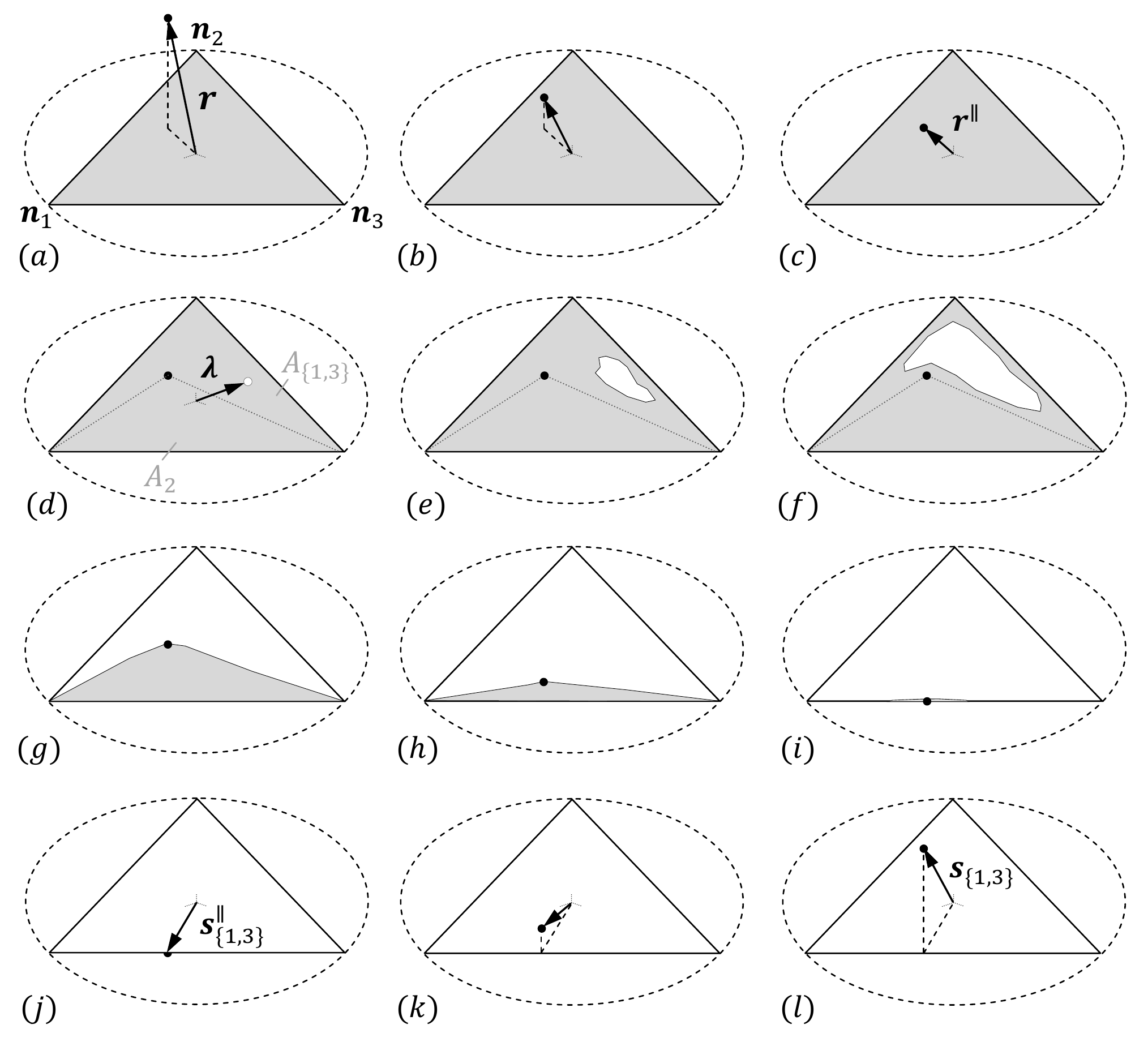}
\caption{The unfolding of the measurement of a degenerate observable $A=a_{\{1,3\}}P_{\{1,3\}} + a_2P_{a_2}$, with $P_{\{1,3\}}=P_{a_1}+P_{a_3}$, which gives rise to two distinguishable outcomes. From (a) to (c), the process is identical to that of the non-degenerate situation described in Fig.~\ref{TheNMeasurement}; (d): the presence of the particle on the membrane gives rise to two regions: $A_2$ and $A_{\{1,3\}}=A_1 \cup A_3$; the initial disintegration of the membrane arises here in $\mbox{\boldmath$\lambda$}\in A_{\{1,3\}}$; (e) and (f): the region $A_{\{1,3\}}$ disintegrates; (g)-(j): following the advanced detachment of the internal anchor point ${\bf n}_2$, the membrane contracts toward the  ${\bf n}_1$-${\bf n}_3$ edge of the $2$-simplex, drawing the particle to position ${\bf s}^\parallel_{\{1,3\}}$; then, following the detachment of the remaining two anchor points, the membrane shrinks toward the particle, without affecting its acquired position; (k) and (l): the deterministic orthogonal reemergence of the particle (here assumed to happen in the same direction as that of the previous immersive process), reaching its final location ${\bf s}_{\{1,3\}}$, representative of the outcome of the quantum measurement. 
\label{TheNMeasurementDeg}}
\end{figure}

Initially, it proceeds in exactly the same way as the measurement of a non-degenerate observable, with the point particle in position ${\bf r}$, at time $t=0$, deterministically moving toward the $(N-1)$-membrane, following the shortest path, thus reaching, at time $t=t_1$, the on-membrane position ${\bf r}^\parallel$ (see Fig.~\ref{TheNMeasurementDeg} (a)-(c) and Eq.~(\ref{deterministicprojection})). 

Once reached position ${\bf r}^\parallel$, described by the convex linear combination ${\bf r}^\parallel=\sum_{i=1}^N r^\parallel_i {\bf n}_i$ (see Eqs.~(\ref{rparallelexpansion}) and (\ref{trans-general})), different from the non-degenerate situation, the point-particle will not generate this time $N$  convex regions, but only $M\leq N$ (not necessarily convex) disjoint regions $A_{I_{k}}=\cup_{i\in I_{k}}A_i$, $k=1\dots,M$, formed by the union of the regions $A_i$ associated with a same subset $I_{k}$ (representative of the eigenspace $P_{I_{k}}{\cal H}$). 

The consequence of this fusion of the regions associated with a same subset $I_{k}$, into a single structure, is that, when a disintegration is initiated in one of them, the process now propagates in all of them, i.e., in the whole $A_{I_{k}}$. In addition to this, the disintegrative process spreads  in a way that the internal anchor points, those which are commonly shared by more than a sub-region of $A_{I_{k}}$, all simultaneously detach before the others, i.e., before those located at the boundaries of $A_{I_{k}}$. 

The above modifications in the functioning of the $(N-1)$-membrane is what operationally distinguishes the measurement of a degenerate observable from a non-degenerate one. Let us see how. As for the non-degenerate situation, at some time $t_2\geq t_1$, the modified $(N-1)$-membrane disintegrates initially in a point {\mbox{\boldmath$\lambda$}}. If {\mbox{\boldmath$\lambda$}} belongs to a region $A_i$ whose index $i$ is in a subset $I_{k}$ which is a singleton, then the process proceeds exactly in the same way as in the non-degenerate situation, as $A_i$ will have no internal anchor points. On the other hand, if $I_{k}$ is not a singleton, the following occurs.

The disintegration of $A_{I_{k}}$ causes the anchor points shared by its sub-regions to detach first, so producing the contraction of the elastic membrane, drawing the particle to the position: 
\begin{equation}
{\bf s}_{I_k}^\parallel =  \sum_{i\in I_{k}} {r^\parallel_i \over \sum_{j\in I_{k}}r^\parallel_j} \,{\bf n}_i,
\label{sIkparallel}
\end{equation}
at time $t_3$, which is a point belonging to the sub-simplex $\tilde{\triangle}_{M_k}$, generated by the $M_k$ unit vectors ${\bf n}_i$, $i\in I_{k}$. Then, also the remaining (exterior) anchor points of $A_{I_{k}}$ simultaneously detach, causing the further shrinking of the $(N-1)$-membrane in the direction of the particle, without affecting its acquired position ${\bf s}_{I_k}^\parallel$ (see Fig.~\ref{TheNMeasurementDeg} (g)-(j)). Assuming that the disintegration happens at time $t=t_2\geq t_1$, and that the contraction of the $(N-1)$-membrane brings the particle to one of the $M$ possible end points ${\bf s}_{I_k}^\parallel$ at exactly time $t_3\geq t_2$, we can modelize this process of collapse by writing:   
\begin{equation}
{\bf r}(t)={\bf r}^\parallel + \frac{t-t_2}{t_3-t_2}\, \Theta(t-t_2) ({\bf s}_{I_k}^\parallel -{\bf r}^\parallel),\quad t\in [t_1,t_3],
\label{deterministicprojectionDeg}
\end{equation}
which generalizes (\ref{deterministicprojection}). 

However, and again differently from the non-degenerate situation, the on-membrane position ${\bf s}_{I_k}^\parallel$ is not yet the final outcome of the measurement process. Indeed, in the degenerate situation (when ${\bf s}_{I_k}^\parallel$ is not a unit vector, i.e., one of the vertex points of $\triangle_{N-1}$), there is an additional process, corresponding to a deterministic movement of emersion of the point-particle from the membrane's hyperplane, bringing it from point ${\bf s}_{I_k}^\parallel$ to the final point ${\bf s}_{I_k}$ (along a rectilinear path orthogonal to $\triangle_{N-1}$; see below), which is the vector representative of the final state of the measurement (see Fig.~\ref{TheNMeasurementDeg} (k) and (l)). This final location ${\bf s}_{I_k}$ of the point-particle, is precisely that representative of the operator-state described by the L\"uders-von Neumann projection formula (\ref{Luder-deg}):
\begin{equation}
\label{Luder-deg-bis}
D({\bf s}_{I_k})\equiv  {P_{I_k} D({\bf r}) P_{I_k}\over {\rm Tr}\, D({\bf r})P_{I_k}},
\end{equation}

In other terms, the degenerate measurement process consists first in a deterministic immersive movement, from position ${\bf r}$ to position ${\bf r}^\parallel$, from time $t=0$ to $t=t_1$, then in the indeterministic selection, at some instant $t_2$, of a disintegration point \mbox{\boldmath$\lambda$}, which gives rise to an almost deterministic interaction, producing the movement of the particle to point ${\bf s}_{I_k}^\parallel$, which is reached at time $t_3$, and finally the process is completed by a  deterministic movement of reemersion, from the on-membrane ``collapsed'' position  ${\bf s}_{I_k}^\parallel$ to the off-membrane position  ${\bf s}_{I_k}$. Assuming that this last process starts at time $t_4\geq t_3$ and ends at time $t_5\geq t_4$, we can write:
\begin{equation}
{\bf r}(t)={\bf s}_{I_k}^\parallel+{t-t_4\over t_5-t_4}\Theta(t-t_4)({\bf s}_{I_k}-{\bf s}_{I_k}^\parallel), \quad t\in [t_3,t_5].
\label{deterministicemersion}
\end{equation}

Again, we stress that the physics governing the dynamics at the level of the $(N-1)$-membrane needs not to be conventional, as this is not an ordinary object, but an effective representation of a measurement context. What is important is that the structure operates in a logical way, so as to produce the effects that we have described. Metaphorically, to explain the difference in functioning of the $(N-1)$-membrane when it is associated with a degenerate measurement, we can think that, once the particle has fallen onto the membrane, so defining the regions $A_i$, the fusion of some of them would be obtained through the application of an additional very reactive substance at their boundary, having the twofold effect of eliminating their separation and causing the simultaneous detachment of the interior anchor points before all the others. 

It is also worth noticing that the particle's movement (\ref{deterministicprojectionDeg}), resulting from the contraction of the $(N-1)$-membrane, drawing it from position ${\bf r}^\parallel\in \triangle_{N-1}$ to position ${\bf s}_{I_k}^\parallel\in  \tilde{\triangle}_{M_k}\subset \triangle_{N-1}$, does not correspond to an orthogonal projection of ${\bf r}^\parallel$ onto the sub-simplex $\tilde{\triangle}_{M_k}$. Indeed, as can be inferred from (\ref{sIkparallel}), the position reached by the particle on  $\tilde{\triangle}_{M_k}$ results from the combined effect of the traction exerted by the anchor points of the $(N-1)$-membrane, corresponding to the vertexes of $\tilde{\triangle}_{M_k}$.

Having said this, we have now to show that the mechanism we have described reproduces the quantum measurement of a degenerate observable. For what concerns the probabilistic predictions, this is straightforward, as the transition probability  ${\cal P}({\bf r}\to {\bf s}_{I_k})$ is equal to the transition probability ${\cal P}({\bf r}^\parallel\to {\bf s}_{I_k}^\parallel)$, which in turn is equal to the probability ${\cal P}(\mbox{\boldmath$\lambda$}\in A_{I_k})$ that region $A_{I_k}$ disintegrates first. Being $A_{I_k}$ the union of $M_k$ \emph{disjoint} regions, we  have ${\cal P}(\mbox{\boldmath$\lambda$}\in A_{I_k})=\sum_{i\in I_k}{\cal P}(\mbox{\boldmath$\lambda$}\in A_{i})$, and considering that we have already shown, in the previous section, that ${\cal P}(\mbox{\boldmath$\lambda$}\in A_{i})={\rm Tr}\, D({\bf r}) P_{a_i}$, we obtain: 
\begin{equation}
{\cal P}({\bf r}\to {\bf s}_{I_k})= \sum_{i\in I_k}{\cal P}(\mbox{\boldmath$\lambda$}\in A_{i})=\sum_{i\in I_k}{\rm Tr}\, D({\bf r}) P_{a_i}= {\rm Tr}\, D({\bf r}) P_{I_k}.
\label{calPbfrto}
\end{equation}

The other aspect we need to check is the consistency of the membrane's mechanism from the viewpoint of the projection postulate. Here we need to observe the following: the projection formula (\ref{Luder-deg-bis}), although it describes a ``collapse'' from state $D({\bf r})$ to state $D({\bf s}_{I_k})$, it cannot be put into direct correspondence with the process of collapse of the $(N-1)$-membrane. This was already the case in the non-degenerate situation, as the membrane's collapse was preceded by the deterministic immersion of the particle. Here, in addition to that, we also have the additional emersion of the particle, following the membrane's collapse. Therefore, formula  (\ref{Luder-deg-bis}) describes the combination of these three processes, which reduce to only two in the non-degenerate situation. It is however possible to show that the L\"uders-von Neumann projection formula, if applied to the on-membrane vector ${\bf r}^\parallel$, does correctly describe the change of  state induced by the collapse of the membrane. In other terms, we do have:
\begin{equation}
D({\bf s}_{I_k}^\parallel)= {P_{I_k} D({\bf r}^\parallel) P_{I_k}\over {\rm Tr}\, D({\bf r}^\parallel)P_{I_k}}.
\end{equation}
To show this, we start by observing that: 
\begin{equation}
D({\bf r}^\parallel)= {1\over N}(\mathbb{I} + c_N\, {\bf r}^\parallel\cdot {\bf \Lambda}) = {1\over N}(\mathbb{I} + c_N\, \sum_{i=1}^Nr_i^\parallel \,{\bf n}_i\cdot {\bf \Lambda})=\sum_{i=1}^Nr_i^\parallel  {1\over N}(\mathbb{I} + c_N\, {\bf n}_i\cdot {\bf \Lambda})=\sum_{i=1}^Nr_i^\parallel  P_{a_i},
\label{Drparallel=riP}
\end{equation}
where for the penultimate equality we have used $\sum_{i=1}^Nr_i^\parallel =1$. Therefore:
\begin{equation}
{\rm Tr}\, D({\bf r}^\parallel)P_{I_k} =\sum_{i=1}^Nr_i {\rm Tr}\,P_{a_i}P_{I_k}= \sum_{j\in I_{k}}r^\parallel_j.
\label{traceDrparallelPIk}
\end{equation}
It also follows from (\ref{Drparallel=riP}) that $P_{I_k} D({\bf r}^\parallel) P_{I_k}=\sum_{i=1}^Nr_i^\parallel  P_{I_k} P_{a_i}P_{I_k}=\sum_{i\in I_{k}}r_i^\parallel P_{a_i}$, so that:
\begin{equation}
{P_{I_k} D({\bf r}^\parallel) P_{I_k}\over {\rm Tr}\, D({\bf r}^\parallel)P_{I_k}} = {\sum_{i\in I_{k}}r_i^\parallel P_{a_i}\over \sum_{j\in I_{k}}r^\parallel_j}= {1\over N}\left(\mathbb{I} + c_N\, \sum_{i\in I_{k}} {r^\parallel_i \over \sum_{j\in I_{k}}r^\parallel_j}\, {\bf n}_i\cdot {\bf \Lambda}\right),
\end{equation}
which is precisely $D({\bf s}_{I_k}^\parallel)$, in view of (\ref{sIkparallel}). 

It is worth mentioning that although the projection formula only describes the change of the state of the point-particle on the membrane (the measurement \emph{per se}), the entire scheme is perfectly consistent, as we have the equality ${\cal P}(D({\bf r})\to D({\bf s}_{I_k})) =  {\cal P}(D({\bf r}^\parallel) \to D({\bf s}_{I_k}^\parallel))$, that is, ${\rm Tr}\, D({\bf r})P_{I_k}  = {\rm Tr}\, D({\bf r}^\parallel)P_{I_k} $, which follows from (\ref{trans-general}), (\ref{calPbfrto}) and (\ref{traceDrparallelPIk}). 

To complete our description of the measurement of a degenerate observable, we have to show that the final deterministic ``upward'' movement (\ref{deterministicemersion}), through which the particle emerges from the membrane's potentiality region, follows  a rectilinear path perpendicular to $\triangle_{N-1}$, as it was perpendicular to $\triangle_{N-1}$ its initial immersive path, which was the consequence of the attractive property of the substance forming the $(N-1)$-membrane. This emersive movement can be thought as being caused by the ``hole'' left by the membrane, which would produce a sort of opposite, repulsive effect on the point-particle.  For this, we define the vector $\Delta{\bf s}_{I_k}\equiv {\bf s}_{I_k} - {\bf s}_{I_k}^\parallel$, and as per (\ref{transitiongeneralNxN}), we  write:
\begin{equation}
{\cal P}(D({\bf s}_{I_k})\to P_{a_i})={1\over N}\left[1+(N-1)\, {\bf s}_{I_k}\cdot {\bf n}_i\right]= {N-1\over N} \, \Delta{\bf s}_{I_k}\cdot {\bf n}_i + {\cal P}(D({\bf s}_{I_k}^\parallel)\to P_{a_i}).
\end{equation}
Since ${\cal P}(D({\bf s}_{I_k})\to P_{a_i})={\cal P}(D({\bf s}_{I_k}^\parallel)\to P_{a_i})$, we must have $\Delta{\bf s}_{I_k}\cdot {\bf n}_i =0$, for all $i=1,\dots,N$, meaning that $\Delta{\bf s}_{I_k}$ is perpendicular to $\triangle_{N-1}$. Thus, we can write, in all generality: ${\bf s}_{I_k} ={\bf s}_{I_k}^\perp+{\bf s}_{I_k}^\parallel$. This means that the particle emerges from the membrane by taking that same path it would have taken, in the opposite direction, if ${\bf s}_{I_k}$ was the vector representative of the initial state to be measured, in relation to the same measurement.

We conclude this section by observing that, if $M=1$, i.e., if the measurement is trivial, in the sense that all regions of the $(N-1)$-membrane are fused into a single region, then the collapse of the $(N-1)$-membrane will leave the particle in ${\bf r}^\parallel$, so that in this case the final emersive movement would be exactly the opposite of the initial immersive movement, so producing the trivial transition ${\bf r}\to {\bf r}^\parallel\to {\bf r}$. On the other hand, if $M=N$, i.e., if the experiment is non-degenerate, then ${\bf s}_{I_k}$ is a unit vector and ${\bf s}_{I_k} ={\bf s}_{I_k}^\parallel$.

\section{The tripartite measurement process}
\label{tripartite}

In the previous sections, we were able to demonstrate that, for an arbitrary number of dimensions $N$, it is always possible to represent a general (possibly degenerate) quantum measurement process within the unit real ball $B_1(\real^{N^2 -1})$, in a way that it results from the combination of three different sub-processes. The first one,  deterministic, is a \emph{decoherence-like} process which can be understood as that phase of the measurement during which the entity is brought as close as possible to the measuring system, where the term ``close'' is here to be understood not in a ``spatial sense,'' but in an ``experimental sense.'' Once this first sub-process is completed, and  the entity has established a full contact with the measuring system, the second sub-process, indeterministic, can unfold. This is that phase of the measurement that we have called the ``naked'' measurement, or the measurement \emph{per se}, and corresponds to  the \emph{actualization of a potential interaction}, producing a change of the state of the entity, genuinely unpredictable, which results from the fluctuations which are intrinsic to the measurement context, and  cannot be controlled by the experimenter. Finally, once this \emph{collapse-like} sub-process has occurred, a third and final sub-process, if appropriate, takes also place, corresponding to a  deterministic, \emph{purification-like} process, through which the entity takes again some (experimental) distance from the measurement context. 

The above tripartite description expresses in specific and precise terms the intuitive idea that a measurement is, in a sense, a process through which the measured entity has first (1)  to approach and enter into contact with the measuring system, then (2) interact with it in some way, and (3) finally separate again from it. In our hidden-measurement modelization, this corresponds to the above mentioned three phases, which can be modelized by the following three-step movement of the point-particle ($0\leq t_1\leq t_2\leq t_3\leq t_4\leq t_5$):
\begin{equation}
{\bf r}(t)=\left\{\begin{array}{l} 
{\bf r}+{t\over t_1}({\bf r}^\parallel - {\bf r}), \quad t\in [0,t_1]\quad {\rm (decoherence)}\\ 
{\bf r}^\parallel + \frac{t-t_2}{t_3-t_2}\, \Theta(t-t_2)({\bf s}_{I_k}^\parallel -{\bf r}^\parallel) ,\quad t\in [t_1,t_3]\quad {\rm (collapse)}\\
{\bf s}_{I_k}^\parallel+{t-t_4\over t_5-t_4}\Theta(t-t_4)({\bf s}_{I_k}-{\bf s}_{I_k}^\parallel), \quad t\in [t_3,t_5]\quad {\rm (purification)}
\end{array}
\right.
\label{tripartitemovement}
\end{equation}

We already mentioned that the details  on how the vector ${\bf r}(t)$, representative of the operator-state $D({\bf r}(t))$,  evolves inside the ball, during the measurement process, are not specifically imposed by the model, so that the piecewise linear trajectories described in (\ref{tripartitemovement}) are only dictated by a choice of simplicity. What is however prescribed by the model, are the three transitions:
\begin{equation}
D({\bf r})\longrightarrow D({\bf r}^\parallel) \Longrightarrow D({\bf s}_{I_k}^\parallel) \longrightarrow D({\bf s}_{I_k}),
\label{threeprocesses}
\end{equation}
where the double arrow indicates that the change of state is indeterministic. About the specific times these transitions occur, this is another aspect that the model doesn't specify. We don't know, for instance, if the disintegration of the $(N-1)$-membrane happens as soon as the point-particle enters into contact with it (so that $t_2=t_1$), or if there is a  waiting time before the disintegration is initiated (so that $t_2>t_1$); we don't know if the disintegration and contraction of the $(N-1)$-membrane is an instantaneous process (so that $t_3=t_2$), or a process that takes some finite time to be accomplished (so that $t_3>t_2$). Similarly, we don't know what is the duration, if any, of the two deterministic, immersive and emersive processes, and if the latter is initiated as soon as the collapse is terminated (so that $t_4=t_3$), or if, again, there is a waiting time before it starts (so that  $t_4>t_3$).

In other terms, we can't deduce from the model if the three distinct sub-processes it describes truly develop in time, and how they do so, or if they are just a way to mathematically describe a unique physical transition which would take place instantaneously or, if not instantaneously, however not necessarily according to such three-step modality. This clearly remains an open question, related to the problem of establishing if the extended sphere-model that we have here presented, provides not only an explanation of quantum probabilities, but also an extension of the quantum formalism, with a finer temporal description of the measurement process. 

This finer description would be in accordance with the view that, being a measurement  a physical process, during which an entity is acted upon by another entity, it should unfold in time in some structured way, although the entire process could very well take place (for all practical purposes) in a too brief time period. Is this structure, in general, described by the three transitions (\ref{threeprocesses})? This, as we said, is an open question, which however is in principle testable. Indeed, we can imagine situations where a measurement is initiated at time $t=0$, and then, after a time interval $\tau$, a second \emph{different} measurement would be performed on the same entity, before the completion of the first measurement. Then, by varying $\tau$ from $0$ to a value greater than $t_5$, and analyzing the $\tau$-dependence of the obtained statistics of outcomes, it should be possible, at least in principle, to probe the unfolding of the first measurement, determining in this way if it is really a process formed by different sub-processes, with different realization times. 

As we know, standard quantum mechanics doesn't say anything about the state of the system between the pre-measurement state $D({\bf r})$, and the post-measurement state $D({\bf s}_{I_k})$: it doesn't specify if there exist intermediary states, hidden within a single ``quantum jump.'' Our hidden-measurement model, on the contrary, predicts that the measurement process produces a continuous change of the state of the entity, passing through the two intermediary states $D({\bf r}^\parallel)$ and $D({\bf s}_{I_k}^\parallel)$, so that the apparently instantaneous quantum jump could possibly be an approximate way to describe a more structured process, which in turn could also be an approximate way to describe an even more complex process, probably differing from measurement to measurement, depending on the nature of the measured entity and of the measuring system.

Of course, an experiment testing the above risks not to be easy to perform, as even if a quantum measurement process does truly unfold in time, according to the above tripartite structure, its effective duration could be too short to allow us to discriminate its predictions from those of standard quantum mechanics. Also, it is not clear how to implement in practice two different measurements in an extremely rapid succession, on a same entity, with the first measurement preserving the physical integrity of the entity, and with the second measurement interrupting the first measurement before its completion. This is a difficult problem of quantum tomography, and  we  invite the experimentalists to  take up the challenge of devising possible viable protocols.

With regard to experiments checking the general validity of the hidden-measurement paradigm, it is worth mentioning a negative result obtained some years ago by Durt et al.~\cite{Durtetal2002}, where  the temporal statistics of successive measurements was analyzed with the intent of bringing out possible \emph{memory effects} in the measuring device, due to the presence of fluctuations. The idea at the basis of the experiment (performed with an atomic interferometer) was to consider the measuring device as an entity whose state would constantly fluctuate in time, thus producing different measurement interactions, when combining with a microscopic entity. Thus, by performing a same measurement on different entities prepared in the same state, in a time sequence which would be more rapid than the fluctuations in the measuring device, one expected to observe correlations between these successive measurements, due to the fact that they would be more likely associated with a same measuring interaction, and therefore to a same outcome (this because the measuring system would remain ``frozen'' in a given state during a typical time interval, characteristic of the fluctuations). 

As we said, these experiments were not conclusive, as temporal correlations could not be observed. This can  be explained in different ways. First of all, it is of course possible that  hidden-measurements, although explaining the emergence of quantum probabilities at the formal level, they would not be implemented in the physical reality, at the microscopic level. Secondly, it is also possible that the fluctuations in the measuring system, although real, would just be too rapid to be detected with our current technology. A third explanation is that it would be incorrect to think of the hidden-measurements as resulting from  fluctuations which are internal to the measuring system, varying continuously in time, independently of its interaction with the measured entity. 

Indeed, it is also possible to think, and probably more correct so, of these fluctuations as resulting from the interplay between the measuring system and the measured entity, in the sense that the process of actualization of a potential interaction would be more akin to what happens during a typical symmetry breaking process. Just to give an elementary example, consider a press bending a cylindrical stick vertically planted on the ground. When pushing on the stick along its longitudinal axis, the press will cause its flexion in an a priori unpredictable direction, so breaking the initial rotational symmetry and creating an ``outcome-direction'' that was only potentially present prior to the ``press-experiment.'' We certainly cannot consider the internal fluctuations of the press, prior to its action on the stick, as being responsible for the selection of a measurement interaction, producing the final outcome-direction. Indeed, the fluctuations which are responsible for the creation of the latter are those which only manifest in the moment the press pushes on the stick, and which result from the complex dynamics produced by their unstable contact. 

The above picture is also strongly suggested by our extended Poincar\'e model. Indeed, each measurement corresponds in the model to a different membrane, as is clear that each measurement produces the destruction (collapse) of the membrane. In other terms, the entire ``potentiality region'' of contact between the measured entity and the measuring apparatus has to be recreated at each measurement, and consequently there cannot be correlations between two successive measurements, even if performed in extremely rapid succession. However, if the second measurement is different from the first, and is performed on the same entity when the first measurement is still unfolding, then, as explained above, the statistics of outcomes could differ from what predicted by standard quantum mechanics, due to the ``movement'' of the entity's state inside the generalized Bloch sphere.  

In that respect, we also emphasize, as anticipated in Sec.~\ref{Operator-states}, that our measurement model strongly suggests that operator-states also admit in quantum mechanics an interpretation as \emph{pure} states, as is clear that our tripartite description of the measurement requires  the pre-measurement vector-state to penetrate the unit ball and reach the internal hypersurface of the $(N-1)$-membrane, transforming in this way into a strict operator-state. Within the paradigm of the model, this passage from a vector-state, on the surface of the ball, to an operator-state, inside the ball, cannot be interpreted as the passage from a pure state to a mixture of states. 

In other terms, in accordance with the difficulty mentioned in Sec.~\ref{Operator-states}, regarding a unique interpretation of operator-states (i.e., density operators)  as statistical mixtures, we can put forward the idea that the set of pure states, in quantum mechanics, is not  formed only by the vector-states (the rays of the Hilbert space), but also, and more generally, by the operator-states. Without entering into the details, let us mention that this completion of the set of pure states of quantum theory would also be suggested by how joint entities are formed in quantum mechanics, in connection to the existence of non-product states. Indeed, it can be shown that it is by accepting that also operator-states can describe pure states, that one can save the physical principle affirming that: if a physical entity is the joint entity of two sub-entities, then it exists at a certain moment if and only if the two sub-entities also exist at that moment, where by ``existing'' it is meant here that they are in one and only one pure state (see~\cite{Aerts2000} and the references cited therein). 

That said, we observe that there is a clear formal correspondence between the decoherence-like process described by the immersion of the point-particle inside the ball, to reach the on-membrane position ${\bf r}^\parallel$ (which causes the off-diagonal elements of the operator-state, in the measurement's basis, to gradually vanish) and the entity's immersion in the environment, as described in \emph{decoherence theory}~\cite{Schlosshauer2005}. Indeed, in view of (\ref{trans-general}), we can write: 
\begin{equation}
D({\bf r}^\parallel)  = \sum_{i=1}^N {\cal P}(D({\bf r})\to P_{a_i}) P_{a_i},
\label{decoheredstate}
\end{equation}
which is precisely the reduced density operator associated, in decoherence theory, with a physical entity once the environment has been traced out  and the decoherence condition employed, so as to produce the zeroing of the off-diagonal terms. 

Despite this formal correspondence, there are fundamental interpretational differences between our approach and that of decoherence theory. First of all, in our model  (\ref{decoheredstate}) is not an approximated state, but the exact state of the entity, whereas in decoherence theory it  corresponds to an approximate description, i.e., to a so-called \emph{improper mixture}, as the superposition does not really disappear, but gets just ``diluted'' into the environment. So, in decoherence theory one still needs to explain how the state of the entity will finally collapse, and this is the reason why the non superficial adherents to decoherence theory are compelled to complete their explanatory scheme by resorting to a many-worlds picture~\cite{Everett}.

This difficulty becomes particularly evident  in our model when we observe that (\ref{decoheredstate}) is just a pre-collapsed state, and not a post-collapsed one, and this independently of the fact that it would constitute an exact or approximate description. In other terms, in our model the decoherence process only describes a preparation stage, through which the entity enters in a deeper (experimental) contact with the measurement context. This deterministic and non-unitary process produces a state, described by the operator (\ref{decoheredstate}), which although \emph{appears} as a classical mixture, cannot and should not be interpreted as such. Indeed, it doesn't describe a situation of ignorance regarding the actual state of the entity, but a situation in which the entity does have a well defined state, but has not yet completed its interaction with the measuring system. In other terms, (\ref{decoheredstate}) describes a state of potentiality (with respect to the possible outcomes of the measurement), whose actualization can only arise through the subsequent selection of a measurement interaction. 

In a nut shell: if decoherence is unable to solve the measurement problem, it is because it only describes the first element of a tripartite process. Also, the completion of the decoherence scheme doesn't require the unlikely hypothesis of the existence of many-worlds, hidden within our reality, the much more reasonable hypothesis of existence of many-measurements, hidden within our measurements, being for this more than sufficient.

\section{Non-spatiality}
\label{Nonspatiality}

Before going deeper in our analysis of the measurement process, by also considering, in the next sections, the possibility of non-uniform mixtures of hidden-measurement interactions, we want  to provide in this section some additional explanations regarding the implications of the hidden-measurement interpretation for what regards the nature of our observational processes, and that of the  entities populating the microscopic layer of our physical reality.  

First of all, we emphasize that the hidden-measurement mechanism is a very natural one when we observe how we interact with the ordinary objects of our everyday life, and interpret these interactions as experiments testing the actuality of certain properties. As a very simple but instructive example, consider the property of \emph{solidity}, to be understood in the most ordinary sense. We all know that  macroscopic objects are more or less \emph{solid}, depending on the material with which they are made,  their shape, etc. And each time an object falls from our hands to the ground, without knowing it we are testing the object's \emph{solidity}. If, when it falls, it doesn't break, we can say that it passed the solidity-test, and that the solidity-property has been  confirmed, whereas if it breaks, it is the opposite property, that of non-solidity (or fragility), which has found confirmation. 

Of course, one thing is to fall from half a meter, and another thing is to fall from two meters; one thing is to fall on a soft mat, and another thing is to fall on a marble floor. In other terms, there exist different properties of solidity, depending on the way they are operationally defined, and consequently experimentally tested. But even when the height and nature of the floor are (conventionally and consensually) defined with great precision, this doesn't guarantee that a given object will always be either solid or non-solid, in the sense that the outcome of the measurement of the solidity-observable would always be predictable in advance. For instance, if the object in question is non-spherical, and/or made of a inhomogeneous material, its orientation relative to the floor, at the moment it is dropped, i.e., its initial state when the solidity-measurement is performed, will certainly influence the outcome.

We can easily imagine that certain orientations will always give, with certainty, either a successful (the object doesn't break) or unsuccessful (the object breaks) result. These orientations can be considered as \emph{eigenstates} of the solidity-observable. But no doubts there are also orientations for which the outcome remains genuinely indeterminate, in the sense that it can only be predicted in statistical terms, and it is correct to  think of these initial orientations as \emph{superposition states}, relative to that specific measurement. This not because they would correspond to an object which, nonsensically, would be at the same time solid and non-solid, but in the sense that both possibilities would be \emph{available} to be actualized during the measurement process. 

Let us explain why it is so, as this is a key point in the understanding of the view that we have proposed in the present work.  When the object is prepared in a superposition state, its orientation expresses a condition of \emph{instability} with respect to the solidity-measurement, in the sense that such orientation will cause the object, when dropped, to hit the floor in a critical point of its structure, such that infinitesimal variations around this point will either result in the object breaking, or remaining intact. Now, considering that a human operator, when dropping an object to the floor, cannot \emph{control} all aspects of the interaction between her/his hands and the object, so as to avoid infinitesimal fluctuations that would slightly change the impact point, it is clear that each measurement of the solidity observable will be, technically speaking, a \emph{different measurement}, even though, apparently, they all look like the same measurement. 

The existence of this multiplicity of measurement processes, hidden within a same measurement, is of course without consequence if they are all \emph{equivalent}, in the sense that they all produce the same outcome, given a same initial state. But not all initial states, as we said, are of this kind. Some are very sensitive even to the slightest change in the measurement interaction, which by slightly altering the impact point can produce a totally different outcome. So, even though these measurement interactions are associated with \emph{deterministic} (or almost deterministic) processes, the way they are selected will be in general non-deterministic, and that's why the solidity-measurement can literally create the property it measures, unless the entity was prepared in an eigenstate. 

It is worth mentioning that the hidden-measurement explanation is also a very natural one in the modelization of human cognitive processes, as  in part demonstrated by the recent success of quantum models of cognition and decision~\cite{BusemeyerBruza2012,Aerts2009,AertsSassoli2014a,AertsSassoli2014b}. In this ambit, the subdivision of the measurement (decision) process in different (deterministic and indeterministic) phases, corresponds to what is subjectively perceived by a human subject when confronted with  a \emph{decision context}, which always involves a deterministic phase, during which the \emph{conceptual entity} subjected to the decision is first brought into the ``right position,'' for a choice to be made, and only then the indeterministic process corresponding to the decision \emph{per se} (the naked measurement) can unfold, in a way which is also many times perceived as a sort of ``random breaking mechanism,'' releasing a ``cognitive tension.'' 

We shall not enter into the details of these interesting cognitive mechanisms, nowadays intensively investigated, as this would clearly go beyond the scope o the present article, and we refer the reader to~\cite{BusemeyerBruza2012,Aerts2009,AertsSassoli2014a,AertsSassoli2014b}, and the references cited therein. Our purpose here was just to point out that both in our interactions with the macroscopic objects of our everyday life, and with the more abstract conceptual entities of our mental reality, the hidden-measurement explanation appears to be a powerful one, and this certainly adds plausibility to its applicability also in the description of our interactions with the microscopic entities, in the modern physics' laboratories. 

One of the difficulties in adopting the hidden-measurement explanation, is that many physicists, because of their classical training, are led to understand a measurement as a process of pure \emph{discovery}, and not, also, as a process of \emph{creation}. This ``classical prejudice'' is what has probably prevented us, for quite a long time, from recognizing that many of our observations are irreducibly invasive and indeterministic processes. The indeterministic aspect is formalized in our extended sphere-model by the unstable membrane, whose disintegration is beyond the control power of the experimenter, whereas the invasive aspect is represented by the membrane's attraction, collapse and repulsion (when absent), which produces the change in the position of the point particle, representative of the entity's state,  inside the unit ball. 

When we say that a quantum measurement is to be understood as an intrinsically invasive process, this should not be  understood, however, in the reductive sense  of a \emph{disturbance} which would only follow from the fact that, in practice, it is  difficult to measure something about a system without altering it in some way. A typical example is the checking of the pressure of an automobile tire, which is notoriously difficult to do without letting out some of the air in the process, thus altering the very physical quantity which was meant to be measured. This typology of effects of disturbance is certainly relevant in many domains of physics, but can usually be reduced by improving the observational techniques. The invasiveness we are here referring to, in relation to quantum measurements, is in fact of a much more fundamental nature, being incorporated in the very protocols which operationally define the properties to be measured, so that it cannot be eliminated, not even in principle~\cite{Sassoli2014}.

But there is more. Consider for instance the measurement of the spin of a microscopic entity. If the spin in question is a spin $1/2$, then one can always say that, prior to the measurement, the entity did possess a spin orientation, i.e., that it was in a spin eigenstate (up or down) relative to some (possibly unknown) direction, and that the measurement only produced a transition from that eigenstate to another eigenstate, so that the process is in principle interpretable as a ``classical disturbance,'' where a \emph{possessed} value of a physical quantity (here the spin orientation) is changed in another \emph{possessed} value, as a consequence of the interaction with the measuring device.

However, as emphasized many times in this article, this is just a ``pathology'' of the two-dimensional situation, where each point of the Bloch sphere is representative of a bona fide vector-state. Things dramatically change when one considers entities of dimension greater than $N=2$. Since the spin operators do not form anymore a basis (with the unit operator)  for all hermitian matrices beyond the two-dimensional case, not all states of a spin-$s$ entity, with $s\geq 1$, will  necessarily correspond to spin eigenstates. This means that there exist states such that the entity does not possess a specific spin orientation, and therefore such orientation would be \emph{literally created by the measurement process}. In other terms, a ``quantum disturbance'' cannot be interpreted in classical terms, as in general it doesn't only change the value of a physical quantity, but it also creates a value when no prior value was possessed by the entity, before the measurement. This is the crux of the measurement problem, sometimes referred to as the \emph{objectification problem}, or also as the \emph{observer effect}~\cite{Sassoli2013}, which should be better named \emph{instrument-effect},  to avoid possible misunderstandings regarding what is effectively producing the objectification. 

So, the hidden-measurement interpretation, which we emphasize is just a natural extension of the standard quantum formalism, although on the one hand it solves the conundrum of the quantum measurement problem, on the other hand it asks us to abandon some of our classical prejudices regarding the true nature of the physical entities populating the microscopic layer of our reality. Now, for as long as it is only question of discrete observables, like spin observables, the situation may appear still acceptable to many physicists. However, the same discourse also  applies to continuous observables, like the position observable. 

If a position measurement is a process during which a spatial position (or more generally a localization inside a spatial region) is literally created, and not simply discovered, or perturbed in a classical sense, then we must accept that entities like electrons, protons, neutrons, and even more complex atomic structures (when not organized into macroscopic objects), pass most of their time in a \emph{non-spatial} condition, i.e., they are typically \emph{non-spatial entities}. In other terms, the hidden-measurement interpretation pushes the quantum mystery from the problem of understanding the measurement process, to that of understanding the nature of a non-spatial entity, and how the different spatial and non-spatial entities can relate and organize.

A possible approach toward the solution of this problem has been recently indicated by one of us, in what is called the \emph{conceptuality interpretation of quantum mechanics}. In this approach,  quantum entities are understood as entities having a \emph{conceptual nature}, with varying degrees of abstractness (or concreteness)~\cite{Aerts2009a,Aerts2010}, and their interaction regime with the macroscopic measuring systems are considered as being of the \emph{language type}. Being a ``conceptual entity'' does not mean, however, being a ``human concept.'' It simply means that a quantum entity and a human concept would share a same ``way of being,'' which in essence would be conceptual, in the same way as an electromagnetic wave and a sound wave, despite being very different physical entities, do share the same ``undulatory way of being.'' 

The conceptuality interpretation solves, among other things, the ``non-spatiality problem,'' as conceptual entities and their combinations, because of their nature, need not being ``inside space,'' although they are able to enter space when they reach their most possible concrete (objectual) state (Heisenberg uncertainty principle being then the consequence of the fact that a conceptual entity cannot be simultaneously very abstract and very concrete). Again, it is beyond the scope of the present article to analyze the important notion of non-spatiality~\cite{{Sassoli2014,Aerts1998b,Aerts1999,Sassoli2013d,Aerts1990,Sassoli2012}}, and go into the merits of the views it may originate. With this brief mention of the conceptuality interpretation, we just wanted to point out that once we abandon some of the prejudices of the past, regarding the nature of a physical entity, new explanatory scenarios  become available, and viable.

In this regard, we observe that there is still a considerable amount of resistance, among physicists, in acknowledging that non-locality would be a manifestation of non-spatiality, and in renouncing the idea that our physical reality must necessarily be staged in a three-dimensional theater, or a four-dimensional one, if also relativistic effects are considered. Let us mention, as a paradigmatic example, the approach carried out by David Bohm, today known as \emph{Bohmian mechanics}~\cite{Bohm1952a,Bohm1952b}, which historically was the  first serious attempt  to look for an aspect of the microphysical reality that could explain the emergence of quantum indeterminism, consistently with the no-go theorems, which however was limited by the constraint of obtaining (without success) a spatial description of intrinsically non-spatial entities.  

In the Bohmian model, a quantum entity is described as a point-particle having always a well defined position and momentum. Contrary to our model, the Bohmian point-particle is not just an abstract representation of the state of the quantum entity: it is, in a very literal sense, the quantum entity. The trajectory that such entity follows, however, is non-Newtonian, being described by a law of motion which also involves a \emph{quantum potential}, defined by the wave function, which evolves independently from the particle, according to the Schr\"odinger equation.  

The purely deterministic Bohmian model becomes a statistical model by assuming that the initial position of the point-particle is distributed according to the probability density given by the square modulus of the wave function in configuration space, following a so-called \emph{quantum equilibrium hypothesis}. The reason why the model proposed by Bohm was not in conflict with  the no-go theorems, is that in his approach the quantum observables were not anymore considered to be measurable quantities associated with the properties of the point-particle quantum entity, but only mathematical objects encoding statistical information. Indeed, in Bohmian mechanics the only real physical quantities are the relative positions and momenta of the point-particle, which however remain inaccessible to the experimenter, because of the uncertainty principle. 

The classical picture of waves and particles proposed by Bohm faces  a serious problem when one attempts to describe more than a single quantum entity, as in this case the wave function (defining the quantum  potential which guides the movement of the particles) doesn't act any longer in a three-dimensional Euclidean space, but in a configuration space of higher dimension. Bohm  was perfectly aware of this conceptual difficulty, and he considered for this his model not very plausible, but just a preliminary version of yet another model to come~\cite{Bohm1957}.

In retrospect, we can say that the  strength of the approach of Bohm has been its ability to overcome the obstacle of the hidden-variable impossibility proofs, but its weakness has been that of not having paid enough attention to another -- this time really insurmountable -- obstacle: the impossibility of fitting the whole of physical reality inside the ordinary three-dimensional space. Indeed, as demonstrated by the model presented in this article, it is precisely by abandoning the idea that the microscopic entities should necessarily be \emph{spatial entities}, that one can open to the possibility of describing quantum measurements as processes of actualization of potential (hidden) measurement interactions, and have access to what we think are more advanced explanatory schemes.

\section{Non-uniform fluctuations}
\label{Non-uniform-fluctuations}

In the first part of this article, we have introduced a generalized Poincar\' e-Bloch representation~\cite{Arvind1997, Kimura2003, Byrd2003, Kimura2005, Bengtsson2006, Bengtsson2013}), valid for an arbitrary number of dimensions $N$, and then we have extended such generalized representation to also include a full modelization of the measurement process, based on a hidden-measurement mechanism. Also, in the previous section, we have briefly analyzed some of the consequences of the hidden-measurement modelization for our view of the physical reality. 

So far, in our modelization we have only considered a uniform mixture of measurement interactions. In this section, we want to relax this hypothesis that the selection of a hidden interaction \mbox{\boldmath$\lambda$}  necessarily results from a uniform probability density, i.e., from the (relative) uniform (Lebesgue) measure of the subregions $A_i$ of the membrane's simplex $\triangle_{N-1}$. Indeed, if we take seriously the physical picture subtended by our extended Poincar\' e-Bloch sphere-model, there are no a priori reasons to assume that the substance forming the $(N-1)$-membrane would be \emph{uniformly}  unstable, apart of course that this is precisely what one needs to derive the Born rule. In other terms, if on one hand we can say that quantum probabilities are the consequence of uniform fluctuations in the measurement situation, on the other hand we still have to explain why these fluctuations would be uniform. 

This is what we will do in the present and following sections, thus completing our explanation of the origin of quantum probabilities. More precisely, we will show that uniform fluctuations can be explained as resulting from an average over all possible forms of fluctuations, associated with all possible non-uniform probability densities, describing all possible ways a $(N-1)$-membrane can in principle disintegrate. This means that quantum measurements can be understood as measurements of a very general kind, called \emph{universal measurements}~\cite{Aerts1998b,Aerts1999,AertsSassoli2014a,AertsSassoli2014b}, describing conditions of maximal lack of knowledge (or of control), where not only the interaction which is each time actualized would remain unknown (or not controllable), but also \emph{the way} such interaction would be each time selected, among the available ones. 

Let us start by considering the $N=2$ case. As we have seen in Sec.~\ref{HiddenmeasurementsN=2}, the transition probability ${\cal P}({\bf r}\to {\bf n}_i)$ corresponds to the probability ${\cal P}(\mbox{\boldmath$\lambda$}\in A_i)$ that the elastic band initially disintegrates in region $A_i$, $i=1,2$. If the band is made of a uniform substance, the probability is simply given by the ratio $\mu (A_i)/\mu(\triangle_1)$. On the other hand, if the substance forming the band is non-uniform, so that its disintegrability is described by a non-uniform probability density $\rho(\mbox{\boldmath$\lambda$})$, the transition probabilities become: 
\begin{equation}
{\cal P}({\bf r}\to {\bf n}_1|\rho) = \int_{A_1} \rho(\mbox{\boldmath$\lambda$}) d\mbox{\boldmath$\lambda$} = \int_{-1}^{r_1^\parallel -r_2^\parallel} \!\!\! \rho(x) dx,\quad\quad  {\cal P}({\bf r}\to {\bf n}_2|\rho) = \int_{A_2} \rho(\mbox{\boldmath$\lambda$}) d\mbox{\boldmath$\lambda$} = \int_{r_1^\parallel -r_2^\parallel}^{1}  \rho(x) dx,
\label{ptransition2dimnonuniform}
\end{equation}
where we have used ${\bf r}^\parallel = r_1^\parallel {\bf n}_1 +r_2^\parallel {\bf n}_2 = (r_1^\parallel -r_2^\parallel)\,{\bf n}_1$,  the fact that $A_1$ (resp., $A_2$) is the line segment lying between ${\bf r}^\parallel$ and $- {\bf n}_1$ (resp., between ${\bf n}_1$ and ${\bf r}^\parallel$), and we have set $\mbox{\boldmath$\lambda$}= x\, {\bf n}_1$, with $-1\leq x\leq 1$. The probability density $\rho(x)$ obeys the normalization condition $\int_{-1}^1 \rho(x) dx=1$, and (\ref{ptransition2dimnonuniform}) generalizes the quantum probability (\ref{trans-general}), as is clear that for a uniform probability density $\rho_u(x)={1\over 2}$, one finds back the Born rule, ${\cal P}({\bf r}\to {\bf n}_i|\rho_u)= r_i^\parallel$, considering that $r_1^\parallel +r_2^\parallel=1$.

To give some examples of non-uniform  $\rho(x)$, let $\rho(x) = \delta(x-x_0)$, $x_0\in[-1,1]$, which corresponds to a situation where the band can disintegrate only in a single predetermined point $x=x_0$, so that only a single (almost) deterministic interaction -- a so-called \emph{pure measurement}~\cite{AertsSassoli2014a} -- would be available to be selected. As a consequence, the measurement outcomes are described by probabilities which can only take the values $1$ or $0$, depending on the initial state of the system (excluding the special case of an initial state described by a point-particle ``falling'' exactly on the disintegration point $x_0$).  

Another interesting example is that of an elastic band whose disintegrability is described by a double-Dirac distribution: $\rho(x) =  c_1\,\delta(z+1) + c_2\,\delta(x-1) $, $c_1+c_2=1$, $c_1,c_2\geq0$. This corresponds to a situation where the band can only disintegrate in its two end points, so that the outcome probabilities are $ {\cal P}({\bf r}\to {\bf n}_i|\rho) = c_i$, $i=1,2$, and do not depend on the initial state of the entity. This is a typology  of  measurements -- called \emph{solipsistic}~\cite{AertsSassoli2014a} -- which can reveal nothing about the state of the entity before the experiment, as is clear that only the structure of the hidden interactions is relevant to determine the value of the probabilities.

As a third paradigmatic example, consider a $1$-membrane which can uniformly disintegrate only in its central region $[-\epsilon,\epsilon]$,  $\epsilon\in [0,1]$, as per the probability density $\rho_\epsilon(x)={1\over 2\epsilon}\chi_{[-\epsilon,\epsilon]}(x)$, where $\chi_{[-\epsilon,\epsilon]}(x)$ is the characteristic function of  $[-\epsilon,\epsilon]$. A simple calculation yields for the transition probabilities: 
\begin{equation}
{\cal P}({\bf r}\to {\bf n}_1|\rho) =\left\{\begin{array}{ll} 
0, & r_1^\parallel-r_2^\parallel \in [-1,-\epsilon]\\ 
{1\over 2}(1+{r_1^\parallel-r_2^\parallel\over \epsilon}), &r_1^\parallel-r_2^\parallel \in [-\epsilon,\epsilon]\\
1, & r_1^\parallel-r_2^\parallel \in [\epsilon,1]
\end{array}
\right., \quad
{\cal P}({\bf r}\to {\bf n}_2|\rho) =\left\{\begin{array}{ll} 
1, & r_1^\parallel-r_2^\parallel \in [-1,-\epsilon]\\ 
{1\over 2}(1-{r_1^\parallel-r_2^\parallel\over \epsilon}), &r_1^\parallel-r_2^\parallel \in [-\epsilon,\epsilon]\\
0, & r_1^\parallel-r_2^\parallel \in [\epsilon,1]
\end{array}
\right..
\label{epsilonprob}
\end{equation}
In the limit $\epsilon\to 0$, $\rho_\epsilon(x)\to\delta(x)$, and we recover the classical situation of a pure measurement, whose transition probabilities can only be $0$ or $1$, here associated with a band which  can only initially disintegrate in its middle point $x=0$. On the other hand, in the opposite $\epsilon\to 1$ limit, $\rho_\epsilon(x)\to\rho_u(x)$, and we recover the pure quantum situation described by uniform fluctuations, which in the special case of an initial vector ${\bf r}$ representative of a vector-state, gives the well-known spin-${1\over 2}$ transition probabilities: 
\begin{equation}
{\cal P}({\bf r}\to {\bf n}_1|\rho) ={1\over 2}(1+\cos\theta)= \cos^2 {\theta\over 2}, \quad\quad 
{\cal P}({\bf r}\to {\bf n}_2|\rho) =
{1\over 2}(1-\cos\theta)= \sin^2 {\theta\over 2},
\label{epsilon=1}
\end{equation}
where we have used ${\bf r}^\parallel = (r_1^\parallel -r_2^\parallel)\,{\bf n}_1 = \|{\bf r}\|\cos\theta$, and the fact that $\|{\bf r}\|=1$ for a vector-state.

This last example is known as the \emph{$\epsilon$-model}, and has been extensively studied in the past, as it constitutes a very simple paradigmatic example of an intermediary measurement situation which is neither classical nor quantum~\cite{Aerts1998,Aerts1999,Sassoli2013b}. Indeed, as we are now going to prove, when $\epsilon\neq 0,1$, the model can neither be described by a classical Kolmogorovian probability model, nor by a Hilbertian probability model, and corresponds to a genuinely different situation, of intermediate knowledge. 

\begin{figure}[!ht]
\centering
\includegraphics[scale =.9]{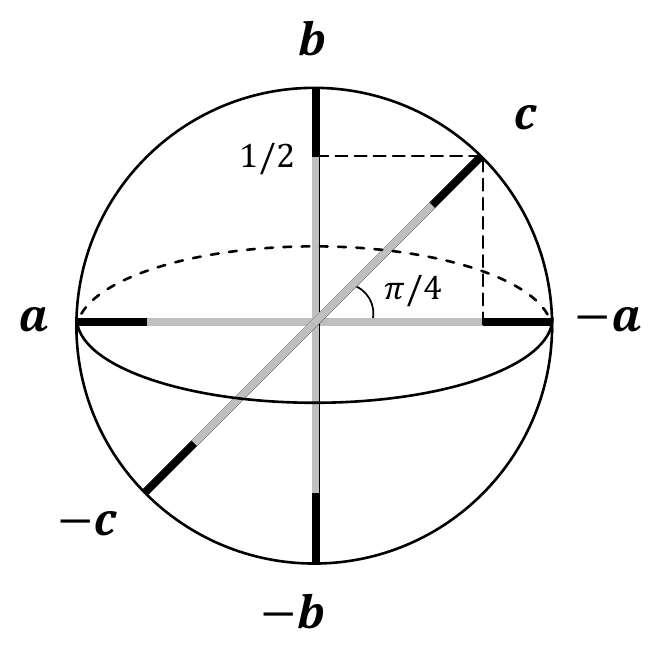}
\caption{The three measurements of the observables $A$, $B$ and $C$, represented by three non-uniform elastic bands oriented along the directions ${\bf a}$, ${\bf b}$ and ${\bf c}$, respectively. The angle between ${\bf c}$ and ${\bf b}$ is ${\pi\over 4}$, and the angle between ${\bf b}$ and ${\bf a}$ is ${\pi\over 2}$. The non-disintegrable segments of the band are in black color, the uniformly disintegrable segments in gray color, and the picture corresponds to the choice $\epsilon = {\sqrt{2}\over 2}$ (so that the disintegrable segments are of length $1$). 
\label{threeMes}}
\end{figure}

We start by showing that the probability model associated with $\rho_\epsilon(x)$ cannot be fitted into a classical probability model. Consider three arbitrary events $a$, $b$ and $c$ (which are the elements of a $\sigma$-algebra of subsets of a sample space), and their complements $\bar{a}$, $\bar{b}$ and $\bar{c}$. Then, according to classical probability theory (obeying Kolmogorovian axioms), if ${\cal P}$ is a probability measure,  the following three equalities must evidently hold: 
\begin{eqnarray}
\label{setEqualities}
&{\cal P}(b\cap c) = {\cal P}(a\cap b\cap c)+{\cal P}(\bar{a} \cap b\cap c)\label{eq1}, \\ 
&{\cal P}(a\cap c) = {\cal P}(a\cap b\cap c)+{\cal P}(a \cap \bar{b} \cap c)\label{eq2},\\
&{\cal P}(\bar{a}\cap b) = {\cal P}(\bar{a} \cap b\cap c)+{\cal P}(\bar{a} \cap b \cap \bar{c}).\label{eq3}
\end{eqnarray}
Eq. (\ref{eq1}) minus (\ref{eq2}) gives: ${\cal P}(\bar{a} \cap b\cap c) = [{\cal P}(b\cap c) - {\cal P}(a\cap c)] + {\cal P}(a \cap \bar{b} \cap c)$, and since the last term is positive, we have: ${\cal P}(\bar{a} \cap b\cap c)\geq [{\cal P}(b\cap c) - {\cal P}(a\cap c)]$. Also, from (\ref{eq3}), we deduce that: ${\cal P}(\bar{a} \cap b\cap c)\leq {\cal P}(\bar{a}\cap b)$, and putting together these two  inequalities, we find: $[{\cal P}(b\cap c) - {\cal P}(a\cap c)]\leq {\cal P}(\bar{a} \cap b\cap c)\leq {\cal P}(\bar{a}\cap b)$, which implies: 
\begin{equation}
\label{setEqualities6}
[{\cal P}(b\cap c) - {\cal P}(a\cap c)]\leq {\cal P}(\bar{a}\cap b). 
\end{equation}
So, if we can prove that  (\ref{setEqualities6}) is violated by the $\epsilon$-model,  we have also proved that it cannot be fitted into a classical probabilistic model. For this, we assume that $\epsilon \in [0,{\sqrt{2}\over 2}]$, and  consider the following three observables: 
\begin{equation}
\label{threeobservables}
\begin{array}{lll}
A=aP_{a}+\bar{a}P_{\bar{a}}, &\quad P_{a}={1\over 2}\left(\mathbb{I} + {\bf a}\cdot\mbox{\boldmath$\sigma$}\right), &\quad P_{\bar{a}}={1\over 2}\left(\mathbb{I} - {\bf a}\cdot\mbox{\boldmath$\sigma$}\right),\\
B=bP_{b}+\bar{b}P_{\bar{b}}, &\quad P_{b}={1\over 2}\left(\mathbb{I} + {\bf b}\cdot\mbox{\boldmath$\sigma$}\right), &\quad P_{\bar{b}}={1\over 2}\left(\mathbb{I} - {\bf b}\cdot\mbox{\boldmath$\sigma$}\right),\\
C=cP_{c}+\bar{c}P_{\bar{c}}, &\quad P_{c}={1\over 2}\left(\mathbb{I} + {\bf c}\cdot\mbox{\boldmath$\sigma$}\right), &\quad P_{\bar{c}}={1\over 2}\left(\mathbb{I} - {\bf c}\cdot\mbox{\boldmath$\sigma$}\right),
\end{array}
\end{equation}
where ${\bf a}$, ${\bf b}$ and ${\bf c}$ are three unit vectors belonging to a same plane, and such that there is an angle of ${\pi\over 2}$ between ${\bf b}$ and ${\bf c}$, and an angle of ${\pi\over 4}$ between ${\bf a}$ and ${\bf b}$, as illustrated in Fig.~\ref{threeMes}. 
It is worth observing that although we have written the above three observables in the standard quantum notation, they have now to be understood as generalized observables (still associated, however, with measurements of the first kind), whose measurements are governed by non-uniform $1$-membranes, described by $\rho_\epsilon(x)$, changing the state of the entity in a different way than the uniform $1$-membranes associated with pure quantum measurements. 

Keeping this in mind, we assume that the point-particle representative of the state of the system is initially located in position ${\bf c}$, corresponding to an eigenstate of observable $C$. Therefore, if we measure $C$, we will obtain the outcome associated with eigenvalue $c$, with certainty, i.e., ${\cal P}_{\bf c}(c) =1$. The probability of obtaining outcome $b$, when we measure $B$, with the particle in state ${\bf c}$, is also equal to $1$, i.e., ${\cal P}_{\bf c}(b) =1$, since the particle orthogonally projects onto the non-disintegrable segment with end point {\bf b}, as one can check on Fig.~\ref{threeMes}. Therefore, given that the particle is initially in state ${\bf c}$, the joint probability that we obtain first outcome $c$, and then outcome $b$, in a sequential measurement, is: ${\cal P}_{\bf c}(c\,{\rm and}\, b)={\cal P}_{\bf c}(b){\cal P}_{\bf c}(c)= 1$. On the other hand, since ${\cal P}_{\bf c}(a) =0$ (see Fig.~\ref{threeMes}), we also have ${\cal P}_{\bf c}(c \,{\rm and}\, a)={\cal P}_{\bf c}(a){\cal P}(c)_{\bf c}= 0$. Finally, considering that when we measure $A$, a particle in position ${\bf b}$ ``falls'' exactly onto the middle point of the band, we have ${\cal P}_{\bf b}(\bar{a}) = {1\over 2}$, and ${\cal P}_{\bf c}(b\,{\rm and}\, \bar{a})={\cal P}_{\bf b}(\bar{a}){\cal P}_{\bf c}(b) = {1\over 2}$. Considering that $[1-0]>{1\over 2}$, we thus obtain:
\begin{equation}
[{\cal P}_{\bf c}(c\,{\rm and}\, b) -{\cal P}_{\bf c}(c\,{\rm and}\, a)]>{\cal P}_{\bf c}(b\,{\rm and}\, \bar{a}),
\label{setEqualities7}
\end{equation}
which is  a  violation of (\ref{setEqualities6}).

Note that, to keep the proof simple, in our counter example we have only used probabilities of sequential measurements, here interpreted as joint probabilities. An alternative \emph{ex absurdum} proof, using instead conditional probabilities, can also be constructed, considering a situation where there is an additional lack of knowledge about the state of the particle, described by a uniform probability distribution. Then, choosing again three suitable measurements, one can show that Bayes' rule for conditional probabilities is violated. The calculation of conditional probabilities, with an additional mixture of states, is however much more involved, and we refer the interested reader to \cite{Aerts1986} and \cite{Aerts1995}.

Let us now show that the sequential probabilities generated by non-uniform $1$-membranes described by $\rho_\epsilon(x)$ cannot be fitted, either, into a Hilbertian probability model. According to (\ref{probA=anondegenerate}), if the initial state of the system is $P_{c}$, we can write, with obvious notation:
\begin{eqnarray}
\label{setEqualities8}
&1= {\cal P}_{\bf c}(c\,{\rm and}\, b) = ({\rm Tr}\, P_{c}P_{b})\, ({\rm Tr}\, P_{c}P_{c}) = {\rm Tr}\, P_{c}P_{b}=|\langle b|c\rangle|^2,\\
&1= {\cal P}_{\bf c}(c\,{\rm and}\, \bar{a}) =  ({\rm Tr}\, P_{c}P_{\bar{a}})\, ({\rm Tr}\, P_{c}P_{c}) = {\rm Tr}\, P_{c}P_{\bar{a}}=|\langle \bar{a}|c\rangle|^2,\label{setEqualities9}\\
&{1\over 2}=  {\cal P}_{\bf c}(b\,{\rm and}\, a) = ({\rm Tr}\, P_{b}P_{a})\, ({\rm Tr}\, P_{c}P_{b}) ={\rm Tr}\, P_{b}P_{a} =|\langle a|b\rangle|^2,
\label{setEqualities10}
\end{eqnarray}
where for (\ref{setEqualities9}) we have used ${\cal P}_{\bf c}(\bar{a})=1$, implying ${\cal P}_{\bf c}(c\,{\rm and}\, \bar{a}) = {\cal P}_{\bf c}(\bar{a}){\cal P}_{\bf c}(c)= 1$, and for (\ref{setEqualities10}) we have used (\ref{setEqualities8}), and the fact that ${\cal P}_{\bf b}(a)={1\over 2}$ implies ${\cal P}_{\bf c}(b\,{\rm and}\, a) ={\cal P}_{\bf b}(a)  {\cal P}_{\bf c}(b)  = {1\over 2}$. From  (\ref{setEqualities8}), we deduce  ${\rm Tr}\, P_{\bar c}P_{b}=0$, so that $\langle \bar{c}|b\rangle =0$; from (\ref{setEqualities9}), we deduce  ${\rm Tr}\, P_{\bar c}P_{\bar a}=0$, so that $\langle a|c\rangle =0$; and (\ref{setEqualities10}) implies  $\langle a|b\rangle \neq 0$. Therefore:
\begin{eqnarray}
\label{setEqualities11}
0\neq \langle a|b\rangle = \langle a|c\rangle\langle c|b\rangle+\langle a|\bar{c}\rangle\langle \bar{c}|b\rangle=0,
\end{eqnarray}
which is clearly a contradiction, showing that the above sequential transition probabilities are not compatible with, and thus cannot be described by, the Born rule. 

Note that it is possible to analyze the above three measurements also for the values $\epsilon\in [{\sqrt{2}\over 2}, 1]$, and show that (\ref{setEqualities7}) continues to hold, which means that for all values $\epsilon\in [0,1]$, the $\epsilon$-model is non-Kolmogorovian. Using Accardi-Fedullo inequalities \cite{AccardiFedullo1982}, it is also possible to show that, in general, no Hilbert space model exists, except for the uniform $\epsilon =1$ case \cite{Aertsetal1999}. 

What the simple example of the $1$-membranes characterized by a probability density $\rho_\epsilon(x)$ already reveals, is the structural richness of those regions of reality -- that we may call the \emph{mesoscopic} regions -- lying somehow in between the purely classical, deterministic descriptions, corresponding to measurement situations with zero lack of knowledge about the interaction between the measuring system and the entity under study ($\epsilon\to 0$), and purely quantum, indeterministic ones, corresponding to situations of maximal lack of knowledge, characterized by a uniform probability density ($\epsilon\to 1$). It also explains why a consistent theory for these mesoscopic regions could not be built within orthodox theories, be them classical or quantum, as the  probability models they are associated with are generally non-Kolmogorovian and non-Hilbertian.

\section{Universal averages}
\label{universal}

In the previous section, we have shown, on the basis of the two-dimensional  $\epsilon$-model example, that when non-uniform probability densities are used to describe how a membrane can possibly disintegrate, one will generally obtain probability models which are non-Hilbertian and non-Kolmogorovian, with the Hilbertian model corresponding only to the very special situation of a uniform probability density $\rho_u(x)$. What we will now show is that as well as the Born rule can be explained in terms of a selection mechanism involving a \emph{uniform} distribution of hidden (almost) deterministic interactions, this same uniform distribution can in turn be explained as resulting from a \emph{universal average} over all possible ways of selecting an interaction, each one described by a different probability density $\rho(x)$. This means that a quantum measurement is interpretable as a \emph{universal measurement}~\cite{AertsSassoli2014a,AertsSassoli2014b}, i.e., as the expression of a process of actualization of potential properties arising at the meta-level. 

To do so, we have to confront with the problem of defining and determining such huge universal average over all possible uncountable $\rho(x)$ -- which will be denoted $\langle\,\cdot\,\rangle^{\rm univ}$ -- without being trapped in insurmountable technical issues related to the foundations of mathematics. The strategy we will follow is similar to what has been historically done in the definition of the \emph{Wiener measure}, which attributes probabilities to continuous-time random walks by considering them as suitable limits of discrete-time processes. Indeed, to define the universal average, we will also proceed by first considering discretized (cellular) probability densities $\rho_{n}(x)$, whose uniform average will be  well-defined, and only in the end take the infinite $n\to\infty$ limit. 

To show that such procedure is consistent, in the sense that it allows to include  all possible probability densities, so that  the obtained average is truly universal, our first step will be to show that all $\rho(x)$ (including the generalized functions) can be approximated by discretized (cellular) probability densities $\rho_n(x)$, in the sense that, given a $\rho(x)$, and the two subsets $A_1=[-1,x_p]$, $A_2=[x_p,1]$, defined by the coordinate $x_p\equiv r_1^\parallel -r_2^\parallel$ of the point-particle on the elastic band, we can always find a suitable sequence of cellular  $\rho_{n}(x)$, such that: 
\begin{equation}
\label{limitntoinfty}
 {\cal P}(x\in A_i|\rho) = \lim_{n\to\infty}{\cal P}(x\in A_i|\rho_n), \quad i=1,2.
\end{equation}
By ``cellular probability density'' we mean a probability density describing a structure made of $n$ regular cells,  tessellating $\triangle_1 \equiv  [-1,1]$ (or more generally $\triangle_{N-1}$, in the general $N$-dimensional situation, as we will explain later). These $n$ cells are only of two sorts: they are such that $\rho_{n}(x)$ is equal to a constant inside them (the same constant for all cells), or such that $\rho_{n}(x)$ is equal to zero inside them, which corresponds, respectively, to the situation of uniformly disintegrable cells and non-disintegrable cells.

Once proven that all $\rho(x)$ can be approximated, with arbitrary precision, in this way, our second step will be that of studying the average over all possible cellular structures (excluding the totally non-disintegrable structures, only made of non-disintegrable cells, as they produce no outcomes in a measurement). Given a  $n \in \mathbb{N}$, the total number of possible $\rho_{n}(x)$ is: $C_{n}^0 + C_{n}^1 + C_{n}^2 + \cdots + C_{n}^{n} -1 = 2^{n}-1$. Thus, for each $n$, we can unambiguously define the  average probability: 
\begin{equation}
\label{average1}
\langle {\cal P}(x\in A_i|\rho_n)\rangle\equiv {1\over 2^{n}-1}  \sum_{\rho_{n}}{\cal P}(x\in A_i|\rho_n),  \quad i=1,2,
\end{equation}
where the sum runs over all possible $2^{n}-1$ cellular probability densities $\rho_{n}(x)$, formed by exactly $n$ cells (excluding the totally non-disintegrable one). Clearly, $\langle {\cal P}(x\in A_i|\rho_n)\rangle$ is the probability for the transition ${\bf r}\to {\bf n}_{i}$, when a cellular $1$-membrane with $n$ cells is chosen at random to perform the measurement, among all possible $n$-cellular $1$-membrane structures. Then, considering that the $\rho_{n}(x)$ are dense in the space of probability densities, in the sense specified above, the third and last step, in the definition of the universal average, consists in taking the infinite-cell limit: 
\begin{equation}
\label{average-limit}
\langle {\cal P}(x\in A_i)\rangle^{\rm univ}\equiv\lim_{n\to\infty} \langle {\cal P}(x\in A_i|\rho_n)\rangle,  \quad i=1,2.
\end{equation}
Such limit, as we will prove, exactly yields the probability ${\cal P}(x\in A_i|\rho_u)$, associated with a uniform probability density $\rho_u(x)$, thus showing that quantum probabilities can be interpreted as being the result of measurements of the universal kind, in the sense that universal measurements, performed on physical entities whose set of states has an Hilbertian structure, produce exactly the same values for the outcome's probabilities as those predicted by the Born rule.

\subsection{Limits of cellular structures}
\label{limit of cellular structures}

In this subsection we show that, given an arbitrary $\rho(x)$, we can always find a suitable sequence of cellular $\rho_n(x)$, such that (\ref{limitntoinfty}) holds. For this, we partition the interval $[-1,1]$ into $n=m\ell$ elementary cells: 
\begin{eqnarray}
\label{elemintcells}
&[-1,-1+{2\over n}], [-1+{2\over n},-1+{4\over n}],\dots,[-1+(j-1){2\over n},  -1+j{2\over n}],\dots,[1-{2\over n},1].
\end{eqnarray}
These elementary cells, of length ${2\over n}$, are in turn contained in $m={n\over \ell}$ larger intervals, of length ${2\over m} = {\sqrt{2}\ell\over n}$, which are the following:
\begin{eqnarray}
\label{cells}
&[-1,-1+{2\over m}], [-1+{2\over m},-1+{4\over m}],\dots,[-1+(i-1){2\over m},  -1+i{2\over m}],\dots,[1-{2\over m},1].
\end{eqnarray}
In other terms, each cell $S_i\equiv [-1+(i-1){2\over m},  -1+i{2\over m}]$ is made of $\ell$ elementary cells:
\begin{eqnarray}
\label{elemcells2}
&\sigma_{i,j}\equiv [-1 + (i-1){2\over m}+(j-1){2\over n},  -1+(i-1){2\over m}+ j{2\over n}],
\end{eqnarray}
so that $S_i=\bigcup_{j=1}^{\ell} \sigma_{i,j}$, $[-1,1] =\bigcup_{i=1}^{m} S_i = \bigcup_{i=1}^{m}\bigcup_{j=1}^{\ell} \sigma_{i,j}$ (see Fig.~\ref{cells}, for an example with $m=4$, $\ell = 5$ and $n=20$).

We assume that the coordinate $x_p$ of the point-particle on the elastic band is such that $x_p\in (-1+(j-1) {2\over m},-1+j {2\over m}]$, for some given $j$. This condition can also be expressed as $(x_p+1){m\over 2}\in (j-1,j]$. Therefore, by definition of the \emph{ceiling function}, $\lceil (x_p+1){m\over 2}\rceil=j$, and we can write: 
\begin{equation}
\label{partition4}
{\cal P}(x\in A_1|\rho)= \int_{-1}^{-1+(j-1){2\over m}} \!\rho(x) dx +\int_{-1+(j-1){2\over m}}^{x_p} \rho(x) dx =\! \sum_{i=1}^{\lceil (x_p+1){m\over 2}\rceil-1}\!\int_{S_i}  \rho(x)dx +r_{m}(x_p|\rho),
\end{equation}
where the rest:
\begin{equation}
\label{restrho}
r_{m}(x_p|\rho)= \int_{-1 +(\lceil (x_p+1){m\over 2}\rceil-1){2\over m}}^{x_p}\rho(x)dx
\end{equation}
tends to zero as $m\to\infty$, considering that 
\begin{equation}
\label{ceilinglimit}
\lim_{n\to\infty}{\lceil nx\rceil \over n}=x.
\end{equation}

At this point, we introduce the following cellular probability density ($n =m\ell$):
\begin{equation}
\label{cellular}
\rho_{m\ell}(z)={\chi_{m\ell}(x)\over \int_{-1}^{1}\chi_{m\ell}(x)dx} ={n\over 2n^d}\,\chi_{m\ell}(x),
\end{equation}
describing a cellular elastic band made of $n=m\ell$ elementary cells, which can only be of two sorts: disintegrable ($d$), or non-disintegrable ($\bar{d}$); see Fig.~\ref{cells}. Here $\chi_{m\ell}(x)$ denotes a step-like function, taking the constant values $1$ (for the disintegrable cells) or $0$ (for the non-disintegrable cells) inside each interval $\sigma_{i,j}$, and $n^d$ is the total number of disintegrable elementary cells of the structure. For a cellular probability density of this kind, (\ref{partition4})  becomes:
\begin{figure*}[!ht]
\centering
\includegraphics[scale =1]{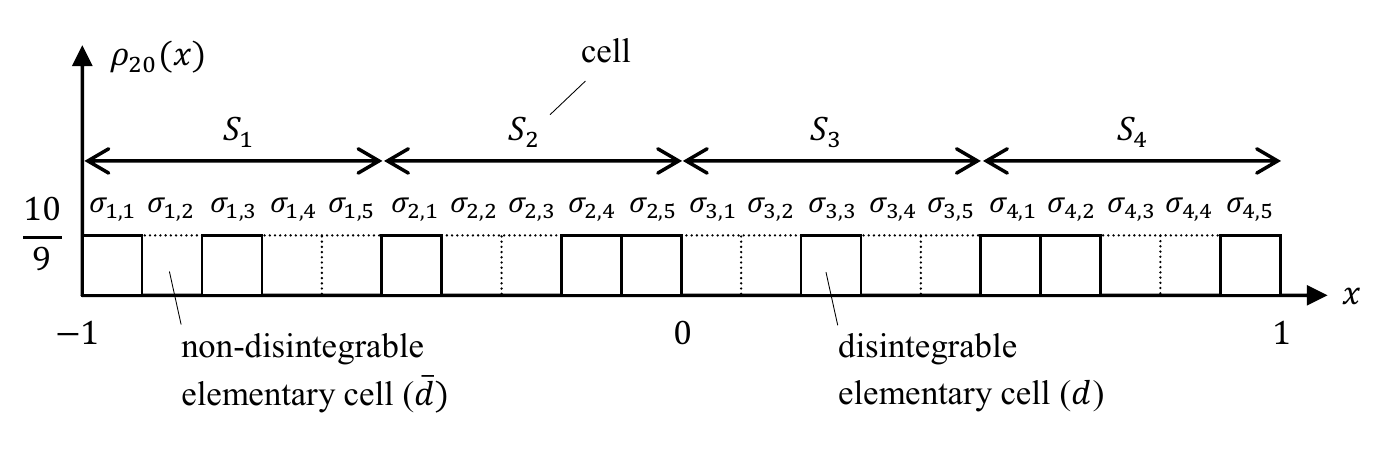}
\caption{An example of a cellular probability density $\rho_{20}(x)$, made of $n=20$ elementary cells, subdivided into four cells, each one made of five elementary cells. Inside the $n^d =9$ disintegrable elementary cells, $\rho_{20}(x)$ takes a constant value, equal to ${10\over 9}$, whereas inside the $n-n^d =11$ non-disintegrable elementary cells, $\rho_{20}(x)=0$. Since each elementary cell has width ${1\over 10}$, we have the normalization: $\int_{-1}^1 \rho_{20}(x) dx=  9 {10\over 9}{1\over 10}=1$.
\label{cells}}
\end{figure*}
\begin{equation}
\label{partition5}
{\cal P}(x\in A_1|\rho_{m\ell})=  \sum_{i=1}^{\lceil (x_p+1){m\over 2}\rceil-1}{n_i^{d}\over n^d} +r_{m}(x_p|\rho_{m\ell}),
\end{equation}
where $n_i^{d}$ is the number of disintegrable elementary cells in $S_i$, and the rest is defined as in (\ref{restrho}). Comparing (\ref{partition4}) with (\ref{partition5}), we obtain:
\begin{equation}
{\cal P}(x\in A_1|\rho) - {\cal P}(x\in A_1|\rho_{m\ell})= \sum_{i=1}^{\lceil mx_1\rceil}\left(\int_{S_i} \rho(x)dx - {n_i^{d}\over n^d}\right)+[r_{m}(x_p|\rho)-r_{m}(x_p|\rho_{m\ell})].\label{difference-m-l}
\end{equation}

All we need to do is to observe that we can always choose $\rho_{m\ell}(x)$ in such a way that ${n_i^{d}\over n^d}\to\int_{S_i} \rho(x)dx$, as $\ell\to\infty$, for all $i=1,\dots,m$. This is so because, for each $i$, the probability $\int_{S_i} \rho(x)dx$ is a real number in the interval $[0,1]$, and rational numbers of the form ${n_i^{d}\over n^d}$, with $0\leq n_i^{d}\leq n^d$, $n^d>0$, are dense in $[0,1]$. Therefore, by a convenient choice of the sequence $\rho_{m\ell}(x)$, taking  the limit $\ell\to\infty$, the sum in (\ref{difference-m-l}) vanishes, and taking the limit $m\to\infty$, also the two rests in (\ref{difference-m-l}) vanish, so that we can conclude that (\ref{limitntoinfty}) holds, i.e., that we can always find a suitable sequence of probability densities $\rho_{n}(x)\equiv \rho_{m\ell}(x)$, describing  structures made of disintegrable and non-disintegrable elementary cells, such that in the infinite-cell limit $n\to\infty$ they produce exactly the same probabilities as $\rho(x)$. Of course, the same reasoning holds true, \emph{mutatis mutandis},  for outcome ${\bf n}_2$, as it can also immediately be deduced from ${\cal P}(x\in A_2|\rho_n)= 1-{\cal P}(x\in A_1|\rho_n)$.

\subsection{Averaging over cellular structures}
\label{Averaging over finite cellular structures}

In the previous subsection we have shown that a measurement performed by means of a non-uniform $1$-membrane characterized by an arbitrary $\rho(x)$, can always be understood as a suitable limit of measurements performed by using discretized structures $\rho_{n}(x)$, having a finite number ${n}$ of disintegrable and non-disintegrable elementary cells, when the number  $n$ of these elementary cells tends to infinity. Our next step is to determine the average probabilities $\langle {\cal P}(x\in A_i|\rho_n)\rangle$, describing measurement situations where the cellular structure is not given a priori, but selected at random among all possible structures with a given number ${n}$ of elementary cells, which we recall can either be of the disintegrable ($d$) or non-disintegrable ($\bar{d}$) kind.

For simplicity, we assume that the point-particle is located between two elementary cells, so that we can write: $A_1\equiv A_1(i)=[-1,-1 +(n-i){2\over n}]$, and $A_2\equiv A_2(i)=[-1 +(n-i){2\over n}, 1]$, for some given $i\in\{0,1,\dots,n\}$, with $A_1(i)$ containing  $n-i$ elementary cells and $A_2(i)$ containing $i$ elementary cells. We want to show that $\langle {\cal P}(x\in A_1(i)|\rho_n)\rangle ={\cal P}(x\in A_1(i)|\rho_{u;n})$, where $\rho_{u;n}(x)$ is the probability distribution characterizing a uniform structure made of $n$ elementary cells all of the disintegrable kind. This will be sufficient to prove (\ref{average-limit}), considering that ${\cal P}(x\in A_1(i)|\rho_{u;n})$ tends to ${\cal P}(x\in A_1(i)|\rho_{u})$, as $n\to\infty$. 

Now, for a uniform cellular structure, we clearly have ${\cal P}(x\in A_1(i)|\rho_{u;n}) ={n-i\over n}$, as is clear that in this case the number of  disintegrable elementary cells in $A_1(i)$ (resp., in the entire interval $[-1,1]$) is equal to the total number $n-i$ (resp., $n$) of elementary cells it contains, so that what we need to prove is: 
\begin{equation}
\label{averagerho2-tris}
\sum_{\rho_{n}}{\cal P}(x\in  A_1(i)|\rho_n)=\left({2^{n}-1} \right) {n-i\over n}.
\end{equation}

The cases $i=0$ and $i=n$ are trivial, since $A_1(0) =[-1,1]$, and $A_1(n)=\{-1\}$, so that ${\cal P}(x\in  A_1(0)|\rho_n)=1$, and ${\cal P}(x\in  A_1(n)|\rho_n)=0$, for all $n$. Thus, we only need to consider the values $i=1,\dots,n-1$. For simplicity, we adopt the following notation: ${\cal P}(i|c_1\dots c_n)\equiv {\cal P}(x\in  A_1(i)|\rho_n)$, where  $(c_1\dots c_n)$, $c_j\in\{d,\bar{d}\}$, denotes the sequence of disintegrable $(d)$  and non-disintegrable $(\bar{d})$ elementary cells characterizing $\rho_n(x)$; see Fig.~\ref{cells}. For $i=1$, (\ref{averagerho2-tris}) becomes: 
\begin{equation}
\label{i=1}
\sum_{(c\cdots)} P(1|c\cdots) = (2^n-1){n-1\over n},
\end{equation}
and we can write:
\begin{equation}
\sum_{(c\cdots)} P(1|c\cdots) =\sum_{(\bar{d}\cdots)} P(1|\bar{d}\cdots)+ \sum_{(d\cdots )} P(1|d\cdots ),
\label{cdots}
\end{equation}
where the first sum in the r.h.s. of the equation runs over all cellular probability densities $\rho_n(x)$ starting with a left non-disintegrable $(\bar{d})$ elementary cell, and the second sum runs over all the $\rho_n(x)$ starting with a left disintegrable $(d)$ elementary cell. We observe that all probabilities in the first sum are equal to $1$, implying that the sum is equal to $2^{n-1}-1$. Also, the second sum can be written as $\sum_{k=0}^{n-1} {k\over k+1} {n-1 \choose k}$, and using a symbolic computational program (like Mathematica, of Wolfram Research, Inc.), one finds the exact identity:
\begin{equation}
\label{Wolfram}
\sum_{k=0}^{n} {k\over k+1} {n \choose k} = {2^n (n-1) +1\over n+1},
\end{equation}
so that (\ref{cdots}) becomes:
\begin{equation}
\sum_{(c\cdots)} P(1|c\cdots) = 2^{n-1}-1 +  {2^{n-1} (n-2) +1\over n} = (2^n-1){n-1\over n},
\end{equation}
which proves (\ref{i=1}). 

To prove (\ref{averagerho2-tris}) for all $i\in\{1,\dots,n-1\}$, we can reason by recurrence. As we have shown that the equality holds for $i=1$, we assume it holds for some $i$, and have to show that this implies that it also holds for $i+1$. We write:
\begin{equation}
\label{sumgeneral}
\sum_{(c \cdots)} P(i+1|c\cdots) =\sum_{(\cdots \bar{d}\cdots)} P(i+1|\cdots \bar{d}\cdots)
+ \sum_{(\cdots d\cdots )} P(i+1|\cdots d\cdots ),
\end{equation}
where the first sum, in the r.h.s. of the equation, runs over all $\rho_n(x)$ having a non-disintegrable $(i+1)$-th elementary cell, and the second sum  runs over all $\rho_n(x)$ having a disintegrable $(i+1)$-th elementary cell. Considering that $P(i+1|\cdots \bar{d}\cdots)=P(i|\cdots \bar{d}\cdots)$, we can write for the first sum:
\begin{eqnarray}
\label{sum-final}
\sum_{(\cdots \bar{d}\cdots)} P(i+1|\cdots \bar{d}\cdots)&=&\!\!\!\sum_{(\cdots \bar{d}\cdots)} P(i|\cdots \bar{d}\cdots) = \!\sum_{(\cdots \bar{d}\cdots)} P(i|\cdots \bar{d}\cdots) +\!\sum_{(\cdots d\cdots)} P(i|\cdots d\cdots)-\!\sum_{(\cdots d\cdots)} P(i|\cdots d\cdots)\nonumber\\
&=&\sum_{(c\cdots)} P(i|c\cdots)-\sum_{(\cdots d\cdots)} P(i|\cdots d\cdots) =  (2^n-1){n-i\over n}-\sum_{(\cdots d\cdots)} P(i|\cdots d\cdots),
\end{eqnarray}
where for the second equality we have added and subtracted the same quantity, and for the last equality we have used (\ref{averagerho2-tris}) and the recurrence hypothesis. Then, (\ref{sumgeneral}) becomes:
\begin{equation}
\sum_{(c \cdots)} P(i+1|c\cdots) =(2^n-1){n-i\over n} +\sum_{(\cdots d\cdots)} \left[ P(i+1|\cdots d\cdots) - P(i|\cdots d\cdots)\right].
\label{difference}
\end{equation}

Denoting $k_i$ the number of disintegrable elementary cells at the right of the $i$-th cell, and $k$ the total number of disintegrable elementary cells, for a cellular probability density of the $(\cdots d\cdots)$ kind, we have $P(i|\cdots d\cdots)={k_i\over k}$, and $P(i+1|\cdots d\cdots)={k_i-1\over k}$, so that the difference of probabilities in (\ref{difference}) is equal to $-{1\over k}$, and is independent of $k_i$. Using the exact identity (which again can be obtained using a symbolic computational program, like Mathematica, of Wolfram Research, Inc.):
\begin{equation}
\label{Wolfram3}
\sum_{k=0}^{n} {1\over k+1} {n \choose k}={2^{n+1}-1\over n+1},
\end{equation}
we therefore obtain
\begin{equation}
\label{sumgeneral3}
-\sum_{(\cdots d\cdots)} {1\over k(\cdots d\cdots)} =-\sum_{k=0}^{n-1} {1\over k+1} {n-1 \choose k}=-{2^{n}-1\over n},
\end{equation}
and inserting (\ref{sumgeneral3}) into (\ref{difference}), we finally obtain:
\begin{equation}
\sum_{(c \cdots)} P(i+1|c\cdots) =(2^n-1){n-i\over n}-{2^{n}-1\over n}=(2^n-1){n-(i+1)\over n},
\label{final}
\end{equation}
which proves that (\ref{averagerho2-tris}) also holds for $i+1$, thus completing the recurrence proof. So, having proven that, for all $n$, $\langle {\cal P}(x\in A_1(i)|\rho_n)\rangle ={\cal P}(x\in A_1(i)|\rho_{u;n})$, in the limit $n\to\infty$ we find the equality: $\langle {\cal P}(x\in A_1)\rangle^{\rm univ}= {\cal P}(x\in A_1|\rho_{u})$, and of course the same holds true for the complementary region $A_2$.

\subsection{$N$-dimensional cellular structures}

The last step is to show that the above proof that ``universal measurements = uniform measurements,'' which was carried out for the $N=2$ case, i.e., for one-dimensional membranes, also holds for general $(N-1)$-membranes. The generalization is in fact straightforward, and here we will just sketch the reasoning, referring the interested reader to~\cite{AertsSassoli2014b} for more details. First of all, we need to show that also non-uniform $(N-1)$-membranes can be approximated by suitable sequences of cellular structures, in the sense that:
\begin{equation}
\label{limitntoinftygeneral}
 {\cal P}(\mbox{\boldmath$\lambda$}\in A|\rho) = \lim_{n\to\infty}{\cal P}(\mbox{\boldmath$\lambda$}\in A|\rho_n), \quad A\subset \triangle_{N-1},
\end{equation}
where $A$ is an arbitrary, not necessarily convex region of $\triangle_{N-1}$. This can be done by  considering a hyperrectangle $R_{N-1}$, containing $\triangle_{N-1}$, and extending the  domain of definition of $\rho(\mbox{\boldmath$\lambda$})$ to $R_{N-1}$, by  setting $\rho(\mbox{\boldmath$\lambda$})=0$, for $\mbox{\boldmath$\lambda$}\in R_{N-1}\setminus \triangle_{N-1}$. Of course, this domain extension will not affect the probabilities, and $A$ can be considered as belonging to the more regular domain $R_{N-1}$. 

Then, one proceeds by writing $R_{N-1}$ as the Cartesian product of one-dimensional intervals, i.e., $R_{N-1} = I_1\times I_2\cdots I_{N-2}\times I_{N-1}$, with each one of these intervals  partitioned in a way similar to what we have done previously with $\triangle_1$. This will produce a tessellation of $R_{N-1}$, in terms of $(N-1)$-dimensional hyperrectangular cells, so that we can write the integral over $R_{N-1}$ as a sum of contributions coming from each cell which is contained in $A$, plus a finite number of rests produced by the ``peripheral'' cells of $A$, so obtaining an expression which generalizes (\ref{difference-m-l}). Thus, proceeding with the same reasoning as for the one-dimensional situation, one can easily conclude that (\ref{limitntoinftygeneral}) holds for an arbitrary number $N$ of dimensions.

Secondly, we need to show that the equality 
\begin{equation}
\label{averageequalitygeneral}
 \langle{\cal P}(\mbox{\boldmath$\lambda$}\in A|\rho_n)\rangle = {\cal P}(\mbox{\boldmath$\lambda$}\in A|\rho_{u;n}), \quad A\subset \triangle_{N-1},
\end{equation}
also holds for an arbitrary number $N$ of dimensions, i.e., that our above recurrence proof can be extended to the multidimensional situation. Again, this is straightforward, considering that cellular structures, of whatever (finite) dimension, for as long as they are made of a finite number of cells, can always be reorganized to form linear structures. For this, it is sufficient to choose a method to enumerate the $i$ elementary cells contained in the complementary region of $A$, place them in order on the left side of a line, then, to follow, do the same with the remaining $n-i$ elementary cells contained in $A$, placing them on the right side of that same line. In this way, one can transform the $(N-1)$-dimensional cellular membrane, made of $n$ elementary cells, into an effective $1$-dimensional structure, made of the same number of elementary cells, and then apply to it the above recurrence proof.

So, we can conclude that the quantum mechanical Born rule, ${\cal P}(D({\bf r})\to P_{I_k})={\rm Tr}\, D({\bf r}) P_{I_k}$, giving the probabilities for the transitions $D({\bf r})\to D({\bf s}_{I_k})$, described by the  L\"uders-von Neumann formula, can be understood, for general $N$-dimensional physical entities, as resulting from a universal averaging over all possible forms of fluctuations in the measurement context, as expressed by the equality: 
\begin{equation}
\label{Born=universal}
\langle {\cal P}({\bf r}\to{\bf s}_{I_k})\rangle^{\rm univ}={\rm Tr}\, D({\bf r}) P_{I_k}.
\end{equation}
In other terms, according to (\ref{Born=universal}), the probabilities of the Born rule of quantum mechanics can be interpreted  as the probabilities of a first-order non-classical theory, describing situations in which the experimenter lacks complete knowledge not only about the interaction between the measuring system and the entity under investigation, but also about \emph{the way} such interaction is selected, every time an outcome is produced.

\section{The infinite-dimensional limit}
\label{The-infinite-dimensional-limit}

Before offering some conclusive thoughts,  we want in this section to briefly explore the status of the hidden-measurement interpretation when the number of dimensions becomes infinite. Indeed, the extended sphere-model that we have presented in the previous sections was formulated for finite-dimensional quantum entities, and it is  natural to ask if the hidden-measurement explanation remains consistent in the limit of infinite-dimensional quantum entities. 

Of course, the problem does not arise if one argues that infinite-dimensional entities do not exist as such, and are just a convenient representation of finite-dimensional entities having a very large number of degrees of freedom. From that perspective, finite quantum mechanics, and consequently also our extended Bloch representation, would be appropriate to describe the reality of whatever quantum entity, at a fundamental level.

Conversely, one can adopt the viewpoint which consists in saying that, apart pure quantum observables, like spin observables, which are intrinsically finite-dimensional, measurements with a finite number of outcomes are only approximations, as quantum entities would be genuinely infinite-dimensional. In practice, these approximations would be inevitable, considering that in all  measurement situations the set of states experimentally accessible is typically finite. But at a fundamental level, being the reality of a quantum entity infinite-dimensional, it is  licit to ask if the hidden-measurement explanation can still be maintained in this case. 

In that respect, we can observe that the hidden-measurement model that we have presented in this article is valid for a general quantum entity and for an arbitrary number of dimension $N$. Therefore, it will necessarily  remain valid also asymptotically, as $N\to\infty$. What we mean to say is that the existence of a hidden-measurement representation for the quantum probabilities of infinite-dimensional entities only depends on the possibility of showing that the physics of infinite-dimensional quantum entities can  be recovered by taking the $N\to\infty$ limit of suitably defined finite-dimensional ones. If this is possible, then a hidden-measurement explanation for quantum probabilities can certainly be advocated also for infinite-dimensional entities. 

It should be mentioned, however, that finite-dimensional quantum mechanics cannot be considered as a straightforward discretization of the corresponding infinite-dimensional continuous  formulation. This can be easily seen by considering the canonical commutation relation between  position and momentum operators: $[Q,P]=i\mathbb{I}$. Such a relation is  impossible in a $N$-dimensional Hilbert space, as is clear that, if $Q$ and $P$ are two finite-rank operators, by the cyclicity of the trace we have ${\rm Tr}\, [Q,P]=0$, whereas ${\rm Tr} \, i\mathbb{I} = iN\neq 0$. So, relations like the canonical commutation one, and others~\cite{Fritz2013}, cannot hold  in the finite-dimensional case, implying, among other things, that different finite-dimensional constructions can a priori be considered to be the finite-dimensional counterpart of an infinite-dimensional entity. 

But as far as we are concerned, it is sufficient for us to show that, given a quantum entity whose operator-state $D$ acts on an infinite-dimensional (separable) Hilbert space ${\cal H}$, and given a projection operator $P$ such that $P{\cal H}$ is an infinite-dimensional subspace of ${\cal H}$, the probability ${\rm Tr}\, DP$ for the transition $D\to  {PDP\over {\rm Tr}\, DP}$ can always be expressed as the limit:
\begin{equation}
\label{limitDP}
{\rm Tr}\, DP = \lim_{N,M\to\infty\atop N\geq M} {\rm Tr}\, D_NP_M,
\end{equation}
where $D_N$ is the operator-state of a $N$-dimensional entity, and $P_M$, $M\leq N$, a projection operator onto a $M$-dimensional subspace of that entity's Hilbert space. Indeed, if (\ref{limitDP}) holds in general, then, considering that each probability ${\rm Tr}\, D_NP_M$ admits a hidden-measurement representation, the same will necessarily be true for the limit probability ${\rm Tr}\, DP$, whose hidden-measurement representation  then coincides with the limit representation of the sequence of probabilities ${\rm Tr}\, D_NP_M$. 

To show this, we write ${\cal H} =P{\cal H}\oplus (\mathbb{I}-P){\cal H}$, and let $\{|a_1\rangle, |a_2\rangle,\dots\}$ and $\{|b_1\rangle, |b_2\rangle,\dots\}$ be two orthonormal bases of $P{\cal H}$ and $(\mathbb{I}-P){\cal H}$, respectively (separable Hilbert spaces admit countable orthonormal bases). We define the finite-rank orthogonal projection operator $P_M\equiv\sum_{i=1}^M|a_i\rangle\langle a_i|$, which converges strongly to $P$, as $M\to\infty$. We also define the finite-rank orthogonal projection operator $P_N\equiv \sum_{i=1}^M|a_i\rangle\langle a_i|+ \sum_{i=1}^{N-M}|b _i\rangle\langle b_i|$, and the ``compressed'' finite-rank operator $D_N\equiv   {P_NDP_N\over {\rm Tr}\, DP_N}$. We observe that $D_N$ is manifestly self-adjoint, of unit trace, and positive semidefinite. In other terms, it is a bona fide operator-state of a $N$-dimensional entity with Hilbert space $P_N{\cal H}$. 

Thus, the probability ${\rm Tr}\, D_NP_M$ admits a hidden-measurement representation, and all we need to show is that ${\rm Tr}\, D_NP_M$ converges to ${\rm Tr}\, DP$, as $N$ and $M$ tend to infinity, with $N\geq N$. For this, we only need to observe that ${\rm Tr}\, DP_N \to {\rm Tr}\, D =1$, as $N\to\infty$, so that ${\rm Tr}\, D_NP_M = { {\rm Tr}\, DP_M \over {\rm Tr}\, DP_N}$ tends  to ${\rm Tr}\, DP_M$, as $N\to\infty$, which in turn tends to ${\rm Tr}\, DP$ as $M\to\infty$. 

The above show that we can always understand the transition probabilities of an infinite-dimensional entity as the limit of a sequence of probabilities associated with finite-dimensional entities, so that a hidden-measurement interpretation of quantum probabilities can also be attached, in principle, to infinite-dimensional entities. Of course, the above argument is an indirect one, as it does not construct explicitly the asymptotic hidden-measurement representation. A more explicit construction showing the possibility of extending the hidden-measurement explanation to experiments with an infinite set of outcomes was given some years ago by Coecke~\cite{Coecke1995b}. In his derivation, and different from what we have done in this article, the dimension of the space of the \mbox{\boldmath$\lambda$}-variables, defining the available hidden-interactions, was taken to be equal to the interval $[0,1]$, independently of the number $N$ of outcomes.

Therefore, there is no doubt that the explanation of quantum probabilities in terms of a lack of knowledge about a deeper reality constituted by the hidden-measurements can be maintained also for infinite-dimensional physical entities. However, if we can confidently affirm that, for the finite-dimensional situation, our extended Bloch sphere constitutes the optimal setting for representing the hidden-interactions, what would be the optimal representation for infinite-dimensional systems remains, we believe, an open problem.

\section{Concluding remarks}
\label{Concluding}

The longstanding measurement problem of quantum mechanics is still considered today, by most physicists, to be unresolved. The problem contains two different issues. The first one, usually non-controversial,  is our limitation in directly probing a quantum measurement process, to see what really goes on, ``offstage,'' when it is executed. The second issue, still controversial, is about explaining the nature of the indeterminism which is subtended by a quantum measurement, independently of our (in)ability to directly access the process (apart its outcomes).   

In this article, we have proposed what we think is a convincing and very general solution to the measurement problem of quantum mechanics, in terms of a hidden-variable modelization where the variables are not associated with the states of the entity under investigation (as this would go against the no-go theorems), but with the interactions between the entity and the measuring system, which are actualized in different ways at every run of the measurement, because of the presence of irreducible fluctuations (so that a quantum measurement would  in fact be a multi-measurement, expression of an average over different measurements, which we have called a universal measurement).

Although the hidden-measurement interpretation was already proposed by one of us in the eighties of last century, it has only today reached a rather complete formulation, that we have tried to convey for the first time in this article, in a self-consistent way. It is our hope that this more general and transparent presentation (both mathematically and conceptually) will be able to promote a non-negligible spread of the universal ideas it contains.

Albeit for the time being, in microphysics, the existence of the hidden-interactions remains hypothetical, we know it is possible to construct real macroscopic machines whose behavior is genuinely quantum, or quantum-like, and for which their existence is certainly not a hypothesis, but a fact~\cite{Aerts1986,Aerts1993,Aerts1998b,Aerts1999,Sassoli2013b,Sassoli2013c,Sassoli2013d}. As we mentioned in Sec.~\ref{Nonspatiality}, this is also the case in human cognition experiments, where these interactions result from subconscious ``non-logical'' intrapsychic processes, which although cannot be easily discriminated at the conscious level, not for this can be considered less real~\cite{BusemeyerBruza2012,Aerts2009,AertsSassoli2014a,AertsSassoli2014b}. Let us also mention that the hidden-measurement paradigm has allowed to explain the origin of the violation of Bell's inequalities, challenging also in this case the widespread belief that quantum structures would  only be present at the microscopic level of our reality~\cite{Aerts1984,Aerts1990,Aerts1991,Aertsetal2000,Sassoli2014b}. 

The above is important to emphasize, as the majority of interpretations of quantum probabilities rely on assumptions which remain admittedly very speculative, whereas the hidden-measurement interpretation has the advantage that we already know that it validly describes the quantum-like behavior of macroscopic systems (including human minds), when they are acted upon according to specific protocols. On the other hand, we don't have concrete examples of parallel universes, quantum potentials, etc. This, in our opinion, is an important added value of the interpretation, which can claim to base its explanations on the existence of concrete models, implementing all the concepts put in place.

Also, independently of one's personal position regarding the reality of the hidden-measurement mechanism, the extended Bloch representation  that we have presented in this work, which we plan to further investigate in the future, has the considerable advantage of making a measurement perfectly visualizable, and therefore understandable, as a mechanistic-like process which takes place in the Euclidean $\real^{N^2-1}$ space. Visualizing (and more generally imaging) the unfolding of a quantum process is something which is usually believed to be impossible to do: another preconception that our model clearly falsifies. 

Interestingly, we can also use this visualizability of the measurement process to elucidate, in turn, the structure of the set of states of a quantum entity. To give an example, in Sec.~\ref{Transition} we have proved that, given two unit vectors ${\bf n}_i$ and ${\bf n}_j$, representative of two orthonormal vector-states, there cannot be any vector representative of a vector-state between them, on the same plane where they belong. To understand the reason of this unexpected constraint, let us consider Fig.~\ref{Singletriangle}, which shows, for the $N=3$ case, a $2$-membrane (an equilateral triangle) representative of a measurement. If there cannot be unit vectors representing states lying between the three vertex-vectors, this  means that, on that plane, the only vectors which are representative of states are those belonging to the membrane's triangle.

To understand why it is so, we observe that every vector representative of a state has to be able to produce outcomes in a measurement. For this, it has either to be already a vector on the membrane's triangle representing the measurement in question, or, in case it would not, be a vector orthogonally projectable onto it. Now, if we consider a unit vector different from the three vertex-vectors of the measurement's triangle, obviously it will not belong to the latter. However, if it belongs to the same plane of the measurement's triangle, it will not be projectable onto it. In other terms, such a vector would not be able to produce outcomes in the measurement represented by that specific membrane's triangle, and consequently cannot be a bona fide state. 

As a last remark, we emphasize that the model we have presented does not only constitute an interpretation of the mathematical formalism of quantum mechanics: it also constitutes an extension of such formalism. This because, first of all, our extended sphere model, contrary to the Hilbertian model, allows for a full representation of the measurement process, in terms of uniform $(N-1)$-membranes. If such representation is taken seriously, then our model suggests that also density matrices (operator-states) should be generally considered as pure states, and this is an experimentally testable prediction that cannot be made by standard quantum mechanics. 

The model constitutes an extension of the Hilbertian formalism also because it allows to describe more general typologies of measurements (still of the first kind), associated with non-uniform $(N-1)$-membranes, representative of measurement contexts in which the effective fluctuations are non-uniform. These non-uniform fluctuations may give rise to non-Kolmogorovian and non-Hilbertian probability models, describing experimental situations characterized by intermediate conditions of lack of knowledge, or control, associated with less \emph{robust} statistics of outcomes, described by generalized Born rules~\cite{AertsSassoli2014b}. Quantum mechanics, in that respect, appears to be a first order non-classical theory, describing experimental situations of maximal lack of knowledge, where all possible ways of actualizing a hidden-interaction are allowed by the experimental protocols. 

Finally, the model extends the Hilbertian formalism as it distinguishes in a measurement what comes from the structure of the set of states, and what comes from the ``potentiality region'' of contact between the measured entity and the measuring system. The former is a geometric-like, deterministic aspect, which strongly depends on the nature of the entity under investigation, whereas the latter is a generally indeterministic aspect, describing how the ``actual breaks the symmetry of the potential,''  depending on how we act on the entity in question. Standard quantum mechanics only describes the specific situation of entities whose states have a Hilbertian structure, and whose measurements are characterizable in terms of effective uniform fluctuations. However, more general structures for the set of states, and more general forms of fluctuations, can be defined, and could better describe certain layers of our vast reality, and how we experience and experiment them.

\appendix 

\section{The volume of simplices}
\label{simplexes}

The $(N-1)$-simplex $\triangle_{N-1}$, generated by the $N$ normal (but non-orthogonal) vectors $\{{\bf n}_1,\dots,{\bf n}_N\}$ has $N$ vertices and  ${N(N-1)\over 2}$ edges. To determine its  Lebesgue measure $\mu(\triangle_{N-1})$, we can proceed as follows: we define the $(N+1)\times (N+1)$ matrix $B =(b_{ij})$, where $b_{ij}\equiv \| {\bf n}_i-{\bf n}_j\|^2$. Then, considering the matrix $\hat{B}$ obtained from $B$ by bordering $B$ with a top row $(0,1,\dots,1)$ and a left column $(0,1,\dots,1)^T$, from Cayley-Menger's determinant formula we have:
\begin{equation}
\mu(\triangle_{N-1}) ={\sqrt{|\det \hat{B}|}\over 2^{N-1\over 2}(N-1)! }.
\end{equation}
Considering that $b_{ij}={2N\over N-1}\delta_{ij}$, this gives: 
\begin{equation}
\mu(\triangle_{N-1}) ={\sqrt{N \left({2N\over N-1}\right)^{N-1}}\over 2^{N-1\over 2}(N-1)! } ={\sqrt{N-1}\over (N-1)!}\left({N\over N-1}\right)^{N\over 2}\!\!\!\!\!\! .
\label{areaformula}
\end{equation}

Let us also calculate the radius $r_N$ of the inscribed ball. For this, we can use:
\begin{equation} 
\mu(\triangle_{N-1}) =r_{N}\, {\mu(\partial\triangle_{N-1})\over N-1},
\end{equation}
which is the generalization of the standard formula giving the area of a triangle as the product of the radius of the incircle times the semiperimeter.  Considering that $\triangle_{N-1}$ is made of $N$ sub-simplexes $\tilde{\triangle}_{N-2}$ of dimension $N-2$, we can write:
\begin{equation} 
\mu(\triangle_{N-1}) =r_{N}\, {N\over N-1} \mu(\tilde{\triangle}_{N-2}).
\label{formulatrianglebis}
\end{equation}
To calculate $\mu(\tilde{\triangle}_{N-2})$, we observe that $\tilde{\triangle}^i_{N-2}$ is the simplex generated by the $N-1$ vectors $\{{\bf n}_1, \dots, {\bf n}_{i-1},{\bf n}_{i+1}, \dots{\bf n}_{N}\}$, which doesn't have the same Lebesgue measure as the simplex $\triangle_{N-2}$ (thus explaining the ``tilde'' in the notation). Indeed, as explained in Sec.~\ref{Transition}, the length of the edges of $\triangle_{N-1}$ (the length of the difference between two vertex-vectors) varies with $N$, and is equal to $\sqrt{2N\over N-1}$. This means that the sub-simplexes $\tilde{\triangle}^i_{N-2}$, generated by the above mentioned $N-1$ vectors, have edges of length $\sqrt{2N\over N-1}$, that is, the same length as the edges of $\triangle_{N-1}$, whereas a simplex $\triangle_{N-2}$ has edges of length $\sqrt{2(N-1)\over N-2}$. Therefore, we have:
\begin{eqnarray}
\mu(\tilde{\triangle}^i_{N-2}) &=& \left(\sqrt{2N\over N-1}\over \sqrt{2(N-1)\over N-2} \right)^{\!\! N-2}\!\!\!\!\!\!\!\!\!\mu(\triangle_{N-2}) =\left({\sqrt{N(N-2)}\over N-1} \right)^{\!\! N-2} \!\!\!\!{\sqrt{N-2}\over (N-2)!}\left({N-1\over N-2}\right)^{N-1\over 2}\\
&=& {\sqrt{N-1}\over (N-2)!}\left(N\over N-1\right)^{\!\!{N-2\over 2}}\!\!\!\!\!\!\!\!\!.
\label{triangletilde}
\end{eqnarray}
Replacing (\ref{triangletilde}) into (\ref{formulatrianglebis}), we thus obtain 
\begin{equation} 
\mu(\triangle_{N-1}) =r_{N}{N\over N-1}{\sqrt{N-1}\over (N-2)!}\left(N\over N-1\right)^{\!\!{N-2\over 2}}\!\!\!\!\!\!\!\!\! .
\label{formulatriangletris}
\end{equation}
Finally, comparing (\ref{formulatriangletris}) with (\ref{areaformula}), after a short algebra one finds: $r_{N}={1\over N-1}$.

\end{document}